\documentclass[aps,pre,twocolumn,amsmath,amssymb,aps,floatfix, longbibliography,superscriptaddress]{revtex4-1}
\usepackage[latin9]{inputenc} % shouldn't need if file created with UTF 8 encoding
\usepackage{graphicx}
\usepackage{dcolumn}
\usepackage{bm}
\usepackage{amssymb}
\usepackage{amsmath}

\begin{document}

\title{Interactions between two adjacent convection rolls in turbulent Rayleigh-B{\'e}nard convection} 

\author{Eric Brown}
\affiliation{Eric Brown Labs LLC, New Haven, CT 06511}
\affiliation{Department of Physics, Southern Connecticut State University, New Haven, CT 06515}
\email{ericmichealbrown@gmail.com}
\author{Dandan Ji}
\affiliation{Department of Physics, Yale University, New Haven, CT 06520}

\date{\today}

\begin{abstract}
We seek to develop a low dimensional model for the interactions between horizontally adjacent turbulent convection rolls.  This was tested in Rayleigh-B{\'e}nard convection experiments with two  adjacent cubic cells with a partial wall in between.  Observed stable states include both counter-rotating and co-rotating states for Rayleigh number $7.6\times10^7 <$ Ra $ <3.5\times10^9$ and Prandtl number 6.41.  The stability of each of these states and their dynamics can be modeled low-dimensionally by stochastic ordinary differential equations of motion in terms of the orientation, amplitude, and mean temperature of each convection roll.  The form of the interaction terms is predicted based on an effective turbulent diffusion of temperature between the adjacent rolls, which is projected onto the neighboring rolls with sinusoidal temperature profiles.  With measurements of a constant coefficient for effective thermal turbulent diffusion, quantitative predictions are made for the nine forcing terms which affect stable fixed points of the co- and counter-rotating states for $5.5\times10^8 < $ Ra $ < 3.5\times10^9$.  Predictions are found to be accurate within a factor of 3 of experiments.  This suggests that the same turbulent thermal diffusivity that describes macroscopically averaged heat transport also controls the interactions between neighboring convection rolls.  The surprising stability of co-rotating states is due to the temperature difference between the neighboring rolls becoming large enough that the heat flux between the rolls stabilizes the temperature profile of aligned co-rotating states.  This temperature difference can be driven with an asymmetry, for example, by heating the plates of the two cells to different mean temperatures.  When such an asymmetry is introduced, it also shifts the orientations of the rolls of counter-rotating states in opposite directions away from their preferred orientation, which is otherwise due to the geometry of the cell.  As the temperature difference between the plates of the different cells is increased, the shift in orientation increases until the counter-rotating states become unstable, and only co-rotating states are stable.  At very large temperature differences between cells, both the counter-rotating and predicted co-rotating state become unstable --  instead we observe a co-rotating state with much larger temperature difference between the rolls that cannot be explained by turbulent thermal diffusion.  Spontaneous switching between co-rotating and counter-rotating states is also observed, including in nominally symmetric systems.  Switching to counter-rotating states occurs mainly due to cessation (a significant weakening of a convection roll), which reduces damping on changes in orientation,  allowing the orientation to change rapidly due to diffusive fluctuations.  Switching to co-rotating states is mainly driven by smaller diffusive fluctuations in the orientation, amplitude, and mean temperature of rolls, which have a positive feedback that destabilizes the counter-rotating state.

\end{abstract}

\maketitle

%PhySH: fluid dynamics: RBC, turbulence, stochastic differential equations

\section{Introduction}

%challenge - overview 
While turbulent flows are often thought of as irregular and erratic, large-scale coherent flow structures are commonplace in turbulence.  An example is convection rolls driven by buoyancy in natural convection.  Such structures and their dynamics can play a significant role in heat and mass transport.   A particular challenge that is the focus of this manuscript is to develop a model for how these large-scale flow structures interact with each other, for example to result in neighboring convection rolls that are counter-rotating or co-rotating.  

%Model system
We investigate this in the model system of turbulent Rayleigh-B\'enard convection.  In Rayleigh-B\'enard convection, a  fluid is heated from  below and cooled from above to generate buoyancy-driven flow \cite{AGL09,LX10}. This system exhibits convection rolls which are robust large-scale coherent structures that retain a similar organized flow structure over a long time. For example, in containers  of aspect ratio near 1, a large-scale circulation (LSC) forms. This LSC consists of localized blobs of coherent fluid known as plumes.  The plumes collectively form a single convection roll in a vertical plane  that can be identified by averaging over the flow field or timescales longer than the circulation period \cite{KH81}.  This LSC spontaneously breaks the symmetry of symmetric containers, but turbulent fluctuations cause the LSC orientation $\theta_0$ in the horizontal plane to meander spontaneously and erratically, and allow it to sample different orientations to recover the symmetry over long times \cite{BA06a}.  While the LSC exists nearly all of the time, on rare occasions these fluctuations lead to spontaneous cessations  followed by reformation of the LSC \cite{BA06a, XX07}. The LSC exhibits oscillation modes \cite{HCL87, SWL89, CGHKLTWZZ89, CCL96, TSGS96, CCS97, QYT00, QT01b, NSSD01, QT02, QSTX04, SXT05,TMMS05}, which in circular cylindrical containers consists of twisting and sloshing \cite{FA04, XZZCX09, ZXZSX09, BA09}, and  at some aspect ratios a jump-rope-like mode \cite{VHGA18, HSA22}.  %Earth's Coriolis force causes a  rotation of the LSC orientation \cite{BA06b,ZLW17, SLZ16}. 
%sim cessations: MVE11

%geoemetry-dependence
 The qualitative behavior of the LSC depends on the geometry of the cell, which is necessary to account for if we are to understand the interactions between neighboring rolls in some geometry.  In containers with rectangular horizontal cross-sections, the preferred alignment of the LSC orientation $\theta_0$ is along the  longest diagonals, and the LSC orientation can spontaneously switch between adjacent corners \cite{LE09, SBHT14, BJB16, FNAS17, GKKS18, VSFBFBR16, VFKSSV19}.  A regular oscillation of $\theta_0$ can occur between nearest-neighbor diagonals in a non-square rectangular cross-section \cite{SBHT14}.   In a cubic container, the oscillation structure of the LSC corresponds to an advected oscillation mode with one oscillation per LSC turnover period \cite{JBB20}, and an oscillation in the shape of the temperature profile that does not occur in a circular cross-section cell \cite{JB20b}.
 %A similar oscillation in $\theta_0$ was found in a cubic cell where the preferred orientation aligns with grooves in the top and bottom plates \cite{FNAS19}, which -- similar to the tilted cell -- can at least qualitatively be described as an oscillation around a preferred orientation at a potential minimum.  
 
 %low D
In principle, the Navier-Stokes equations describe fluid flow, but they are impractical to solve for such complex turbulent flows, so low-dimensional models are desired. It  has long been  recognized that the states and dynamics of a single large-scale coherent structure are similar to those of low-dimensional dynamical systems  models \cite{Lo63} and stochastic ordinary differential equations   \cite{BA07a, TB07, TMS14, RMBM15}.  Low-dimensional  models can potentially describe parameters of stable states, dynamics such as oscillation modes, and behavior of transitions between states.  While low-dimensional models lack detail of smaller structures, they have the advantage of being much simpler to solve and understand the behavior because they involve simpler equations -- such as ordinary differential equations instead of partial differential equations.  

%comparison of models - single roll
There are several low-dimensional models for single roll LSCs. Early models tried to characterize flow reversals in simplified flow in a two-dimensional plane  \cite{SBN02, Be05, FGL05} %similar to Lorenz' famous model for chaos atmospheric convection \cite{Lorenz63}
However, these two-dimensional models could not characterize more complex three-dimensional dynamics with motion in $\theta_0$ such as reorientations and oscillation modes. 
  %In one case, an approximate analytical model was applied to ellipsoidal containers, which predicted an oscillation with a geometry-dependent restoring force \cite{RPTDGFL06}, however it relied on solutions to flows in ellipsoidal containers, so is not straightforward to generalize to flows in other geometries.  
Some models are obtained by transforming high-dimensional fluid velocity field data (usually from direct numerical simulation) and reducing it to lower-dimensional models consisting of a few highest-energy Fourier modes or eigenmodes.  These models have been able to characterize the detailed shape of the LSC and its dynamics, including flow reversals in two dimensions \cite{CV11,PS15} spontaneous corner-switching and oscillations in cubic cells \cite{GKKS18, VFKSSV19}, and the twisting, sloshing, and jump-rope oscillation modes of the LSC \cite{HSA22}.   Because these models are obtained from high-dimensional data, they are descriptive in higher detail, but the models are not formulated in such a way to  predict flows when detailed data is not already available.  We desire a low-dimensional modeling technique that can be predictive and generalizable to other systems with more limited or no experimental input required.

%BA model
We build off the low-dimensional model of Brown \& Ahlers, where model terms are derived from approximations of the Navier-Stokes equations \cite{BA08a}.   The model consists of a pair of stochastic differential equations for the LSC, in terms of the orientation $\theta_0$ and amplitude $\delta$.  The model of Brown \& Ahlers and its extensions have successfully described most of the known dynamical modes of the LSC in including the meandering, cessations, and twisting and sloshing oscillation modes described above \cite{BA08a, BA08b, BA09,ZLW17, SLZ16}, with the exception of the jump rope mode. The combination of twisting and sloshing oscillation modes  \cite{FA04, XZZCX09, ZXZSX09, BA09} can alternatively be described in this model as a single advected oscillation mode, with two oscillation periods per LSC turnover period in a circular cross-section \cite{BA09}, or with one oscillation period per LSC turnover period in a cubic cross-section which is excited by a potential $V_g$ due to the shape of the container \cite{JBB20}.  This potential $V_g$ can be predicted as a function of container cross-section geometry without experimental input \cite{BA08b, SBHT14, JB20a}.  This same  potential $V_g$ also explains the preferred orientation along diagonals of a rectangular cross-section container, the oscillations between diagonals \cite{SBHT14}, and the stochastic switching between diagonals \cite{LE09, BJB16, FNAS17, GKKS18, VSFBFBR16, VFKSSV19}.  This model requires experimental measurements of two diffusivity parameters that characterize the strength of turbulent fluctuations in $\theta_0$ and the LSC strength, but these are relatively simple parameters that can be obtained from short term measurements of a single state.  Predictions of oscillation frequencies, average rates of stochastic switching between states, widths of probability distributions, and state boundaries are typically accurate within a factor of 3 \cite{BA07a, BA08a, BJB16, JB20a}, but can be more accurate when more fit parameters are used \cite{AAG11}.  

%interacting convection rolls
 While systems of aspect ratio close to one tend to have a single convection roll, horizontally extended convection systems tend to consist of multiple convection rolls, each  of aspect ratio on the order of one, arranged side-by-side, typically counter-rotating relative to their neighbors \cite{BES10, PSSL12}.   Counter-rotating behavior is claimed to be prevalent in nature, for example, in textbook pictures of convection rolls in the atmosphere, in the convection layer of stars, or planetary cores.  However, a few  simulations have been able to produce co-rotating states in horizontally adjacent rolls in non-turbulent convection, with lateral heating \cite{PMELB04},  %This corresponded to a state that is more like one large convection roll with flow stronger near the outer walls remaining in the same direction along the entire top and bottom plates, with weaker vertical flow columns in the middle.  
 or with an inclination relative to gravity of 0.01 rad  \cite{MBAK19}.    Vertically extended systems  often have counter-rotating rolls stacked on top of each other \cite{ZTS20}.  A turbulent convection experiment with two fluids, one on top of the other, found a convection roll in each fluid with two stable states, one where the rolls are counter-rotating, and one where the rolls are co-rotating \cite{XX13}, with rare stochastic switching between the two states.  A variation on stacked co-rotating rolls was observed in simulations of a non-turbulent vertical channel where  vertical flow in opposite directions occurred along opposite side walls.  In this case, the co-rotating rolls are forced by the opposite vertical flows on either side \cite{GSPXQT13}.  

Two-dimensional theories using linear stability analysis in the non-turbulent regime have been able to predict the existence of stable counter- and co-rotating states. In one case, a lateral heating was shown to provide a forcing to produce stable co-rotating states in horizontally neighboring rolls \cite{PMELB04}.  Another model considered vertically stacked rolls, and showed stable counter-rotating states where the coupling was dominated by viscous forces, as well as stable co-rotating states  where the coupling was mainly through vertical buoyancy forces \cite{PB03}.  Since these earlier models were focused on  two-dimensions \cite{PB03, PMELB04}, they can identify co-rotating and counter-rotating states, but they are missing the orientation of rolls, which is important because stability in three dimensions requires stability in the orientation as well as the flow strength, and many dynamics of rolls involve changes in orientation \cite{FA04, BA06a, BA06b, VHGA18, AYSSHVE22}.  These previous models also focused on linear stability analysis of laminar flow, but many natural flows are turbulent, well beyond the range where linear stability analysis rigorously applies.  It has yet to be determined whether three-dimensional systems of interacting turbulent convection rolls can be captured by low-dimensional models, which is the primary goal of this work. 

 %specific problem
  We hypothesize that the low-dimensional model of Brown \& Ahlers \cite{BA08a,BA08b, BJB16, JB20a, JBB20} can be extended to systems of multiple convection rolls using an existing set of ordinary differential equations of motion for each roll, and adding interaction terms to the equations of motion for each neighboring roll.   We seek to test whether a low-dimensional model can predict or describe the preferred states (e.g.~counter-rotating and/or co-rotating), their parameter values (e.g.~orientation, flow strength), and their dynamics (e.g.~how does switching between states occur).  We also seek to understand the physical origin of the interaction terms, in particular due to turbulent thermal diffusion across the interface between the rolls so that a general predictive model can be made.  This is in the spirit of low-dimensional models with heat transport in one dimension that have already proven effective, for example, to explain an oscillation in the shape of the temperature profile with an orientation-dependent vertical heat transport \cite{JB20b}, and to explain thermal waves with a height-dependent vertical heat transport \cite{UKMSS22}.   In principle, if we can write equations for convection rolls with interaction terms for a neighbor, results from two convection rolls can be extended to systems of more rolls assuming there are no significant interaction terms that involve three or more rolls.

%summary
The remainder of this manuscript is organized as follows.     Section \ref{sec:methods} explains the experimental apparatus, and methods used to characterize the LSC. Section \ref{sec:modelreview} summarizes the existing low-dimensional model of Brown \& Ahlers for a single LSC to use as a starting point.  Section \ref{sec:preferredstates} describes initial observations of counter-rotating and co-rotating states of two neighboring rolls.  Section \ref{sec:interactionmodel} presents a physical derivation of a model for neighboring roll interactions based on effective turbulent thermal diffusion.  This model is tested in Sec.~\ref{sec:interactionmodeltest} for the symmetric case where the control of both cells is nominally identical.  Section \ref{sec:DeltaTm} extends this model to the cases with asymmetric forcing on the two cells due to  a  difference in mean temperature of the plates, to better understand co-rotating states, and to test some of the model terms.  Section \ref{sec:switching} presents observations of spontaneous switching between co-rotating and counter-rotating states, and explains how these can be understood with the model.

\section{Methods}
\label{sec:methods}

\subsection{Experiment setup}

\begin{figure}
\includegraphics[width=0.375\textwidth]{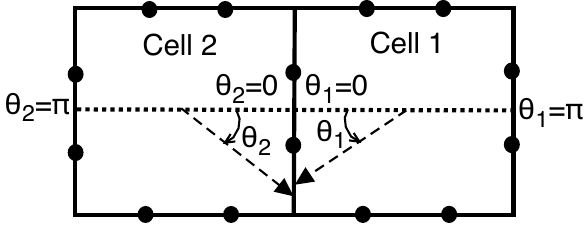} 
\caption{A schematic of the  experimental setup to measure interactions between two neighboring convection rolls, viewed from above.  The LSC orientation in each cell is measured as the angle between the flow direction of the LSC near the bottom plate (dashed line) relative to the axis going through the center of both cells (dotted line), such that aligned counter-rotating states correspond to $\theta_1=\theta_2$, and aligned co-rotating states correspond to $\theta_1 = \theta_2+\pi$ rad.  
 Thermistor locations on the sidewall are indicated by solid circles.
%The length of the circulation plane $D(\theta_0)$ across a horizontal cross-section determines the model potential
}
\label{fig:setup}
\end{figure}   

%summary
The experimental apparatus is the same one used in \cite{JB20a}, with the relevant details and modifications presented here. 
The apparatus consists of two adjacent cells as illustrated in Fig.~\ref{fig:setup}.  Each cell is nearly cubic with  $H=20.32$ cm and horizontal lengths $L=20.02\pm0.03$ cm.   Since two rolls may be preferred at an aspect ratio 2 without a wall in between \cite{PSSL12}, %2D sim 
 the apparatus is modified from \cite{JB20a} with a partial opening in the middle wall to enforce a flow with two convection rolls that can interact with each other.

%plates
To control the temperature difference $\Delta T$ between the top and bottom of each cell, water is circulated through top and bottom plates.   Each top and bottom plate is controlled by its own temperature-controlled water bath so that $\Delta T$ could be controlled in each cell more precisely, and a difference $\Delta T_m$ could be imposed between the mean plate temperatures of the two cells when desired. The plates are aluminum, with double-spiral water-cooling channels as in \cite{BFNA05}, except that each plate has its own double-spiral, and the inlet and outlet of each plate were adjacent to minimize the  spatial temperature variation within the plates.   The baths pump water in a pattern symmetric around the middle wall to minimize temperature asymmetries between the two cells.  Each plate  has 5 thermistors  to record $\Delta T$ and $\Delta T_m$, with one thermistor at the center and four on the diagonals, halfway between the center and each corner of the plate.   The standard deviation of temperatures within each plate is $0.005\Delta T$  \cite{JB20a}.  The top and bottom plates are parallel within 0.06$^{\circ}$.

\begin{figure}
\includegraphics[width=0.35\textwidth]{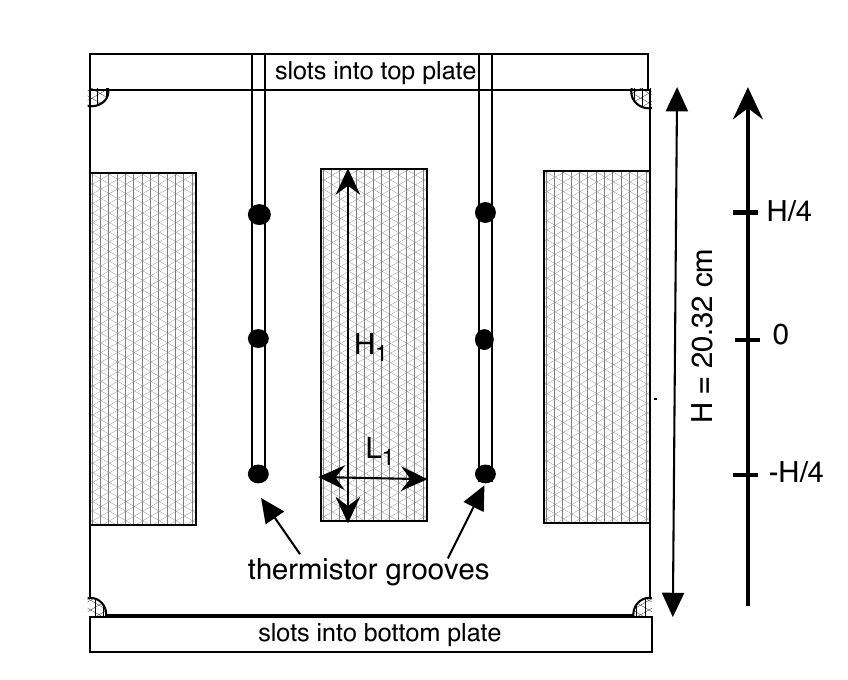} 
\caption{A schematic of the middle wall (in between the two cells).  Cross-hatched sections indicate cutouts.  Half of the area of the middle wall is cut out to enforce a flow with two convection rolls while allowing interaction between the neighboring convection rolls.  Thermistor locations are indicated by solid circles.
}
\label{fig:mid_wall}
\end{figure}   

 %sidewalls 
The sidewalls of the cubic cells are plexiglas to thermally insulate the cell from the surroundings.  The outer sidewalls have a thickness of 0.55 cm, and are further insulated from the room by foam insulation.  The middle wall is shared in between the two flow cells.   The middle wall has a thickness of $0.91$ cm to thermally insulate the cells from each other, and extends 2.2 cm into the gap between the top and bottom plates to thermally insulate the plates of different cells from each other.   The middle wall has gaps in it to allow flow in the two cells to interact with each other.  Three sections of height $H_1=13.6\pm0.1$ cm by $L_1=4.9\pm0.1$ cm are cut out of the middle wall, as shown in Fig.~\ref{fig:mid_wall}.  A quarter circle with radius $0.35\pm0.05$ cm was cut out of each corner of the cell  to allow gas bubbles to escape the cell during a degassing process to prepare the working fluid for experiments.  This results in a fraction $A=0.50$ of the middle wall open to allow interaction between the two cells. All of the cutouts are symmetric around the mid-height of the cell and symmetric from left to right. 

%modifications
Two vertical grooves were cut in the middle wall as shown in Fig.~\ref{fig:mid_wall} to place thermistors and run their wiring out the top of the cell.  The remaining area of the grooves was filled with epoxy to keep the walls as flat as possible, as explained in \cite{JB20a}.  Detailed internal dimensions are given in \cite{JB20a}, where it is shown that their effect on the symmetry of the flow is negligible compared to the effects of the cubic shape, as well as compared to effects of non-uniformity of the heating and cooling plate temperatures. 
%The middle wall used for reported data is referred to as the symmetric middle wall. 
%The second version of the middle wall is referred to as the top half, as it covers the top half of the interface only, standing on two steel rods of diameter 0.32 cm. 

  %insulation 
  The apparatus was further insulated from the room  as in \cite{BFNA05} by surrounding it with 5 cm thick closed-cell foam,  which itself was surrounded on the sides by a copper shield with water   circulating through a pipe welded to the shield.  The circulating water temperature was set to match the mean temperature $\langle T_0 \rangle$ averaged over the two cells, with a standard deviation of $0.006^{\circ}$ K. The shield was surrounded by another layer of 2.5 cm thick open-cell foam.

%thermistors
To measure the LSC, thermistors were mounted in the sidewalls as in \cite{BA06b}.   There are three rows of thermistors: at heights $+H/4$, $0$, and $-H/4$ relative to the mid-height of the cell, as shown in Fig.~\ref{fig:mid_wall}.  In each row, eight thermistors are equally spaced in the angle $\theta$ around the mid-plane, as illustrated in Fig.~\ref{fig:setup}.  The coordinate $\theta$ in the horizontal plane is measured relative to the axis going through the center of both cells, as illustrated in Fig.~\ref{fig:setup}.    The LSC orientation $\theta_1$ when corresponding to the flow direction near the bottom plate when viewed from above is measured counter-clockwise in cell 1 and $\theta_2$ is measured clockwise in cell 2, so that counter-rotating states correspond to $\theta_1=\theta_2$ when the orientation vectors align head-to-head, and co-rotating states correspond to $\theta_1 = \theta_2+\pi$ rad when they align head-to-tail.

%tilt
%Tilt angles $\beta$ were measured with a digital protractor placed on the top plate with 5 repetitions along both perpendicular axes, which results in a standard deviation of the mean of $0.03^{\circ}$ on $\beta$.   
The cell was leveled in the direction perpendicular to the axis going through the center of both cells, with an uncertainty of $0.03^{\circ}$.  For some experiments, the cell was intentionally tilted by an angle $\beta$ relative to the level cell along the axis going through the center of both cells to introduce a forcing from buoyancy that breaks the symmetry of the two cells.
% measured with digital level

%parameter values
The working fluid was degassed and deionized water with mean temperature of 23.0$^{\circ}$C, for a Prandtl number $Pr = \nu/\kappa = 6.41$ where   $\nu=9.36\times10^{-7}$ m$^2$/s is the kinematic viscosity and $\kappa=1.46\times10^{-7}$ m$^2$/s is the thermal diffusivity. The Rayleigh number is given by $Ra= g\alpha \Delta T H^3/\kappa\nu$ where $g$ is the acceleration  of gravity, and $\alpha=0.000238$ K$^{-1}$ is the thermal expansion coefficient.  %Unless otherwise specified, we report measurements at $\Delta T=19.0$ K  for $Ra = 2.7\times10^9$. %$\Delta T=4$ K  for $Ra = 5.7\times10^8$. 

%calibration
We calibrated thermistors at five mean temperatures from 21$^{\circ}$ C to 25$^{\circ}$ C.   %The calibrations were run with four baths -- one for each plate in each cell --
The calibrations were run with a small $\Delta T=0.04$ K to enhance mixing. The uncertainty on temperature measurements is 1.9 mK for the sidewall thermistors and 1.2 mK for the plate thermistors, adding in quadrature the contributions from the standard deviation  in thermistor temperatures during calibration runs, and the root-mean-square differences between the mean temperatures and the calibration fit.

\subsection{Obtaining the LSC parameters}
\label{sec:lsc}

 The LSC orientation $\theta_i$, amplitude $\delta_i$, and mean temperature $T_{0,i}$ are the main parameters used to characterize the LSC.  They were obtained using the same methods as \cite{BA06a}, but are now labeled with indices $i=1,2$ to differentiate the two cells.  As the LSC moves hot fluid from near the bottom plate up one side and cold fluid from near the top plate down the other side, a temperature difference is detected along  a horizontal direction at the mid-height of the container.  We fit the thermistor temperatures at the middle row in each cell to the function 
\begin{equation}
T_i = T_{0,i} + \delta_i cos(\theta - \theta_i)
\label{eqn:tempprofile}
\end{equation}
to get the orientation $\theta_i$, temperature amplitude $\delta_i$, and spatial mean temperature $T_{0,i}$ of the LSC roll.  To obtain a time series, these fits are done at every measured time step, which is typically 7 s.  

%errors
Due to the frequent failure of thermistors in the interior of the cell, we only report data from the sidewall thermistors, leaving only six thermistors in each cell at the middle row.  
%Using earlier datasets where we had 8 good thermistors at the middle row in a single cubic cell \cite{JB20a}, we find that the rms difference between when using 6 thermistors instead of 8 is  0.25 rad in $\theta_0$ and 0.032 K in $\delta$ at $\Delta T=4^{\circ}$ C. Peaks of probability distributions shift by up to 0.6 rad,
For example, at $\Delta T=19$ K, where the time-averaged amplitude $\bar\delta_i$ is 0.42 K in a counter-rotating state, this fit results in average uncertainties on instantaneous measurements of 0.18 rad  on $\theta_i$ and 0.06 K on $\delta_i$  from fitting Eq.~\ref{eqn:tempprofile}.  Since the uncertainties on individual thermistor measurements are only 1.9 mK, and deviations from the sinusoidal profile due to oscillation of the structure are only about 6\% of the amplitude \cite{JB20b}, the largest contribution to these random uncertainties is large turbulent fluctuations around the mean temperature profile. 
%These errors are comparable to the standard deviations of $\delta$ or $\theta$ in a single-peaked $p(\theta_i)$, which are 0.12 rad and 0.03 K in counter-rotating state, 0.31 rad and 0.05 K in co-rotating state -- resolution of width of distributions in question.
The small amplitude of $\Delta T=0.04 $ K during calibrations is expected to produce an LSC with mean amplitude $\bar\delta_i= 2.9$ mK based on an extrapolation of measured values of $\bar\delta_i$ \cite{JB20a}, which introduces a systematic error of 2.9 mK on measurements of $\delta_i$.   Using Eq.~\ref{eqn:tempprofile}, this error propagates to a systematic error on $\theta_i$ of 0.007 rad at $\Delta T=19$ K where $\bar\delta_i =0.42 $ K, or an error on $\theta_i$ of 0.024 rad  when $\bar\delta_i=0.12$ K at $\Delta T=4 $K.
  
%corrections for spikes -fixed
%Thermistor 306 exhibits negative spikes in temperature.  I implemented a smooth where a raw temperature measurement of a sidewall thermistor differed from the averaged of the preceding and following timestep by a threshold, then it was replaced by that average.  A threshold of 0.07 K was enough to prevent false corrections in datasets from 200910 to 200917.  The remaining uncertainty on thermistor 306 of 7 mK propagates to a 3 mK error on $\delta$, so is not larger than other measurement errors.

\section{Review of model for a single LSC, i.e.~without neighboring roll interactions}
\label{sec:modelreview}

In this section we summarize the model of Brown \& Ahlers \cite{BA08a}, which we use as a baseline of comparison to the behavior of a single-roll LSC.  The model consists of a pair of stochastic ordinary differential equations,  using the empirically known,  robust LSC structure as an approximate solution to the Navier-Stokes equations to obtain equations of motion for parameters that  describe the LSC dynamics.  The effects of fast, small-scale  turbulent fluctuations are separated from the slower, large-scale coherent motion when  obtaining this approximate solution,  then added back in as a stochastic term in the low-dimensional model.   The flow strength in the direction of the LSC is represented by the temperature amplitude $\delta_i$, which is proportional to the mean flow speed in the LSC.   The equation of motion for $\delta_i$ is
\begin{equation}
\dot\delta_i = \frac{\delta_i}{\tau_{\delta}} - \frac{\delta_i^{3/2}}{ \tau_{\delta}\sqrt{\delta_0}} + f_{\delta}(t) + \dot\delta_{\kappa,i}\ .
\label{eqn:delta_model}
\end{equation}  
The first  forcing term on the right side of the equation corresponds  to buoyancy, which strengthens the LSC.  The second term is a non-linear damping approximated for a boundary-layer dominated flow, which weakens the LSC.    $\delta_0$ is the stable fixed point value of $\delta_i$ where buoyancy and damping balance each other, and $\tau_{\delta}$  is a damping timescale for changes in the strength of the LSC.    $f_{\delta}(t)$ is a stochastic forcing term representing the effect of small-scale turbulent fluctuations and is modeled as Gaussian white noise with diffusivity $D_{\delta}$.  Thus, $\delta_i$ tends to exhibit strong fluctuations around the stable fixed point value of $\delta_0$.  $\dot\delta_\kappa$ is a placeholder for forcing terms due to the interaction between neighboring rolls, to be derived in Sec.~\ref{sec:interactionmodel}.

The equation of motion for the LSC orientation $\theta_0$ is
\begin{equation}
\label{eqn:theta_model}
\ddot{\theta}_i = - \frac{\dot{\theta}_i\delta_i}{\tau_{\dot{\theta}}\delta_0} - \nabla V_g(\theta_i) + f_{\dot{\theta}}(t) + \ddot\theta_{\kappa,i} \ . 
\end{equation}   
The first term on the right side of the equation is a damping term which comes from the advective term of the Navier-Stokes equations, where  $\tau_{\dot{\theta}}$ is a damping time scale for changes of orientation of the LSC. $f_{\dot{\theta}}$ is another stochastic forcing term with  diffusivity $D_{\dot\theta}$.  $V_g$ is a potential which represents the pressure of the sidewalls acting on the LSC, and is a function of the geometry of the cell, so $-\nabla V_g(\theta_i)\equiv -\partial V_g/\partial\theta_i$ is the forcing due to this geometric potential.  %Equation \ref{eqn:theta_model} is mathematically equivalent to diffusion in a potential landscape $V_g(\theta_i)$. 
 $\ddot\theta_\kappa$ is another placeholder for forcing terms due to the interaction between neighboring rolls, to be derived in Sec.~\ref{sec:interactionmodel}.

Since the terms of Eqs.~\ref{eqn:delta_model} and \ref{eqn:theta_model} were derived from the Navier-Stokes equations, functional predictions for $\delta_0$, $\tau_{\delta}$ and $\tau_{\dot\theta}$ exist \cite{BA08a, JB20a}.  The diffusivities $D_{\delta}$ and $D_{\dot\theta}$ have so far been measured from data \cite{BA08a, JB20a}.

%cubic potential
The geometric potential $V_g(\theta_i)$ can be expressed as a function of the $\theta$-dependent diameter of the cell, and thus can be calculated for convex cell geometries. $V_g(\theta_i)$ is predicted to be inversely proportional to the diameter across the cell as a function of LSC orientation $\theta_0$, and thus the lowest potentials are aligned with the longest diagonals \cite{BA08b}.  In the case of a cubic cell, this corresponds to four potential minima, one for each corner of the cell, which correspond to preferred orientations of the LSC.   This was tested in the same apparatus for a single LSC in which there was an insulating middle wall which totally isolated the roll from the neighboring cell \cite{JB20a}.

To characterize interaction terms, we will look for differences from these baseline results from Eqs.~\ref{eqn:delta_model} and \ref{eqn:theta_model}.

\section{Observations of preferred states}
\label{sec:preferredstates}

\begin{figure}
\includegraphics[width=0.485\textwidth]{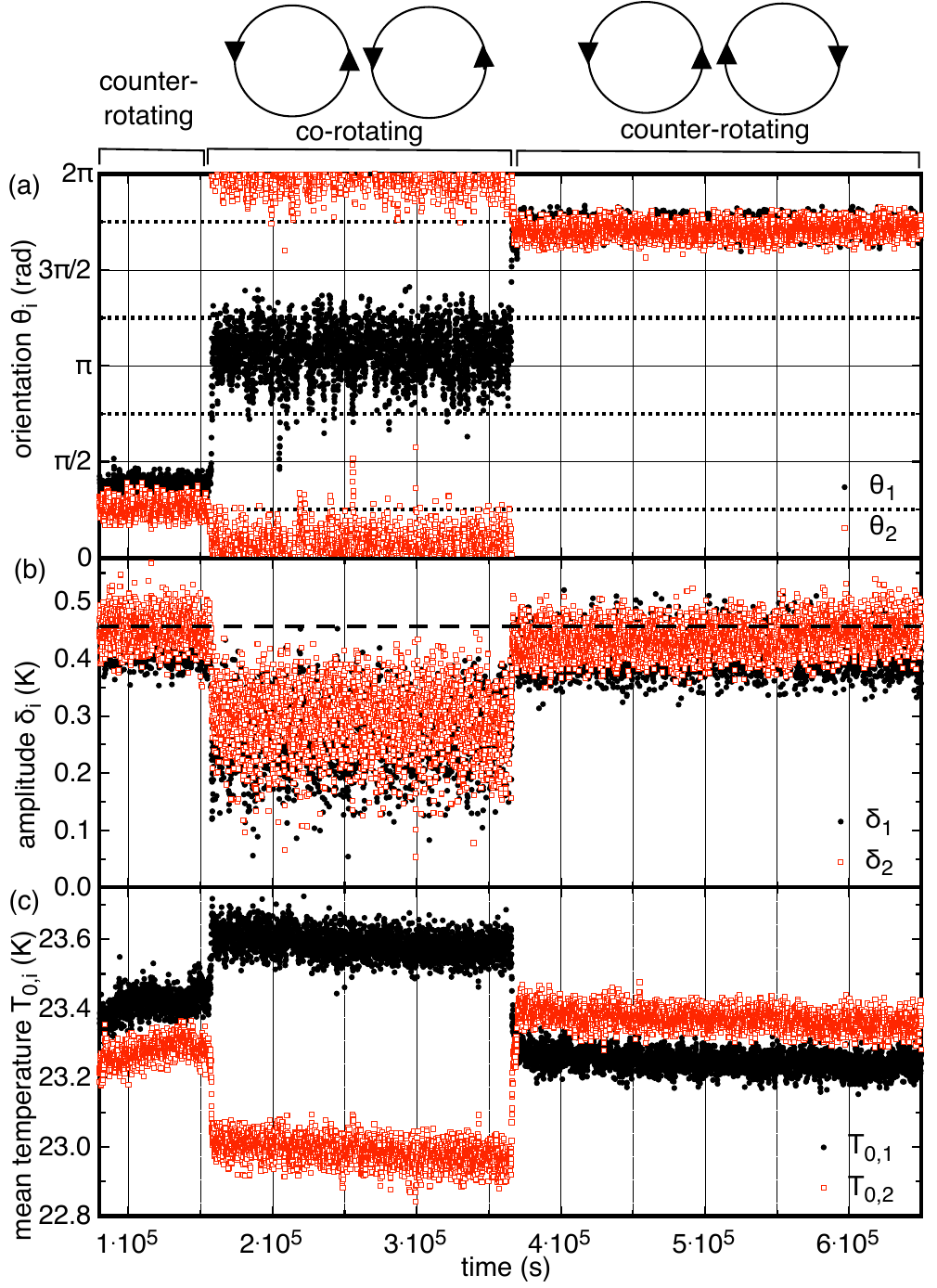} 
\caption{(color online) An example time series for two neighboring rolls.  (a) LSC orientation $\theta_i$.  Horizontal solid lines correspond to alignments parallel to walls, and dotted lines correspond to diagonals. (b) LSC amplitude $\delta_i$.    Dashed line: the mean amplitude for a single LSC in a cubic cell with no neighboring roll \cite{JB20a}. (c) LSC mean temperature $T_{0,i}$.   Side views of the corresponding co-rotating and counter-rotating states are illustrated above.  Both co-rotating and counter-rotating states are found to be stable, with spontaneous switching between the two states.    Co-rotating states are found to have a reduced $\bar\delta_i$, and introduce a large temperature difference between the mean temperatures $T_{0,i}$ of the neighboring rolls.  
}
\label{fig:switchingexample}
\end{figure}

%example 201111
Before developing complicated models, we categorize the observed states of behavior to identify what solutions a good model should have.  An example time series of the LSC orientation $\theta_i$, amplitude $\delta_i$, and mean temperature $T_{0,i}$ lasting 7 days is shown in Fig.~\ref{fig:switchingexample} at Ra $= 2.7\times 10^9$ ($\Delta T=19$ K).  To show the long time series, only 1 in every 10 data points is shown in this figure. The system is seen to spontaneously switch between two distinct types of states.  Before 160,000 s and after 370,000 s, there are fluctuations around a mean state with $\theta_1\approx\theta_2$.  This corresponds to a  counter-rotating state with the orientation vectors at $\theta_i$ aligned head-to-head, with flow in the same direction adjacent to the interface between the rolls, as illustrated at the top of Fig.~\ref{fig:switchingexample}a.   Counter-rotating states are usually aligned along a diagonal, which is the preferred state for a single roll in a cubic cell, and in these examples is at $\pi/4$ rad before 160,000 s, and $7\pi/4$ rad after 370,000s.   From 160,000-370,000 s, there are fluctuations around a mean state in which $\theta_1$ and $\theta_2$ instead differ by $\pi$ rad, with stable values of $\theta_1\approx 0$ and $\theta_2\approx \pi$ rad.  This corresponds to a co-rotating state where both rolls are rotating in the same direction, with counter-flow adjacent to each other at the interface between the cells, as illustrated at the top of Fig.~\ref{fig:switchingexample}a. 

%delta
The LSC amplitude $\delta_i$ also varies with the flow state, as shown in Fig.~\ref{fig:switchingexample}b.  For comparison, the mean amplitude $\delta_0$  for a single, non-interacting roll is drawn as a dashed line \cite{JB20a}.  In the counter-rotating state, the mean amplitude  is 9\% smaller than $\delta_0$, while in the co-rotating state, the mean amplitude is 38\% smaller than $\delta_0$.  This indicates the interaction between neighboring rolls reduces the temperature amplitude $\delta$, especially in the co-rotating state.  While strong turbulent fluctuations in both states hide any detailed patterns that might appear within a state, the widths of the distributions of $\theta_i$ and $\delta_i$ around their mean values in a state are both narrower in the counter-rotating state, suggesting the stabilizing forces are stronger in the counter-rotating state.  In the case of $\delta_i$, the non-linear damping term in Eq.~\ref{eqn:delta_model} produces less of a stabilizing force when $\bar\delta_i$ is smaller, which may account for the smaller width of $\delta_i$ in the co-rotating state.

%T0
The LSC mean temperature $T_{0,i}$ also varies with the flow state, as shown in Fig.~\ref{fig:switchingexample}c.  $T_{0,i}$ acquires a large systematic offset between the two cells in the co-rotating state, such that the cell whose colder side is at the interface ($\theta_i = \pi$ rad) acquires a higher mean temperature $T_{0,i}$.  The fact that $T_{0,i}$ has any systematic change is notable, since for a single LSC, no patterns in $T_0$ were found or reported due to cessations \cite{BA08a} or changes in preferred orientation \cite{JB20a}, except a small periodic modulation due to the jump-rope oscillation mode of the LSC \cite{HSA22}. 

%generalizations from more time series
To identify the possible states over a wide parameter space, we carried out experiments at different values of control parameters, including a temperature difference $\Delta T$  between the top and bottom plates in the range $0.53$ K $\le \Delta T \le 24.4$ K ($7.6\times10^7\le$ Ra $\le 3.5\times10^9$), a difference $\Delta T_m$ between the mean plate temperatures of each cell as large as $\Delta T$, and tilt angles $\beta$ ranging from $0 \le |\beta| \le 10^{\circ}$.  Experiments typically lasted about 1 day, and switching between counter-rotating and co-rotating states occurred on average about once every other day. The dynamics of this switching will be discussed more thoroughly in Sec.~\ref{sec:switching}. 

\begin{figure}
\includegraphics[width=0.475\textwidth]{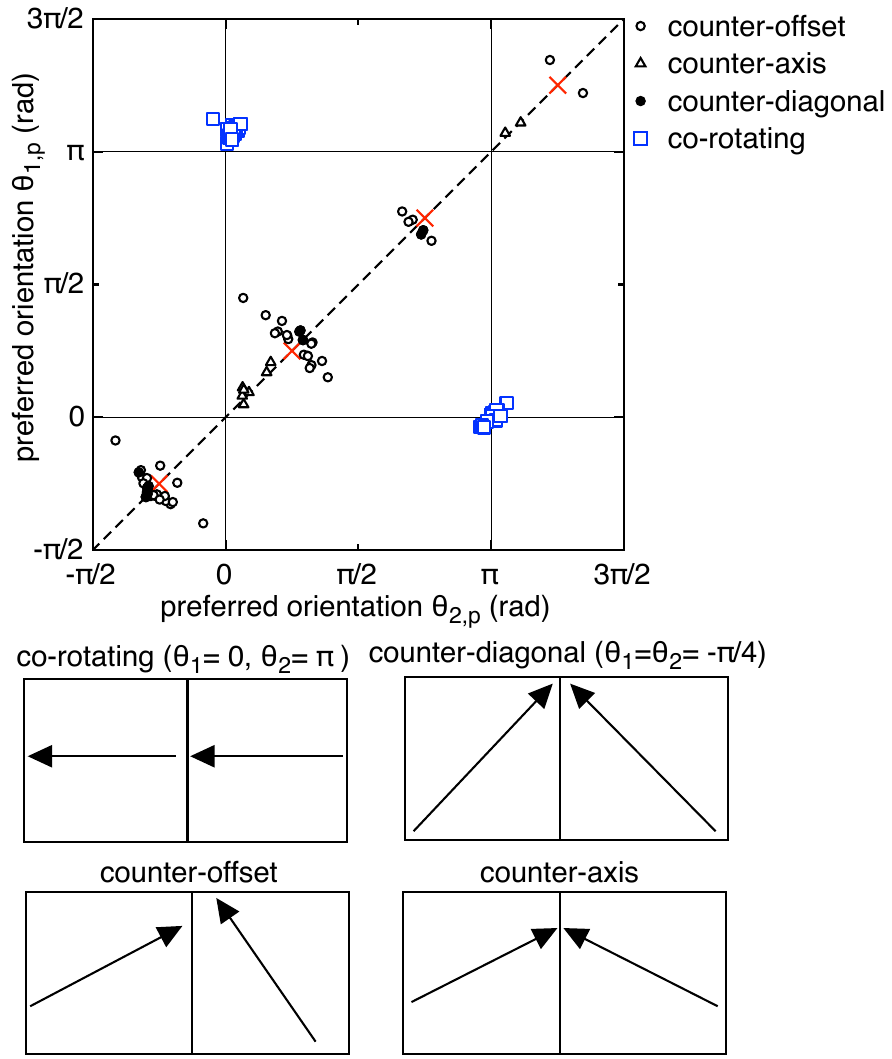} 
\caption{Scatter plot of the preferred orientation  $\theta_{1,p}$ of cell 1 vs.~the preferred orientation $\theta_{2,p}$ of cell 2  from a wide range of experiments.  
Counter-rotating states are defined by $\theta_{1,p}\approx \theta_{2,p}$, near the dashed line.  Solid circles: counter-diagonal states which align near a diagonal of the cell (red Xs).  Open triangles: counter-axis states which align in between the corner and  the axis between the two cells, due to some unintended asymmetry of the setup.    Open circles: counter-offset states defined by a small offset between $\theta_1$ and $\theta_2$,  which occur when a slight asymmetric heating is introduced with $\Delta T_m/\bar\delta >0.1$ or tilt $|\beta|>0.03^{\circ}$.   Open squares: co-rotating states, which align with the line passing through the center of both cells.  Lower panels: Example illustrations of different states are shown, viewed from above, with the arrow indicating the direction of flow in the LSC near the bottom plate at orientation $\theta_i$ in each cell.
}
\label{fig:thetap1_thetap2}
\end{figure}

%scatter plot of preferred orientations
 To characterize the preferred orientations of the states, a state diagram is made by plotting the values of the preferred orientation $\theta_{p,1}$ of cell 1 against the preferred orientation $\theta_{p,2}$ of cell 2 in Fig.~\ref{fig:thetap1_thetap2}.   Preferred orientation were obtained from fits of the peaks of probability distributions of $\theta_i$.   Experiments with switching between clear preferred orientations were divided up into separate datasets based on the preferred orientation to identify the preferred stable states.  Counter-rotating states are defined by  $\theta_{1,p} \approx\theta_{2,p}$  (dashed line in Fig.~\ref{fig:thetap1_thetap2}) such that they both flow in the same direction at the interface between the rolls.  In the most symmetrically-driven cases, where $|\Delta T_m|/\delta_0 < 0.1$ and $|\beta|<0.03^{\circ}$ (solid circles in  Fig.~\ref{fig:thetap1_thetap2}),  usually both cells are aligned near a diagonal (illustrated as the red Xs in Fig.~\ref{fig:thetap1_thetap2}) which we refer to more specifically as counter-diagonal states.  This matches the preferred diagonal state of a single cell \cite{LE09, BJB16, FNAS17, GKKS18, VSFBFBR16, VFKSSV19, JB20a}.  The alignment of the two rolls in a counter-rotating state with $\theta_{1,p} \approx\theta_{2,p}$ suggests there is a stable forcing in $\theta_1-\theta_2$. 

%counter axis
However, in a few nominally symmetric cases we found counter-rotating states with preferred orientations roughly halfway between a diagonal and on-axis alignment.  In some of these cases, there were switching events between the two corresponding symmetric preferred orientations on either side of $\theta=0$ or $\pi$ rad.   This alignment suggests there is a component of force that aligns the flow of the two cells in opposite directions toward the axis going through $\theta=0$ and $\theta=\pi$ rad, which combines with the geometric forcing $-\nabla V_g(\theta_i)$ to produce intermediate preferred orientations.
Since this forcing likely comes from an unintended asymmetry of our setup, these counter-axis states will not be the focus of the work, although they are discussed in Sec.~\ref{sec:counteraxis}.

 %Delta Tm
With a slight asymmetric forcing added to the basic counter-diagonal state with either $0.1 < |\Delta T_m|/\delta_0 < 1.7$ or $0.03^{\circ} < |\beta| < 2^{\circ}$, we find asymmetric offsets from the basic counter-rotating case where $\theta_{1,p}$ and $\theta_{2,p}$ shift in opposite directions away from the corner (open circles in  Fig.~\ref{fig:thetap1_thetap2}), so we refer to these as counter-offset states. 
 
%co
Stable co-rotating states are always found to be aligned with one of $\theta_1$ and $\theta_2$ aligned nearly with 0,  and the other nearly with $\pi$ rad (open squares in Fig.~\ref{fig:thetap1_thetap2}).  Co-rotating states are clustered around these orientations regardless of the value of $\Delta T_m$ and $\beta$.  While we are able to find co-rotating states some fraction of the time for any parameter values in our experiment, when a large enough asymmetry is introduced, either $|\Delta T_m|/\delta_0 \stackrel{>}{_\sim} 1.7$ or $|\beta| \stackrel{>}{_\sim} 2^{\circ}$, we find only co-rotating states and no more counter-offset or other counter-rotating states.  Since these preferred orientations $\theta_{p,i}=0$ or $\pi$ rad are along the axis between the center of both cells, this indicates the interaction between neighboring rolls adds a significant forcing stable in $\theta_i-\theta_{p,i}$, with $\theta_{p,i}=0$ or $\pi$ rad that can be dominant over the geometric potential $V_g(\theta_i)$ to align both rolls along this axis in co-rotating states.

\section{Model for neighboring roll interaction: effective turbulent thermal diffusion}
\label{sec:interactionmodel}

\subsection{Forcing terms $\dot\delta_{\kappa,i}$ and $\ddot\theta_{\kappa,i}$ due to neighboring roll interactions}
%motivation
%We initially considered viscous drag as the main interaction force between neighboring convection rolls by modifying the derivation of the viscous drag terms \cite{BA08a} at the interface between rolls in the model Eqs.~\ref{eqn:delta_model} and \ref{eqn:theta_model}. However, we found this could not explain the stability of co-rotating states (see discussion in Sec.~\ref{sec:drag} for an explanation).   Instead, we find that heat transport by a turbulent thermal diffusion between neighboring rolls is able to explain the stability of co-rotating states and other observations.

\begin{table}
\begin{tabular}{lll}
 \hline\hline
  & definition &   type  \\ %& co  ($6\tau_{\delta}$) \\ 
 \hline
$L$ & cell height & controlled\\ 
$A$ & fraction of interface open & controlled\\ 
 $\kappa$ & thermal diffusivity & controlled\\ 
  $T_{m,i}$ & mean temp.~of plates & controlled\\
 $\Delta T_m$ & temp.~diff.~of adjacent plates & dependent  ($\Delta T_{m,i}$)\\ 
 $\theta_i$ & LSC orientation & measured \\
 $\delta_i$ & LSC amplitude & measured \\
 $T_{0,i}$ & LSC mean temperature &measured  \\
   $\Delta T_0$ & temp.~diff.~between LSCs & dependent ($\Delta T_{0,i}$)\\
 $\delta_0$ & stable LSC amplitude & known \cite{JB20a}\\
 $\tau_{\delta}$  & damping timescale of $\delta_i$ & known \cite{JB20a}\\
 $\tau_{\theta}$ & damping timescale of $\theta_i$ & known \cite{JB20a}\\
$D_{\delta}$  & turbulent diffusivity for $\delta_i$ & known \cite{JB20a}\\
$D_{\dot\theta}$  & turbulent diffusivity for $\theta_i$ & known \cite{JB20a}\\
$D_{T}$  & turbulent diffusivity for $T_{0,i}$ & not tested\\
Nu & Nusselt number & known \cite{FBNA05}\\ \
  $\kappa_t$ & turbulent thermal diffusivity & fit\\ 
 %$\theta_{p,i}$ & geometric... & measured \\
%$\omega_r^2$ geometric...
%\tau_T & & dependent($\kappa_t$)\\
 \hline\hline
\end{tabular} 
\caption{Parameters that are relevant to the model of turbulent thermal diffusion for the interaction between neighboring rolls.   Parameters referred to as `known' have known values from previous experiments with a single LSC.  The `measured' parameters are those that describe the LSCs.  The `dependent' parameters are useful in results expressed in terms of differences between neighboring rolls and cells.  The parameter $\kappa_t$ is unknown, and is the single new fit parameter.
}
\label{tab:modelparameters}
\end{table}

%model summary
The model for the neighboring roll interaction terms is derived in detail in the appendix, and we summarize the physical mechanisms and resulting equations here.  We start with the assumption that Eqs.~\ref{eqn:delta_model} and \ref{eqn:theta_model} describe a baseline model for each LSC, and that we only need to derive terms for additional forcing due to neighboring-roll interactions.  The forcing on the temperature profile along the interface between neighboring convection rolls is assumed to come from the turbulent diffusion of heat across the interface with effective turbulent thermal diffusivity $\kappa_t$, which is assumed to be uniform and constant. The turbulent thermal diffusion represents the enhancement of heat transport by eddies relative to the thermal diffusivity $\kappa$ due to advection, analogous to a turbulent viscosity or eddy viscosity.   This formulation is mathematically analogous to boundary layer approximations with thermal diffusion in an interfacial mixing layer.  We include the unknown mixing layer thickness in the value of $\kappa_t$, which will be a fit parameter, so we can use the known cell size $L$ as the lengthscale in the thermal diffusion equation.    Because of the middle insulating wall blocking half of the interface, the heat transport acts over the exposed fractional area $A=0.50$ of the interface between the two cells.    The model terms are calculated by applying this turbulent thermal diffusion in a $\theta$-dependent heat equation for heat flux in the direction perpendicular to the interface between neighboring cells.
The resulting forcing terms (as calculated in appendix A) are:
\begin{multline}
\dot\delta_{\kappa,1} =  \frac{A\kappa_t}{\pi L^2}\bigg[ \sqrt{2}(T_{0,2}-T_{0,1})\cos\theta_1 \\
+\delta_2\sin(\theta_1-\theta_2)\sin(2\theta_1)/2\\
+[\delta_2\cos(\theta_1-\theta_2) - \delta_1][\pi/4+\cos(2\theta_1)/2] \bigg] \\
- \frac{\sqrt{2}\dot T_{0,1}}{\pi}\cos\theta_1\ .
\label{eqn:deltaforcing}
\end{multline}
and
\begin{multline}
\ddot\theta_{\kappa,1} = - \frac{A\kappa_t}{\pi L^2 \tau_{\dot\theta}\delta_0}\bigg[ \delta_2\sin(\theta_1-\theta_2)[\pi/4-\cos(2\theta_1)/2]\\
+[\delta_2\cos(\theta_1-\theta_2) - \delta_1]\sin(2\theta_1)/2
+\sqrt{2}(T_{0,2}-T_{0,1})\sin\theta_1 \bigg]\\
+ \frac{\sqrt{2}\dot T_{0,1}}{\pi\delta_1}\sin\theta_1 
\label{eqn:thetaforcing}
\end{multline}
%where $\Delta T_0 \equiv T_{0,2}- T_{0,1}$.  
For brevity, we write all equations in this section for the forcing on cell 1 only, as the equations for cell 2 are identical other than an exchange of the subscripts 1 and 2 in each equation. 
%Combining thermal diffusion forcing with the previous model
These forcings $\dot \delta_{\kappa,1}$ and $\ddot\theta_{\kappa,i}$ can be inserted directly to the existing  stochastic equations of motion for a single LSC (Eqs.~\ref{eqn:delta_model}  and \ref{eqn:theta_model}, respectively).  

%Since Eq.~\ref{eqn:theta_model} is in terms of $\ddot\theta_1$, but we derived a forcing in $\dot\theta_1$ (Eq.~\ref{eqn:thetaforcing}), we assume the forcing in $\dot\theta_1$ can be treated as the forcing in the overdamped limit, corresponding to an offset in the $\dot\theta_1$-term of Eq.~\ref{eqn:theta_model}, so that 
%\begin{equation}
%\ddot\theta_{\kappa,1} = \frac{\dot\theta_{\kappa,1}\delta_1}{\tau_{\dot\theta}\delta_0} \ .
%\label{eqn:ddottheta}
%\end{equation}

%effect of including $\bar\delta/\delta_0$:
%forcing weaker in co, but for all terms equally in theta equation
%co-stabilizing forcing weaker (less stable), but more stable on axis when also including $\bar\delta^2$ on $\omega_r^2$
%during cessation, forcing to fixed points also weakens if delta-factor included
%correlation time in theta larger in co-rotating state -- only for destabilizing term and term with unknown sign

\subsection{Equation of motion for $T_{0,i}$}
%motivation
The model for a single LSC did not include an equation of motion for the mean temperature $T_{0,i}$ because it was observed to have trivial behavior. With neighboring convection cells, we observe a systematic difference between  $T_{0,2}$ and $T_{0,1}$ (Fig.~\ref{fig:switchingexample}c), and this difference may drive dynamics in $\theta_i$ and $\delta_i$ according to Eqs.~\ref{eqn:thetaforcing} and \ref{eqn:deltaforcing}.  Thus, we need an equation of motion for $T_{0,i}$.  

In an equation of motion for $T_{0,i}$, we should also expect to have some diffusive fluctuations driven by turbulence, analogous to Eqs.~\ref{eqn:delta_model} and \ref{eqn:theta_model}, so we include a fluctuation term $f_T(t)$ with diffusivity $D_T$.   The deterministic part of the equation for $T_{0,1}$ is assumed to be due to the same net heat flux between the cells that contributes to the forcings $\dot\delta_{\kappa,i}$ and $\ddot\theta_{\kappa,i}$.  An additional vertical heat transport from each of the top and bottom plates to each LSC  is calculated using the standard boundary layer approximation in terms of the Nusselt number Nu.
% calculated by integrating both sides of Eq.~\ref{eqn:turbulentdiffusion} over $d\theta/\cos^2\theta$.  The denominator of $\cos^2\theta$ accounts for the relative interface area at each infinitesimal $d\theta$ to satisfy conservation of energy. %\footnote{Such a factor is not included in Eqs.~\ref{eqn:fouriertransform} and \ref{eqn:fouriertransformdelta}, because the quantities $\delta_0_{\kappa,i}$ and $\theta_{\kappa,i}$ do not scale with thermal energy.  In any case, the factor of $\cos^2 \theta$ in the integral only produces a 2% correction on the relative size of the 2 terms of Eq.~\ref{eqn:Tforcing}.   
%The term of Eq.~\ref{eqn:turbulentdiffusion} with $A$ due to the neighboring roll interaction only contributes in the interaction region $|\theta| < \pi/4$ rad so is only integrated over that range.  For simplicity of later equations, we will solve for aspect ratio 1 ($H=L$).  % and define $B\equiv \mbox{Nu} \kappa/\kappa_t$ as the ratio of vertical to horizontal turbulent thermal diffusivities at aspect ratio 1.  
This results in (as calculated in Appendix B):  
%\begin{multline}
%\dot T_{0,1} = f_T(t) + \frac{4\mbox{Nu}\kappa}{L^2}(T_{m,1}-T_{0,1}) + \frac{A\kappa_t}{2 L^2}\times\\
%\int_{-\pi/4}^{\pi/4} \frac{T_{0,2}-T_{0,1} + \delta_2\cos(\theta-\theta_2) - \delta_1\cos(\theta-\theta_1)}{\cos^2\theta}d\theta \ .
%\label{eqn:dotTintegral}
% \end{multline}
 %Using the same substitution as before for $\cos(\theta-\theta_2)$, and integrating, results in
 \begin{multline}
\dot T_{0,1} = f_T(t)+\frac{4\mbox{Nu}\kappa}{L^2}(T_{m,1}-T_{0,1})+\\
\frac{A\kappa_t}{L^2}\left[T_{0,2}-T_{0,1}+ 0.88(\delta_2\cos\theta_2 - \delta_1\cos\theta_1)\right] \ .
\label{eqn:Tforcing}
 \end{multline}
%page 64
While this equation could be plugged into Eqs.~\ref{eqn:deltaforcing} and \ref{eqn:thetaforcing}, we find it more insightful to leave Eqs.~\ref{eqn:deltaforcing} and \ref{eqn:thetaforcing} as is for a more direct physical interpretation of their dependence on $T_{0,2}-T_{0,1}$ at the stable fixed points where $\dot T_{0,i}=0$.

\section{Testing the model for symmetric cases ($\Delta T_m=0$)}
\label{sec:interactionmodeltest}

\subsection{Forcing terms on $\ddot\theta_i$ in $\theta_1-\theta_2$ and $\theta_i-\theta_p$}
\label{sec:thetastability}

In Sec.~\ref{sec:preferredstates}, the observed counter- and co-rotating states suggest forcing terms on $\ddot\theta_i$ that are stable in $\theta_1-\theta_2$ and $\theta_i-\theta_{i,p}$, respectively.  In this section, we test the predictions for these forcing terms  from Eq.~\ref{eqn:thetaforcing} for counter- and co-rotating states, starting with the simpler special case of $\Delta T_m=0$.

\subsubsection{Stable fixed points and linearized forcing}  

Qualitatively, Eq.~\ref{eqn:thetaforcing} has stable fixed points that are linearly stable for both the observed counter-rotating states ($\theta_1=\theta_2)$, and co-rotating states ($\theta_1 =0$ and $\theta_2=\pi$ rad or $\theta_1=\pi$ rad and $\theta_2=0$), as shown in Appendix C and D, respectively, confirming that these states are predicted by the model.

In the counter-rotating state, a linear expansion of Eq.~\ref{eqn:thetaforcing} around the stable fixed point at a preferred orientation of the cubic cell (e.g.~$\theta_1= \pi/4$ rad) and assuming $\bar\delta_i=\delta_0$ is  
\begin{equation}
\ddot\theta_{1,counter} \approx -0.79\frac{A\kappa_t}{\pi L^2\tau_{\dot\theta}} (\theta_1-\theta_2) \ .
\label{eqn:thetaforcinglinearcounter}
\end{equation}

%Observed stable fixed points for co-rotating states correspond to $\theta_1 =0$ and $\theta_2=\pi$ rad or $\theta_1=\pi$ rad and $\theta_2=0$,  $\delta_1=\delta_2$, and $\dot T_{0,i}=0$ in steady state.  All four terms of Eq.~\ref{eqn:thetaforcingappendix} are zero in in this case, confirming these are fixed points of Eq.~\ref{eqn:thetaforcingappendix} as well.  There are forcings on $\dot\delta_i$ in Eq.~\ref{eqn:deltaforcing} and $\dot T_{0,i}$ in Eq.~\ref{eqn:Tforcing}, which are addressed in Sec.~\ref{sec:bardelta}.  

In the co-rotating states, assuming constant $\Delta \bar T_{0,co} \equiv \bar T_{0,2}-\bar T_{0,1}$ and $\bar\delta_{co}\equiv\bar\delta_1=\bar\delta_2$, a linear approximation  around $\theta_{p,1} = 0$ or $\pi$ rad for co-rotating states of Eq.~\ref{eqn:thetaforcing} is 
\begin{multline}
\ddot\theta_{1,co} \approx \frac{A\kappa_t\bar\delta_{co}}{\pi L^2\delta_0\tau_{\dot\theta}} \bigg[+0.28(\theta_1-\theta_2-\pi)\\
-\left(\frac{1.41\Delta \bar T_{0,co}}{\bar\delta_{co}}-2\right)(\theta_1-\theta_{p,1}) \ . \bigg]
\label{eqn:thetaforcinglinearco}
\end{multline}

\subsubsection{Multiple linear regression}
\label{sec:regression}

%motivation and methods
To test the functional form of the net forcing around stable fixed points in the Eqs.~\ref{eqn:thetaforcinglinearcounter}, and \ref{eqn:thetaforcinglinearco} added to the stochastic equation of motion (Eq.~\ref{eqn:theta_model}), we carried out a multiple linear regression of the equation
%tried correction with - \Delta\theta_p)$, where $\Delta\theta_p = \theta_{p,1}-\theta_{p,2}$ is a correction for the difference in extrema of the probability distributions.  Didn't make significant difference
\begin{equation}
%\ddot\theta_1 = a\sin(\theta_1-\theta_{p,1}) + b\sin(\theta_2-\theta_{p,2}) + c
\ddot\theta_i = a\sin(\theta_i-\theta_{p,i}) + b\sin(\theta_1-\theta_2) 
\label{eqn:regression}
\end{equation}
with stiffnesses $a$ and $b$.  We insert the sine functions as the simplest way to account for periodicity of the coordinate $\theta$ while remaining linear in the lowest order expansion around the stable fixed points. 

\begin{figure}
\includegraphics[width=0.485\textwidth]{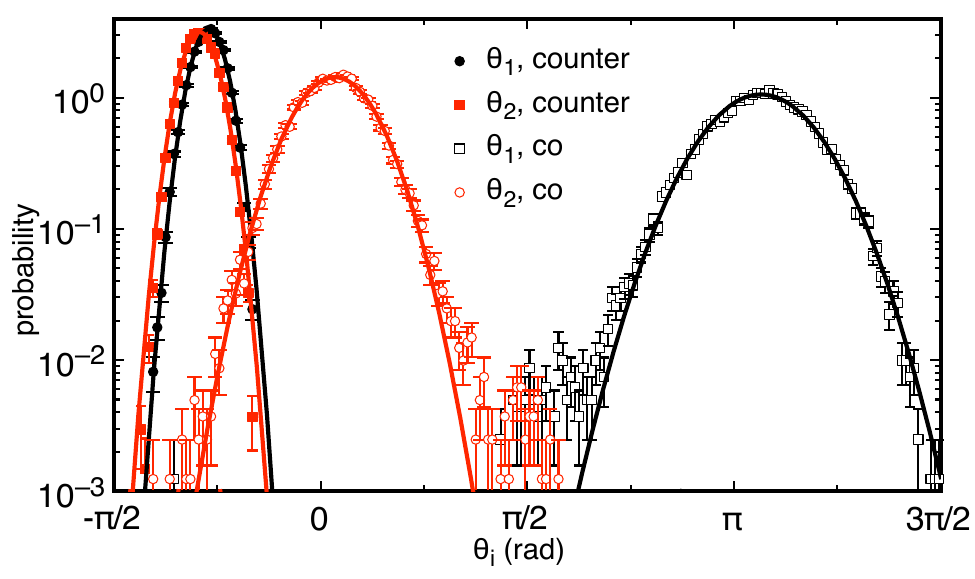} 
\caption{(color online) Probability distribution of the LSC orientation $p(\theta_i)$ in the counter- and co-rotating state.   Solid lines are Gaussian fits corresponding to a quadratic potential and linear forcing in $\theta_i-\theta_{p,i}$, as predicted due to effective turbulent thermal diffusion combined with the geometric potential $V_g(\theta_i)$. 
}
\label{fig:probtheta}
\end{figure}
 
%regression
% when limited to a range of 0.125 rev centered around the peak it leads to significant errors for some short datasets and outliers
To  be able to carry out an accurate linear regression, we first precisely determine the peak locations $\theta_{p,i}$ by fitting the probability distribution $p(\theta_i)$ for each state to a Gaussian function, shown in Fig.~\ref{fig:probtheta} for the same data as Fig.~\ref{fig:switchingexample}.  To calculate the forcing $\ddot\theta_i$, we calculate the 2nd derivative from discrete data as $\ddot\theta_i = [\theta_i(t+\Delta t) - 2\theta_i(t)+\theta_i(t-1)]/(\Delta t)^2$.   We apply a multiple linear regression of Eq.~\ref{eqn:regression} to measured values of $\ddot\theta_i$ to obtain the coefficients $a$ and $b$, using only data with $\delta > 0.5\bar\delta$ to avoid bias during cessations where the acceleration can be much larger than typical due to weaker damping in Eq.~\ref{eqn:theta_model} \cite{BA08a}.   

%results
From the linear regression of Eq.~\ref{eqn:regression}, we generally find $a$ to be negative in the counter-rotating and co-rotating states, confirming that the system is stable against displacements in $\theta_i-\theta_{p,i}$, as predicted due to the geometric potential $V_g$ \cite{JB20a} and Eq.~\ref{eqn:thetaforcinglinearco}, respectively.  We find $b$ negative for cell 1 and positive for cell 2 in counter-rotating states, confirming they are stable in $\sin(\theta_1-\theta_2)$ as predicted in Eq.~\ref{eqn:thetaforcinglinearcounter}.  We also find $b$  positive for co-rotating states in cell 1 and negative in cell 2, which -- due to the flip in the sign of the sine function with a phase shift of $\pi$ rad -- means that co-rotating states are also stable in $\sin(\theta_1-\theta_2)$, corresponding to a forcing to align the orientation vectors head-to-tail in the co-rotating state, but opposite the sign predicted in Eq.~\ref{eqn:thetaforcinglinearco}.   

  %other models 
  If we calculate the forcing  in terms of $\dot\theta_i(\theta_1-\theta_2)$ instead of $\ddot\theta_i(\theta_1-\theta_2)$, similar qualitative results are found, but the strongest forcing is found when $\dot\theta_i$ is calculated with about a 20 s delay after the time that $\theta_1-\theta_2$ is recorded.  This is comparable to the damping timescale $\tau_{\dot\theta}$ that corresponds to the ratio of these two terms in Eq.~\ref{eqn:theta_model}.  This confirms that a forcing in terms of $\ddot\theta_i$ better describes the effects of a neighboring roll interaction, justifying the conversion of $\dot\theta_{\kappa,i}$ into $\ddot\theta_{\kappa,i}$ in Eq.~\ref{eqn:ddottheta}.

\subsubsection{Functional form of forcing  $\ddot\theta_i(\theta_i-\theta_{p,i})$} 
\label{sec:forcinga}

In co-rotating states and counter-diagonal states, we typically find the stiffness $a$ much larger than $b$ from the regression, which indicates that the forcing in $a\sin(\theta_i-\theta_{p,i})$ usually determines the overall stability of the state. For example, for the data in Fig.~\ref{fig:switchingexample},  $a=10b$ in the counter-rotating state, and $a=2.6b$ in the co-rotating state  \footnote{For counter-offset states with $\Delta T_m /\delta_0 \stackrel{>}{\sim} 0.1$, $\theta_{p,1}$ and $\theta_{p,2}$ shift away from the corner, so that the forcing is not necessarily centered on the peak of $p(\theta_i)$.  In these cases the apparent value of $b$ drops significantly even for small $\Delta T_m$.  We attempted to add an extra constant offset term to be fit in the linear regression to fix this.  While the values of $b$ do not systematically drop with small $\Delta T_m$ with a constant offset in the linear regression, the extra regression parameter causes uncertainties to increase to be comparable to fit parameters.  We do not report results for $b$ with $\Delta T_m \ne 0$, and caution about this limitation of the algorithm for states where $\theta_{p,1}$ is offset from $\theta_{p,2}$.}.
   The dominance of the forcing $a\sin(\theta_i-\theta_{p,i})$ provides an opportunity to analyze the functional form of the forcing terms  separately.  Since $a\sin(\theta_i-\theta_{p,i})$ is much larger, then the probability distribution $p(\theta_i)$ shown in Fig.~\ref{fig:probtheta} is determined mostly by this term, so we can first obtain an approximate forcing in $\theta_i-\theta_{p,i}$ while disregarding the forcing in $\theta_1-\theta_2$.   The forcing in $\theta_i-\theta_{p,i}$ can be related to a probability distribution of $\theta_i-\theta_{p,i}$ in the overdamped limit of the stochastic Eq.~\ref{eqn:theta_model} (i.e.~$\ddot\theta_i$ is small compared to the damping and forcing terms) and if $\delta$ is nearly a constant by a Fokker-Planck equation \cite{JB20a}
\begin{equation}
p(\theta_i)  =  \exp\left(-\frac{V(\theta_i)\bar\delta^2}{D_{\dot\theta}\tau_{\dot\theta} \delta_0^2} \right) 
\label{eqn:prob}
\end{equation}
where $V(\theta_i) = V_g(\theta_i) - \int \ddot\theta_{\kappa,i} d\theta_i$ is the net potential from geometric and neighboring roll interaction forces.   A linear forcing in $\theta_i-\theta_{p,i}$ corresponds to a quadratic potential $V(\theta_i)$, and a Gaussian probability distribution.   Each probability distribution $p(\theta_i)$ is fit to a Gaussian function, as shown in Fig.~\ref{fig:probtheta}.   Errors are shown in Fig.~\ref{fig:probtheta} calculated assuming  Poisson statistics.  Errors on $\theta_i$ of 0.18 rad from the fit of the temperature profile (Eq.~\ref{eqn:tempprofile}) divided by the square root of the number of counts are too small to see in Fig.~\ref{fig:probtheta}, but remain significant in the fitting of the data.  Fitting data up to 2.7 standard deviations away from the peak yields reduced $\chi^2$ of 1.4,1.3, 1.2, and 0.9 for the four fits shown in Fig.~\ref{fig:probtheta}, confirming the data are consistent with Gaussian probability distributions in this range.   The Gaussian shape of $p(\theta_i)$ confirms the forcing in $\theta_i-\theta_{p,i}$ is linear around the stable fixed point, confirming the assumption made in Eq.~\ref{eqn:regression}, and matching the prediction of Eq.~\ref{eqn:thetaforcinglinearco} for co-rotating states, as well as the predicted linear forcing for counter-diagonal states due to the geometric potential $V_g(\theta_i)$, which has been previously observed to be linear for a single LSC  \cite{JB20a}.  

%reorientation effect on probability distribution
%There is a slight deviation from the Gaussian distribution in the tails of $p(\theta_i)$ for the co-rotating state in Fig.~\ref{fig:probtheta}.  The tails of $p(\theta_i)$ in Fig.~\ref{fig:probtheta} come mainly from two events, one initiated at $t=204,600$ s and another at $t=255,500$ s, in each case initiated when $\delta$ drops down to 40\% of its mean value, causing a reorientation of the LSC in one cell away from its mean value that lasts for about 600 s.  Such rare reorientations  of the LSC are known to occur when the LSC amplitude $\delta$ drops, resulting in a weaker damping term in Eq.~\ref{eqn:theta_model}, allowing the LSC orientation to be driven further away from the orientation of lowest potential due to the stochastic term \cite{BA08a}.

\subsubsection{Functional form of forcing $\ddot\theta_i(\theta_1-\theta_2)$}
\label{sec:forcingb}

%methods
Since the multiple linear regression in Sec.~\ref{sec:regression} showed that the forcing in $\theta_1-\theta_2$ is smaller than the forcing in $\theta_i-\theta_{p,i}$, then a probability distribution of $\theta_1-\theta_2$ will not reveal the forcing in $\theta_1-\theta_2$ because it will be overwhelmed by the stronger forcing in $\theta_i-\theta_{p,i}$.  Instead we calculate a corrected forcing $\ddot\theta_{i,c} = \ddot\theta_i(\theta_1-\theta_2)-a\sin(\theta_i-\theta_{p,i})$ as a function of $\theta_1-\theta_2$.  This method allows us to correct for the linear dependence on $\theta_i-\theta_{p,i}$ that was confirmed by Fig.~\ref{fig:probtheta}, so that only  $b\sin(\theta_1-\theta_2)$ is expected to remain, based on Eq.~\ref{eqn:regression}.  To do this, we calculate the azimuthal acceleration rate from discrete data as the 2nd order difference $\ddot\theta_i(t) = [\theta_i(t+1)-2\theta_i(t) +\theta_i(t-1)]/(\Delta t)^2$.  The forcing we report with a measurement timestep of $\Delta t=7$ s is consistently about 20\% smaller than runs with a shorter timestep of 2 s, due to the approximation of a second derivative with a non-zero timestep.
For each data point, we subtract $a\sin(\theta_i-\theta_{p,i})$ using the value of $a$ from the regression (Sec.~\ref{sec:regression}).  We  exclude data with $\delta < 0.5\bar\delta$ to avoid biasing data by cessations where the acceleration can be much larger than typical values \cite{BA08a}.  We then bin values of $\theta_{i,c}$ over small ranges of $\theta_1-\theta_2$ to calculate forcing as a function of $\theta_1-\theta_2$ and average $\langle\ddot\theta_{i,c}(\theta_1-\theta_2)\rangle$ in each bin to reduce the contribution of the large stochastic fluctuations in Eq.~\ref{eqn:theta_model}.   The error on the average in each bin is reported as the standard deviation of the mean, assuming the data points are independent.  
 %including error on $\theta_1$ doesn't make a difference in fit
 
 \begin{figure}
\includegraphics[width=0.485\textwidth]{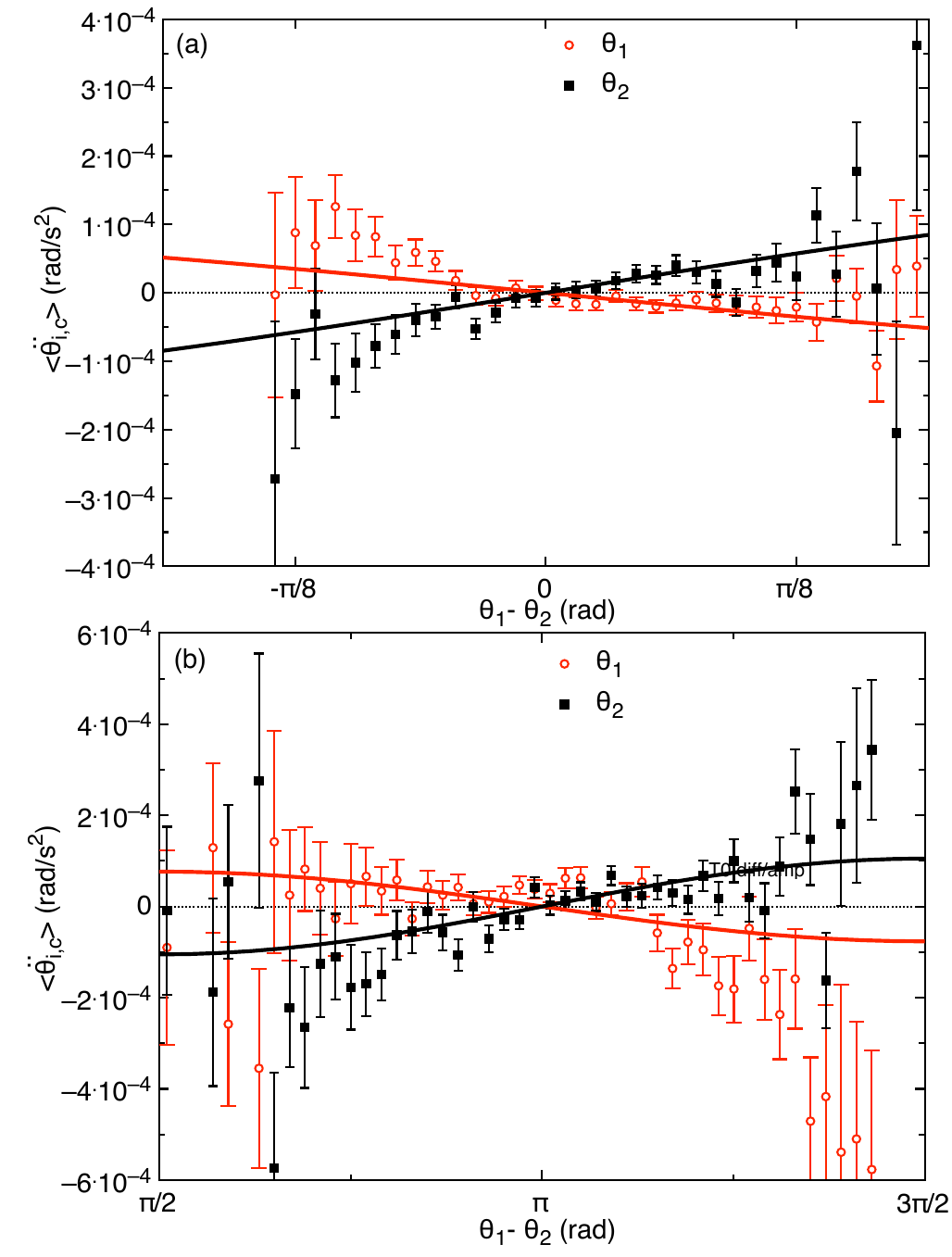} 
\caption{(color online) The average forcing $\langle\ddot\theta_{i,c}(\theta_1-\theta_2)\rangle$ as a function of $\theta_1-\theta_2$, corrected for the forcing in $\theta_i-\theta_{p,i}$. (a) counter-rotating state  (b) co-rotating state. Red circles: cell 1.  Black squares: cell 2.  Data are fit to a sine function $b\sin(\theta_1-\theta_2)$.  For counter-rotating states, the fit is consistent with the data, in qualitative agreement with the prediction for the neighboring roll interaction, while for co-rotating states, this has the opposite sign predicted. 
}
\label{fig:forcing_theta_d2theta}
\end{figure}

%comparison of counterrotating and corotating forcgin:  201111
This corrected forcing $\langle\ddot\theta_{i,c}(\theta_1-\theta_2)\rangle$ is shown in Fig.~\ref{fig:forcing_theta_d2theta} for the same dataset as in Fig.~\ref{fig:switchingexample}.  Data from the counter-rotating state are shown in panel a, and data from the co-rotating state are shown in panel b.  In both states, cell 1 has a negative slope in $\theta_1-\theta_2$, and cell 2 has a positive slope in $\theta_1-\theta_2$, confirming they are stable at their intercepts.  The intercepts are near $\theta_1-\theta_2=0$ for the counter-rotating state, and $\theta_1-\theta_2 = \pi$ rad for the co-rotating state, corresponding to their stable relative orientations (Fig.~\ref{fig:thetap1_thetap2}). 

 %fit
 The corrected forcing  is fit to  $\langle\ddot\theta_{i,c}(\theta_1-\theta_2)\rangle = b\sin(\theta_1-\theta_2)$ where the stiffness $b$ is a fit parameter.  Only bins with at least 10 data points are included in the fit.  The reduced $\chi^2$ values are 1.2 and 1.2 for the counter-rotating data sets, 1.3 and 2.4 for the co-rotating data, indicating the sine function is consistent with the data for the counter-rotating states, but not necessarily for the co-rotating states.  The magnitudes of the values of $b$ are consistent with the values obtained when using the linear regression analysis within a couple of standard deviations of the mean, which is on average 16\% of the mean.  This self-consistency in the magnitudes obtained from both methods confirms the earlier assumptions that the net forcing on the LSC orientation can be represented by Eq.~\ref{eqn:regression}, consistent with the functional forms predicted by $V_g$ and Eqs.~\ref{eqn:thetaforcinglinearcounter}  for counter-rotating states, and the terms are independent enough to use the conditional average in Fig.~\ref{fig:forcing_theta_d2theta}.  The magnitude of $b$ for counter-rotating states will be used to obtain $\kappa_t$ and compared to other model terms in Sec.~\ref{sec:turbulentkappa}.  However, the negative sign of the stiffness $b$ for co-rotating states is inconsistent with the prediction of Eq.~\ref{eqn:thetaforcinglinearco}, which predicted $b>0$ (destabilizing) for co-rotating states.  Since $a>b$, then $b$ is not  responsible for the stability of co-rotating states, but this remains a minor disagreement with the model Eq.~\ref{eqn:thetaforcing}.

\subsubsection{Comparison of forcing term $\ddot\theta_i(\theta_i-\theta_{p,i})$ to the geometric forcing for the counter-diagonal state}

The forcing $-\nabla V_g$ due to the cell geometry in Eq.~\ref{eqn:theta_model} has been observed and predicted to be linear and stable in an expansion of $\theta_i-\theta_{p,i}$ around the potential minima along the diagonals  \cite{JB20a}, which is found to be the preferred orientation for most counter-rotating states.  Thus, it is predicted to be responsible for the measured value of the stiffness $a$ for counter-rotating states.  To determine how much of the measured value of $a$ comes from the geometric potential, we calculated  the forcing $\langle\ddot\theta_i(\theta_i)\rangle$ by averaging values of $\ddot\theta_i$ in bins of $\theta_i$ as in Fig.~\ref{fig:forcing_theta_d2theta} in Sec.~\ref{sec:forcingb} from an experiment with a single LSC in the same apparatus, but with a insulating middle wall that completely blocks the two cells off from each other (i.e.~$A=0$) at Ra $ = 2.7\times 10^{9}$ \cite{JB20a}.  For a single LSC, we find the stiffness $a= -1.02\pm0.11$ mrad/s$^2$, and for neighboring counter-rotating rolls, we find $a= -1.11\pm0.17$ mrad/s$^2$.  The consistency of these values suggest that the stiffness $a$ comes entirely from the geometric forcing $-\nabla V_g$ for counter-rotating states, and  there is no significant neighboring roll interaction contribution to $a$ for counter-rotating states, in agreement with the prediction of Eq.~\ref{eqn:thetaforcinglinearcounter}. %For this comparison, we used a dataset with small $\Delta T_m/\bar\delta<0.01$, as the small but significant $\Delta T_m$ in the dataset in Fig.~\ref{fig:switchingexample} reduces the measured $b$ significantly. 
%We confirmed this consistency for an additional pair of experiments at $\Delta T=18.6$ and a pair of experiments at $\Delta T=3.8$ K. 

%co
For co-rotating states, the stiffness $a$ is expected to come from a competition between the second and third terms of Eq.~\ref{eqn:thetaforcing}, and be reduced by the geometric forcing $-\nabla V_g$ around its unstable fixed point.  Since the source of the stability of co-rotating states is a more fundamental question and involves more parameters, we put this off until Sec.~\ref{sec:stabilityratio} after $\kappa_t$ is fit and more terms of the model are tested.
%\footnote{Note that a previous estimate of the forcing $a\sin(\theta_i-\theta_{p,i})$ based on fitting $-\ln p(\theta_0)$ resulted in a larger $a$ by about a factor of 2 due to the different technique used to obtain $a$ \cite{JB20a}, perhaps because it also relied on separate measurements of diffusivities and relaxation time scales.}

 \subsection{Diffusive fluctuations of $T_0$}

\begin{figure}
\includegraphics[width=0.45\textwidth]{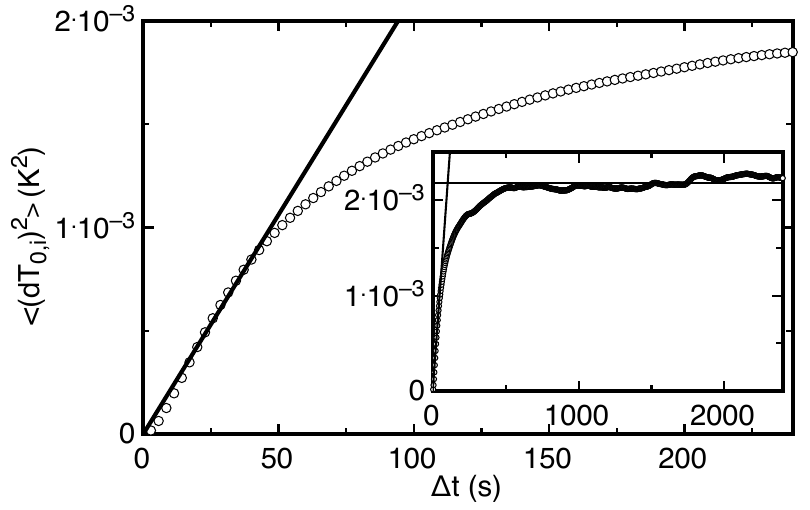} 
\caption{The mean square displacement of $T_{0,i}$.  The initial linear behavior indicates $T_{0,i}$ fluctuates diffusively.  Inset: same data shown over a larger time delay $\Delta t$.  The crossover to a plateau defines the damping timescale $\tau_{T}$ as the intersection of the two limiting fits. 
}
\label{fig:T0diffusivity}
\end{figure}

%diffusivity	
While the stochastic equations of motion for the LSC orientation $\theta_0$ (Eq.~\ref{eqn:theta_model}) and temperature amplitude $\delta$ (Eq.~\ref{eqn:delta_model}) have been tested previously for single convection rolls \cite{BA08a, JB20a}, the equation of motion for the mean temperature (Eq.~\ref{eqn:Tforcing}) is a entirely new.   To test whether the stochastic term representing turbulent fluctuations is diffusive, we measure the mean-square displacement $(dT_{0,i})^2$ of $T_{0,i}$ over different time intervals $d t$.   The mean-square displacement $(dT_{0,i})^2$ is averaged over the two cells and different starting times, and plotted as a function of time interval $d t$ in Fig.~\ref{fig:T0diffusivity}  for a counter-rotating state at Ra $= 2.7\times10^9$.  The apparent linearity in the limit of short time intervals is an indicator of diffusive fluctuations, as we assumed in the model Eq.~\ref{eqn:Tforcing}.  A linear function $\langle(dT_{0,i})^2\rangle= D_T dt$ is fit to the data for $d t < 35$ s to obtain a diffusivity $D_T$.  The time scale of 35 s is approximately the damping timescale $\tau_{\delta}$, which is likely relevant because $\delta_i$ appears in Eq.~\ref{eqn:Tforcing}, so fluctuations in $\delta_i$ may affect changes in $T_{0,i}$ at larger $d t$.  

%damping and plateau
The inset of Fig.~\ref{fig:T0diffusivity} shows the  mean-square displacement $(d T_{0,i})^2$ over a larger range of $d t$.  At large $dt$, the mean-square displacement reaches a plateau $(dT_{0,i})^2 \equiv \tau_{T}D_T$ which defines a damping timescale $\tau_{T}$ as the time of intersection of the two limiting scaling laws.

  \subsection{The effective turbulent thermal diffusivity $\kappa_t$}
 \label{sec:turbulentkappa}
 
 %counterrotating: magnitude of forcing in $\theta_1-\theta_2$ and turbulent diffusivity
Since we introduced the effective turbulent thermal diffusivity $\kappa_t$ as an unknown fit parameter, we do not have first-principles predictions of model terms.  On the other hand, since nearly all of the new deterministic forcing terms in Eqs.~\ref{eqn:deltaforcing}, \ref{eqn:thetaforcing},  and \ref{eqn:Tforcing} are proportional to $\kappa_t$, we can check the consistency of values of $\kappa_t$ required for different forcing terms to fit with data.  %Specifically, we compare the values of $\kappa_t$ required to obtain the forcing $\ddot\theta_{i,c} = b\sin(\theta_1-\theta_2)$ and the damping timescale $\tau_T$ over a range of Ra. 

\begin{figure}
\includegraphics[width=0.475\textwidth]{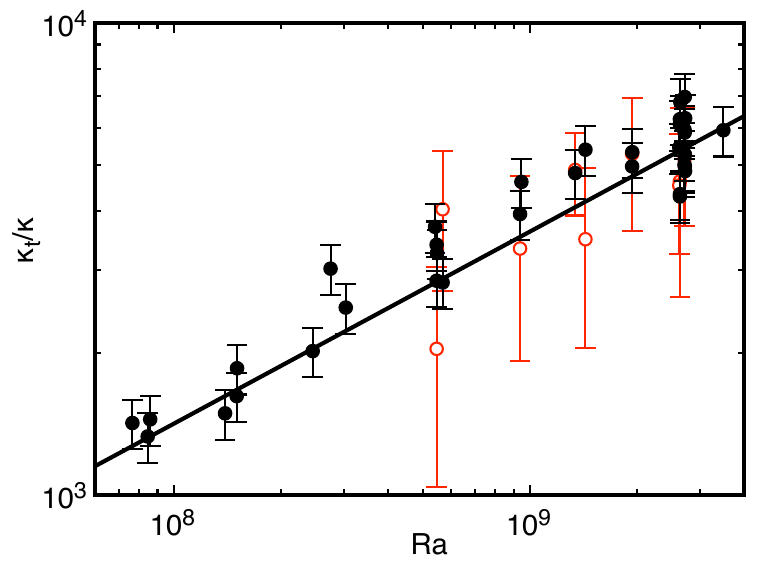} 
\caption{Turbulent thermal diffusivity $\kappa_t$ as a function of Rayleigh number Ra.   Open circles: from measurements of the forcing of $\ddot\theta_{i,c} = b\sin(\theta_1-\theta_2)$ in counter-diagonal states using multiple linear regression.  Solid circles: obtained from the damping timescale $\tau_{T}$ on the mean temperature $T_{0,i}$ from counter-rotating states.  These values are consistent, suggesting a single measurement can parameterize the value of $\kappa_t$ to use for predictions of all model terms that depend on it.
}
\label{fig:turbulentkappa}
\end{figure}

\subsubsection{$\kappa_t$ from $b$}

As one measure of $\kappa_t$, we compare  the measured stabilizing force in $b\sin(\theta_1-\theta_2)$ from  Eq.~\ref{eqn:regression} to the predicted linearized forcing in Eq.~\ref{eqn:thetaforcinglinearcounter} to obtain $\kappa_t/\kappa=4bL^2\tau_{\dot\theta}/A\kappa$.   We use measurements of $b$ from Sec.~\ref{sec:forcingb} for counter-diagonal states with small $\Delta T_m/\bar\delta < 0.1$, and measurements of $\tau_{\dot\theta}$ from the crossover time of limiting scalings of of the mean-square displacement of $\dot\theta_i$ (analogous to $\tau_{T}$) obtained in previous measurements of a single LSC \cite{JB20a}.  Counter-axis states, which are found at $\Delta T\le 2$ K are not included in the analysis because they usually have multiple peaks in the probability distribution of $\theta_i$, making the Gaussian fitting and linear regression algorithm to obtain $b$ unrepresentative of the forcing.  Errors are propagated from errors on $b$ obtained from the linear regression.  These values of $\kappa_t/\kappa$ and shown as a function of Ra in Fig.~\ref{fig:turbulentkappa}.  Unfortunately, the uncertainty on $b$ makes the uncertainties too large to draw strong conclusions about the trend over a small range of Ra from this data alone.

\subsubsection{$\kappa_t$ from the damping timescale $\tau_T$}
\label{sec:turbulentkappa_T0}

For another measure of $\kappa_t$, we use the damping timescale $\tau_T$  for fluctuations in $T_0$ seen in Fig.~\ref{fig:T0diffusivity}.  The coefficient $A\kappa_t/L^2$ of the second term of Eq.~\ref{eqn:Tforcing}  corresponds to an inverse damping timescale $1/\tau_T$ in the limit where $4\mbox{Nu}\kappa/A\kappa_t\ll 1$  (confirmed in Sec.~\ref{sec:DeltaTm}), and assuming the $\delta_i$ terms do not contribute significantly in the counter-rotating state (the $\delta_i$ terms cancel each other out at the stable fixed point).  This relation leads to $\kappa_t/\kappa = L^2/A\kappa\tau_T$, using the measured values of $\tau_T$ from the intersection of the two limiting scaling laws in Fig.~\ref{fig:T0diffusivity}.  The corresponding values of $\kappa_t/\kappa$ vs.~Ra are plotted in Fig.~\ref{fig:turbulentkappa} for counter-rotating states.  Errors are propagated from fits of  $(\Delta T_{0,i})^2$, which were calculated as half the difference between two cells, plus the error from fits of $D_T$, plus half the difference over a fit range from 0 to $0.6\tau_{\delta}$ or $0.8\tau_{\delta}$.  This resulted in an average error of 12\%.   A power law fit to the data obtained from $\tau_T$ yields $\kappa_t/\kappa =  (0.8\pm0.3)Ra^{0.41\pm0.02}$   with a reduced $\chi^2=1.5$. These measurements of $\kappa_t/\kappa$ based on the damping timescale $\tau_T$ are consistent with the data for $\kappa_t/\kappa$ based on measurements of the stiffness $b$, confirming that the same value of $\kappa_t$ drives both the $T_{0,i}$-term in Eq.~\ref{eqn:Tforcing} and the  $\theta_1-\theta_2$-term in Eq.~\ref{eqn:thetaforcing}.

% co-rotating state has some larger values of apparent $\tau_T$, probably due to $\delta$
%$\kappa_t/\kappa =  (75\pm22)Re^{0.64\pm0.04}$ using fit for Re
%201224, 210408: unusually low $\kappa_t$ from $\tau_T$
%increasing fit range by 33\% reduces diffusivity by 3\%
%uncertainty on fit 1.5%

%kappa_t from other measurements
In principle we could plot the forcing for other terms of Eqs.~\ref{eqn:deltaforcing}, \ref{eqn:thetaforcing},  and \ref{eqn:Tforcing}.  However, other measured factors depend on a small difference between two terms.  For example, Eq.~\ref{eqn:deltaforcing} has a difference in terms depending on $\delta_2$ and $T_{0,2}-T_{0,1}$, so that uncertainties on the data and model can heavily bias scaling laws from these small differences.   The values of $\kappa_t$ from measurements of $\tau_T$ remain our most precise method of measuring $\kappa_t$.     Thus, we will use this value of $\kappa_t$ obtained from measurements of $\tau_T$ for predictions of $\kappa_t$ in all other model terms.

\subsection{Stable fixed points of temperature amplitude $\bar\delta$ and mean temperature difference $\Delta \bar T_0$}
\label{sec:bardelta}

\subsubsection{Predictions for $\bar\delta_{co}$ and $\Delta \bar T_{0,co}$ in co-rotating states}  

%The stable fixed points for co-rotating states  in Eqs.~\ref{eqn:deltaforcing} and \ref{eqn:Tforcing} are not trivial solutions like for counter-rotating states.  
%Our observations of co-rotating states in Fig.~\ref{fig:switchingexample} indicate that $\bar\delta_{co} < \delta_0$, and $\Delta \bar T_{0,co}  \ne 0$.  
For co-rotating states where $\theta_1=0$ and $\theta_2=\pi$ ra for $\theta_1=\pi$ rad and $\theta_2=0$, the symmetry of Eqs.~\ref{eqn:deltaforcing} and \ref{eqn:Tforcing}  still requires $\delta_1=\delta_2$ at stable fixed points for co-rotating states, but they do not require $T_{0,1}=T_{0,2}$.  For the symmetric driving case of  $T_{m,2}=T_{m,1}$, in the Boussinesq limit where the mean temperature of the bulk equals the mean temperature of the plates, requires  $(T_{0,2}+T_{0,1})/2 = T_{m,i}$.  Defining the fixed point value of $\delta_i$ in the co-rotating state as $\bar\delta_{co}$, Eq.~\ref{eqn:Tforcing} and its corresponding equation for $\dot T_2$ have stable fixed points ($\dot T_{0,i} =0$) when
 \begin{equation}
\Delta \bar T_{0,co} = \pm \frac{1.76}{1+2\mbox{Nu}\kappa/A\kappa_t} \bar\delta_{co} \ % \approx 1.80 \bar\delta for B=0\ .
\label{eqn:barTco}
 \end{equation}
 where the $+$ sign corresponds to the co-rotating state orientation with $\theta_1=0$ and $\theta_2=\pi$ rad, and the $-$ sign corresponds to swapped orientations.  To obtain the the stable fixed point $\bar\delta_{co}$  in co-rotating states, we evaluate Eq.~\ref{eqn:deltaforcing} at $\theta_1=0$ and $\theta_2=\pi$ rad when $\Delta \bar T_{0,co}>0$ or  $\theta_1=\pi$ rad and $\theta_2=0$ when $\Delta \bar T_{0,co}<0$.  This simplifies Eq.~\ref{eqn:deltaforcing} to 
% \begin{equation}
$( \bar{\dot\delta}_{\kappa,i})_{co} =(A\kappa_t/\pi L^2)\left(1.41|\Delta \bar T_{0,co}|-2.57\delta_i\right)$.
% \label{eqn:deltaforcingcoDeltaT0}$$
% \end{equation}
 The stable fixed point value of $\bar\delta_{co}$ is obtained by using this for $\dot\delta_{\kappa,i}$ in Eq.~\ref{eqn:delta_model}, setting $\dot\delta_i=0$, and linearly expanding around $\delta_i=\delta_0$ to obtain
 \begin{equation}
\bar\delta_{co} \approx \delta_0+ \frac{(2A\kappa_t\tau_{\delta}/\pi L^2)(1.41|\Delta \bar T_{0,co}|-2.57\delta_0) }{1+2.57A\kappa_t\tau_{\delta}/\pi L^2}\ . 
 \label{eqn:bardeltacolinear}
 \end{equation}
 
 Note that a simultaneous evaluation of Eq.~\ref{eqn:barTco} and Eq.~\ref{eqn:bardeltacolinear} would lead to a subtraction of two comparable numbers when $2\mbox{Nu}\kappa/A\kappa_t$ is small, which can result in predictions of differing signs and arbitrarily small magnitudes when including uncertainties from assumptions in the model, so we cannot make accurate predictions of $\bar\delta_{co}$.  Instead, we will use the measured relationships between  $\Delta \bar T_{0,co}$ and $\bar\delta_{co}$ to text for self-consistency of Eqs.~\ref{eqn:barTco} and \ref{eqn:bardeltacolinear}.

 \subsubsection{Measurements of $\bar\delta$ and $\Delta \bar T_0$}
 
\begin{figure}
\includegraphics[width=0.45\textwidth]{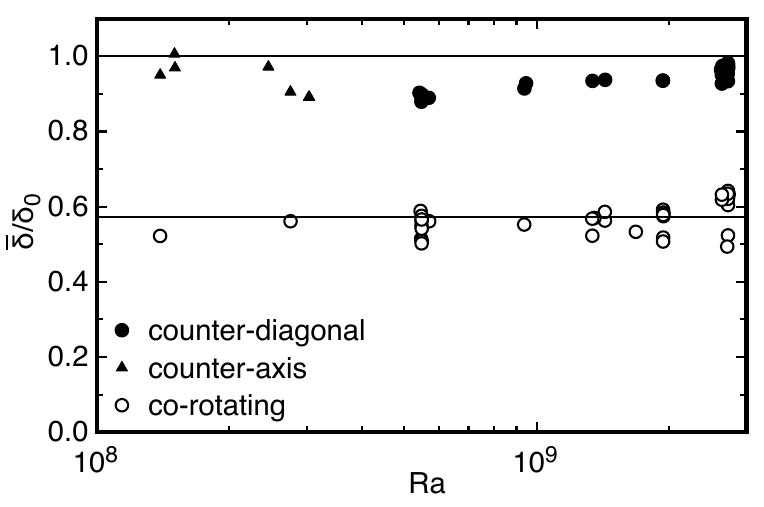} 
\caption{The mean LSC amplitude $\bar\delta$ relative to the value $\delta_0$ for a single LSC, at different Rayleigh numbers Ra.  Solid circles: counter-diagonal states.  Solid triangles: counter-axis states. Open circles: co-rotating states.  The amplitude in the co-rotating state has a consistent average $(0.57\pm0.04)\delta_0$.  The amplitude of counter rotating states is slightly smaller than the predicted value of 1.
}
\label{fig:delta_ra}
\end{figure}

To characterize the change in the mean LSC amplitude $\bar\delta$ due to the interaction between neighboring rolls, we plot values normalized by $\delta_0$ in Fig.~\ref{fig:delta_ra} as a function of Ra.     The average $\bar\delta$ is calculated as the average of $\delta_i$ over both cells and over time while the system remained in one state.  The normalization values of $\delta_0$ are obtained from a fit of a power law to  $\delta_0(Ra)$ from previous experiments with a single LSC in a cubic cell \cite{JB20a}.

 Counter-rotating states consistently have a mean amplitude $\bar\delta$ close to that of the single-cell, with $\bar\delta_{counter}/\delta_0 = 0.94$ with a standard deviation of 0.03, only slightly lower than the predicted value of 1 from Eq.~\ref{eqn:deltaforcing} when $\theta_1=\theta_2$.    This applies to both counter-diagonal and counter-axis states, so it is independent of the preferred orientation of the counter-rotating state.  The slight decrease in $\bar\delta$ from $\delta_0$ may be partly due to fluctuations in $\theta_1-\theta_2$ away from the mean of zero, which can make the third term of Eq.~\ref{eqn:deltaforcing} negative.  Co-rotating states have a smaller mean amplitude with a consistent ratio $\bar\delta_{co}/\delta_0 = 0.57$ with a standard deviation of 0.04.    
  
\begin{figure}
\includegraphics[width=0.45\textwidth]{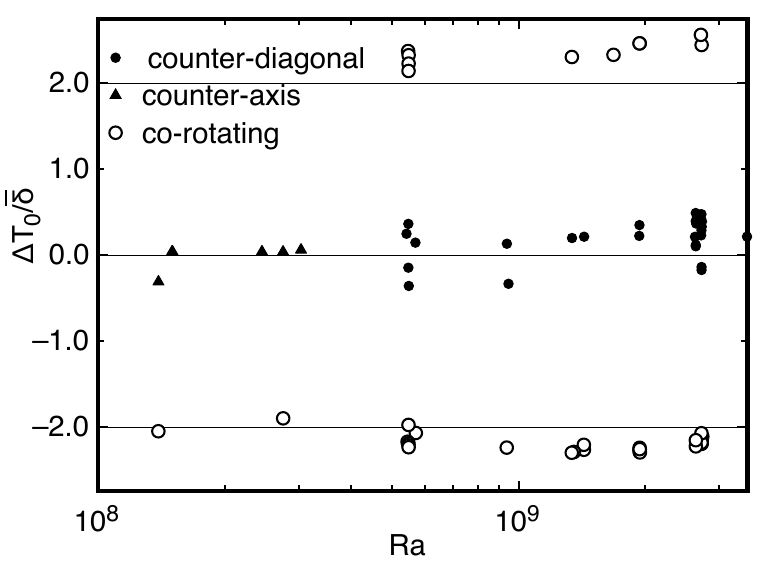} 
\caption{The mean temperature difference between neighboring rolls $\Delta \bar T_0$ normalized by $\bar\delta$ for different Rayleigh numbers.  Solid circles: counter-diagonal states.  Solid triangles: counter-axis states. Open circles: co-rotating states.  Counter-rotating states are scattered around zero.  Co-rotating states have a consistent average $|\Delta \bar T_{0,co}|/\bar\delta_{co} = 2.22\pm0.13$, 30\% larger than predicted due to turbulent thermal diffusion.  Co-rotating states are stable because this temperature difference is large enough to orient their temperature gradients in the same direction as heat flows from one cell to the other. 
}
\label{fig:DeltaT0_ra}
\end{figure}

%\Delta \bar T_0
  We measured $\Delta \bar T_0 \equiv \bar T_{0,2}-\bar T_{0,1}$ as an average over time while the system remained in one state.  Data are normalized by $\bar\delta$ for a sense of scale.  Counter-rotating states are scattered around zero, but the values are much larger than uncertainties on temperature measurements, which will be explained due to small differences in $\Delta T_m$ in Sec.~\ref{sec:DeltaTm}.    Co-rotating states have a consistent average $|\Delta \bar T_{0,co}|/\bar\delta_{co} = 2.22\pm0.13$ (the $\pm$ refers to one standard deviation), 30\% larger than predicted due to effective turbulent thermal diffusion from Eq.~\ref{eqn:barTco} assuming $2B/A\ll 1$.  This is well within the typical errors of this modeling approach of a factor of 3 \cite{BA08a}.
  
    %advection contribution
Although the measured ratio of $|\Delta \bar T_{0,co}|/\bar\delta_{co}$ is close to the prediction, the value  $|\Delta \bar T_{0,co}|/\bar\delta_{co}>2$ has some significance.  It indicates that the cold cell is colder than the hot cell along the entire interface, as well as being colder than the mean of the plates, so turbulent thermal diffusion is expected to provide a net heat flux into the cold cell at this value of $|\Delta \bar T_{0,co}|/\bar\delta_{co}$, regardless of assumptions about how the shape of the temperature profile affects the model terms.  Thus, this magnitude  $|\Delta \bar T_{0,co}|/\bar\delta_{co} >2$ is a violation of the assumption in Eq.~\ref{eqn:Tdot} that $T_{0,i}$ is driven by turbulent thermal diffusion only.  An additional heat transport mechanism is required, which is likely advection of fluid in a coherent flow from one cell to the other.  For example,  a coherent advective flow parallel to the plates has been observed in other co-rotating states \cite{PB03, PMELB04, XX13, GSPXQT13, MBAK19}.  This coherent advection can transport heat from the hot plate of the cold cell directly to the hot cell and from the cold plate of the hot cell directly to the cold cell to balance the diffusive heat transport between cells. 

%self-consistent test of \bar\delta_co
 While we could not accurately predict the value of $\bar\delta_{co}$ from Eq.~\ref{eqn:bardeltacolinear}, we can provide a self-consistency test of the model by using the measured ratio $|\Delta \bar T_{0,co}|/\bar\delta_{co}$ as empirical input into Eq.~\ref{eqn:bardeltacolinear}.  Since the product of $\kappa_t$ and $\tau_{\delta}$ has a mild dependence with $Ra^{-0.17}$, we estimate a value of $\bar\delta_{co}$ at the middle of the range of Ra.   Using empirical input into Eq.~\ref{eqn:bardeltacolinear} of $|\Delta \bar T_{0,co}|/\bar\delta_{co}=2.22\pm0.13$ from Fig.~\ref{fig:DeltaT0_ra}, $\bar\delta_{co}=0.57\delta_0$ from Fig.~\ref{fig:delta_ra}, $\kappa_t/\kappa=3200\pm 400$ from the average and standard deviation of five measurements at Ra $=5.5\times10^8$  in Fig.~\ref{sec:turbulentkappa}, and $\tau_{\delta} = 81$ s from a fit of measurements of a single LSC in a cubic cell \cite{JB20a}, results in $\bar\delta_{co}=(0.83\pm0.07)\delta_0$.  This underestimates the reduction in $\bar\delta$ in the co-rotating state by 60\% with input of $\Delta \bar T_{0,co}$.  The variation of this prediction at the extremes of the measured range of Ra results in a difference of $0.05\delta_0$.

\section{Testing the model with a difference in mean temperature of the plates of the two cells $\Delta T_m$}
\label{sec:DeltaTm}

%transition statement
In this section we extend the predictions of Eqs.~\ref{eqn:deltaforcing}, \ref{eqn:thetaforcing}, and \ref{eqn:Tforcing} and test them in cases where the difference  in the mean temperatures of the plates of the two cells $\Delta T_m \ne 0$.  Controlling $\Delta T_m$ allows testing several terms of these equations which depend on $\Delta \bar T_0$, explains how some asymmetries in counter-rotating states come about due to a small unintentional $\Delta T_m$ in experiments, and helps identify the conditions for stable counter- or co-rotating states.

\begin{figure}
\includegraphics[width=0.475\textwidth]{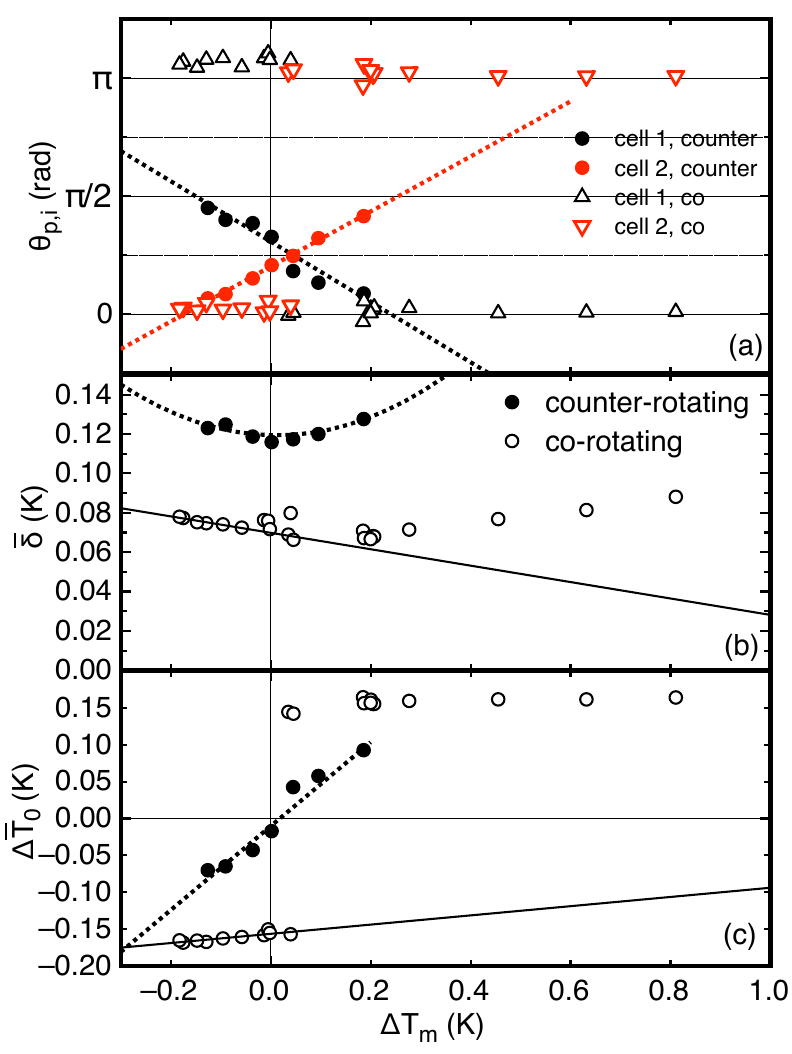} 
\caption{Stable fixed point parameter values with a difference between the mean temperature of the plates of the two cells $\Delta T_m$.     (a) The preferred orientations $\theta_{p,i}$. (b) The mean amplitude $\bar\delta$, averaged over the two cells.  (c) The difference in mean temperature of the two rolls $\Delta \bar T_0$. Solid symbols: counter-rotating states.  Open symbols: co-rotating states.  Solid lines: fits to co-rotating data for $\Delta T_m<0$.  Dotted lines: fits to counter-rotating data.    In the counter-rotating state, a small $\Delta T_m$ drives a temperature difference $\Delta \bar T_0$, which drives a change in preferred orientation $\theta_{p,i}$ in different directions in the two cells.   With a large enough $\Delta T_m$, this forcing destabilizes the counter-rotating state so only the co-rotating state is stable.
}
\label{fig:deltameanplate}
\end{figure}
% all data aat tilt of .012 degrees 201212 at different tilt angle
%results
 Figure \ref{fig:deltameanplate} shows measurements of the stable fixed points $\theta_{p,i}$, $\bar\delta \equiv (\delta_1+\delta_2)/2$, and $\Delta \bar T_0$ for different values of $\Delta T_m$
 at  Ra=$5.5\times10^8$ ($\Delta T=3.8$ K).  A smaller value of Ra was chosen for this set of experiments because switching between states is more frequent at this Ra \cite{BJB16}, making the sampling of different states easier.  A counter-rotating state is found for small $\Delta T_m$, with $\theta_p = \pi/4$ rad in both cells at $\Delta T_m=0.04$ K.  This offset from zero suggests a slight asymmetry in the cells resulting in a horizontal temperature difference that is 1\% of $\Delta T$.  For small $\Delta T_m$, $\theta_p$ varies with  $\Delta T_m$ in opposite directions in the two cells, corresponding to counter-offset states.  $\bar\delta$ and the magnitude of $\Delta \bar T_0$ also increase with $|\Delta T_m|$ for these counter-offset states.   At larger  $|\Delta T_m|$, only co-rotating states are found. 

The different states in  Fig.~\ref{fig:deltameanplate} can be understood qualitatively from Eqs.~\ref{eqn:thetaforcing} and \ref{eqn:Tforcing}.  When $\Delta T_m>0$ and thus $\Delta \bar T_0>0$ from Eq.~\ref{eqn:Tforcing}, the interface between cells is hotter than cell 1, so there will be a forcing on $\theta_1$ towards the interface with the hotter cell according to Eq.~\ref{eqn:thetaforcing}.  $\theta_2$ is forced in the opposite direction as the interface is colder than cell 2.  In the counter-rotating state, this results in the opposite shifts in $\theta_{p,i}$ with increasing $|\Delta T_m|$.  When the forcing in Eq.~\ref{eqn:thetaforcing} becomes large enough as $|\Delta T_m|$ and $|\Delta \bar T_0|$ increase, it pushes $\theta_p$ across an unstable fixed point of Eq.~\ref{eqn:thetaforcing} at $\theta_p=\pi/2$ rad so that only a co-rotating state is stable. In agreement with this prediction, the measured values of $\theta_p$ for counter-offset states  in Fig.~\ref{fig:deltameanplate}a approach but do not cross $\pi/2$ rad, and beyond that value of $\Delta T_m$ only co-rotating states are found.  

%motivation-priority
In principle we could solve the coupled Eqs.~\ref{eqn:deltaforcing}, \ref{eqn:thetaforcing}, and \ref{eqn:Tforcing}, to make direct predictions of how $\theta_p$ and $\Delta \bar T_0$ depend directly on $\Delta T_m$.  However, the predictions end up being due to small differences between terms, which are sensitive to uncertainties in  parameter values.  Since the parameters have uncertainties as large as a factor of 3, resulting model predictions can have much larger uncertainties and even an uncertain sign.   Instead, we evaluate stable fixed point values for one parameter at a time, using other parameters as input, to test the direct dependences of the model and check for self-consistency.

 \subsection{Offset of the preferred orientations $\Delta\theta_p$ in counter-offset states}
\label{sec:thetap_counteroffset}

%counter
In counter-rotating states when $|\Delta \bar T_0| > 0$, the third term of Eq.~\ref{eqn:thetaforcing} gives a forcing on $\ddot\theta_1$ of $-\sqrt{2}A\kappa_t \Delta \bar T_0 \sin\theta_1/\pi L^2\bar\delta\tau_{\dot\theta}$, 
 %\begin{equation}
%\dot\theta_{1,\Delta \bar T_0} = -\frac{\sqrt{2}A\kappa_t \Delta \bar T_0 \sin\theta_1}{2\pi H^2\delta_1} \ .
 %label{eqn:thetaforcingcount.0102er_DeltaTm}
% \end{equation}
 which pushes the orientation $\theta_p$ of the LSC in the colder cell towards alignment with the hotter cell, and pushes $\theta_p$ in the hotter cell away from the colder cell, so that the $\theta_p$ is pushed equally in opposite directions in the two cells.   We define the shift in preferred orientations $\Delta\theta_p \equiv (\theta_{p,2}-\theta_{p,1})/2$. Treating this as a small perturbation away from a counter-diagonal state where $\sin\theta_1\approx (\theta_1-\theta_p)/\sqrt{2}$, $\Delta \theta_p$ is obtained by balancing this forcing from Eq.~\ref{eqn:thetaforcing} with the overdamped forcing obtained in a single cubic cell  $-\nabla V_g = -\omega_r^2(\theta_1-\theta_p)$, where $\omega_r$ is the natural frequency of oscillation in the geometric potential around the corner  \cite{JB20a}.  At lowest order, this leads to a shift in $\theta_p$ of 
 \begin{equation}
\Delta\theta_p=  \pm \frac{A\kappa_t\Delta \bar T_0}{\pi L^2\bar\delta\omega_r^2\tau_{\dot\theta}} \ .
 \label{eqn:theta_stable_DeltaTm}
 \end{equation}
The $\pm$ sign corresponds to the sign of $\sin\theta_{p}$.   %Due to the shift in $\Delta\theta_p$, additional terms in Eq.~\ref{eqn:thetaforcing} become non-zero in higher-order corrections, but these remain less than 30\% of the term proportional to $\Delta \bar T_0$.

%comparison of slope
 %A\kappa/\pi L^2 = 5.64\times10^{-7}$
To test the prediction of $\Delta \theta_p$ from Eq.~\ref{eqn:theta_stable_DeltaTm}, we use as input  $\tau_{\dot\theta}=22$ s , and $\omega_r^2 = (2.1\pm0.4)\times10^{-4}$ s$^{-2}$ from the linear regression of Sec.~\ref{sec:regression}, both from data at Ra=$5.4\times10^8$  for a single LSC in a cubic cell \cite{JB20a} \footnote{This value of $\omega_r^2$ is about half of the value reported in \cite{JB20a}, which was obtained less directly through measurements of the  mean-square displacement of $\dot\theta_0$ divided by the variance  of the probability distribution of $\theta_0$}. We also use as input the measured value $\bar\delta = 0.12$ K, a linear fit of $\Delta \bar T_0 = (0.57\pm0.05)\Delta T_m$ from Fig.~\ref{fig:deltameanplate}c, and $\kappa_t/\kappa=3200\pm 400$ at Ra $=5.5\times10^8$ from Fig.~\ref{fig:turbulentkappa}.  This yields the prediction $\Delta\theta_p = (1.9\pm0.2)\Delta T_m$ rad/K.  A linear fit to $\theta_p$ in Fig.~\ref{fig:deltameanplate}a yields $\Delta\theta_p = (3.7\pm0.4)\Delta T_m$ rad/K, which is within a factor of 2 of the prediction of Eq.~\ref{eqn:theta_stable_DeltaTm}.  This discrepancy also means Eq.~\ref{eqn:theta_stable_DeltaTm} overpredicts the maximum $\Delta \bar T_0$ where counter-offset states are stable by about the same factor, by extrapolating $\Delta\theta_p$ to where it crosses the unstable fixed point of Eq.~\ref{eqn:thetaforcing} at $\theta_i=\pi/2$ rad.

\subsection{Increase of steady-state temperature difference $\Delta \bar T_{0,co}$ with $\Delta T_m$ in co-rotating states}
\label{sec:DeltaT0_DeltaTm_co}
 %co
  In the co-rotating state, Eq.~\ref{eqn:Tforcing} has a stable fixed point solution
 \begin{equation}
\Delta \bar T_{0,co}(\Delta T_m) = \frac{\pm1.76\bar\delta_{co} + (2\mbox{Nu}\kappa/A\kappa_t)\Delta T_m}{1+2\mbox{Nu}\kappa/A\kappa_t} \ . 
\label{eqn:DeltaT0co}
\end{equation}
\noindent The $\pm$ corresponds to the sign of $\Delta T_m$.   Values of Nu are known from fits of the Grossmann-Lohse scaling model \cite{AGL09}. Specifically, at $Ra=5.5\times10^8$,   Nu=55 in a cylindrical cell at the same Ra \cite{FBNA05} (values of Nu in cubic and cylindrical cells are found to agree within 2\% in this range \cite{KX13}).  This corresponds to $\mbox{Nu} \kappa/\kappa_t = 0.017\pm0.002$.  Using this value,  and the fit slope $|\partial\bar\delta_{co}/\partial \Delta T_m|= (0.042\pm0.006)\Delta T_m$ for $\Delta T_m <-0.02$ K, yields a prediction of $|\partial\Delta \bar T_{0,co}/\partial \Delta T_m|= 0.14\pm0.02$, about a factor of 2 larger than the measured slope $|\partial\Delta \bar T_{0,co}/\partial \Delta T_m| = (0.062\pm0.014)$.  However, we note that the trends start to change for values of $\Delta T_m>0.2$, such that the discrepancy becomes a factor of 5 in the range $0.2 < \Delta T_m < 1$.

%In principle, Eq.~\ref{eqn:DeltaT0co} and the measured trends could be used to obtain an indirect measurement of $B$, however due the typical uncertainty in model parameters of a factor of 3, taking a difference between two comparable terms leads to an unphysical negative value of $B$.

 \subsection{Increase of $\Delta \bar T_0$ with $\Delta T_m$ in counter-rotating states}
 \label{sec:DeltaT0_DeltaTm}
 
 %counter -nonideal
While ideal symmetric counter-rotating states correspond to $\Delta T_m=0$, in practice the LSC is very sensitive to asymmetries \cite{KH81,BA06b}.  Even in nominally symmetric counter-rotating states, $\Delta T_m$ is not exactly zero. In our experiment, $\Delta T_m$ is not directly controlled.  Rather, we control the temperature of the water baths pumping water through the top and bottom plates, but the plate temperatures are also affected by heat losses in the piping from the baths to the plates, and the finite conductivity of the aluminum top and bottom plates in our experiments allows them to be coupled to the temperature profile of the LSC.  In practice, this coupling results in a typical change in $\Delta T_m$ between counter-rotating and co-rotating states of $0.002\Delta T$.  That small asymmetry in driving has a large consequence, with a $|\Delta \bar T_0|/\bar\delta$ typically varying by 0.1 in cases where we intended to produce symmetric counter-rotating states, which can lead to significant terms in the model Eqs.~\ref{eqn:deltaforcing}, \ref{eqn:thetaforcing},  and \ref{eqn:Tforcing} in counter-rotating states. In particular, a  general steady-state solution of Eq.~\ref{eqn:Tforcing} for both counter- and co-rotating states that includes variations in $\theta_{p,i}$ and $\bar\delta_i$ with $\Delta T_m$ is
\begin{equation}
\Delta \bar T_0 = \frac{-0.88(\bar\delta_2\cos\theta_{p,2}-\bar\delta_1\cos\theta_{p,1}) + (2\mbox{Nu}\kappa/A\kappa_t)\Delta T_m}{1+2\mbox{Nu}\kappa/A\kappa_t} \ . 
\label{eqn:DeltaT0}
\end{equation}

\begin{figure}
\includegraphics[width=0.4\textwidth]{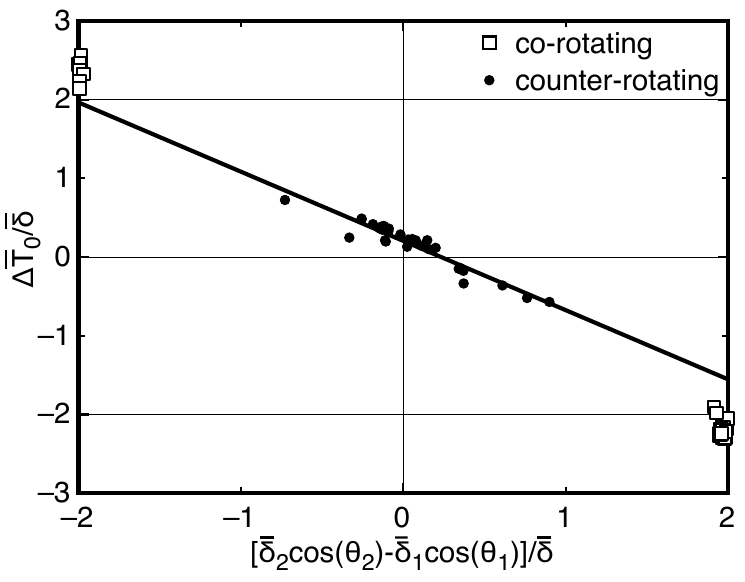} 
\caption{Steady-state values of the difference in mean temperature $\Delta \bar T_0$ as a function of parameters predicted in Eq.~\ref{eqn:DeltaT0}.  The linear fit to counter-rotating state data is consistent with the predicted scaling,  confirming the validity of the detailed functional form of the neighboring roll interaction terms in Eq.~\ref{eqn:Tforcing} and suggesting that many of the asymmetries observed in counter-rotating states are explained by small differences in $\Delta T_m$.
}
\label{fig:DeltaT0_Ddeltacostheta}
\end{figure}

To test the $\theta_{p,i}$-dependent terms of Eq.~\ref{eqn:DeltaT0}, we plot measured values of $\Delta \bar T_0/\bar\delta$ as a function of measured values of $[\bar\delta_2\cos\theta_{p,2}-\bar\delta_1\cos\theta_1]/\bar\delta$ in Fig.~\ref{fig:DeltaT0_Ddeltacostheta}.   The normalization of both axes by $\bar\delta$ allows us to include all of our counter-rotating data at different Ra on the same scale to better interpret the magnitude of the asymmetry.  Co-rotating state data with $\Delta T_m/\bar\delta < 1.3$ are shown to be tightly clustered because $\delta_2\cos\theta_{p,2}-\delta_1\cos\theta_1= 2\bar\delta_{co}$ consistently in co-rotating states with small $\Delta T_m$.  For counter-rotating data, a linear function plus a constant is fit to the data in Fig.~\ref{fig:DeltaT0_Ddeltacostheta}.  The fit yields a slope of $-0.88\pm0.05$ plus a constant offset of $0.20\pm0.02$.  The offset must be the result of some undetermined asymmetry in the nominally symmetric system.  The predicted slope in Fig.~\ref{fig:DeltaT0_Ddeltacostheta}  is $-0.93$, obtained by rearranging Eq.~\ref{eqn:DeltaT0} to isolate $\Delta \bar T_0$, $\Delta T_m = \Delta \bar T_0/0.57$ from a fit of Fig.~\ref{fig:deltameanplate}c, and the measured value $\mbox{Nu}\kappa/\kappa_t=0.017\pm0.002$.  The predicted slope is within a standard deviation of the fit slope of $-0.88\pm 0.05$ in Fig.~\ref{fig:DeltaT0_Ddeltacostheta}, so this confirms the validity of Eq.~\ref{eqn:DeltaT0}, and thus the detailed functional form of the third term of Eq.~\ref{eqn:Tforcing}.     Since Eqs.~\ref{eqn:deltaforcing}  and \ref{eqn:thetaforcing} for $\dot\delta_{i,\kappa}$ and $\dot\theta_{i,\kappa}$ depend directly on $T_0$, the small uncontrolled differences in $\Delta T_m$ and resulting changes in $\Delta \bar T_0$  likely account for much of the scatter in plots such as Fig.~\ref{fig:thetap1_thetap2}, and corresponding asymmetries in Fig.~\ref{fig:switchingexample}.

\subsection{Trends of $\bar\delta_i$ with  $\Delta T_m$}
 \label{sec:delta_DeltaTm}
 
\subsubsection{Co-rotating states}
%co
In the co-rotating state, $\bar\delta$ increases linearly with $\Delta T_m$ in Fig.~\ref{fig:deltameanplate}b.  This change is predicted from the linear expansion in Eq.~\ref{eqn:bardeltacolinear} since $\Delta \bar T_{0,co}$ also increases linearly with $\Delta T_m$.  Using the fit $\partial \Delta \bar T_{0,co}/\partial\Delta T_m = (0.061\pm0.014)$ in the co-rotating state for $|\Delta T_m|<0.2$ K from Fig.~\ref{fig:deltameanplate}c, $\tau_\delta=81$ s from a cubic cell with a single LSC  at the same Ra $=5.5\times10^8$ \cite{JB20a}, and $\kappa_t/\kappa=3200\pm400$ from the fit in Fig.~\ref{fig:turbulentkappa} yields a prediction $|\partial \bar\delta_{co}/\partial\Delta T_m|= (0.019\pm0.004)$, a factor of 2 smaller than the measured slope $|\partial \bar\delta_{co}/\partial\Delta T_m| = 0.042\pm 0.006$ for $\Delta T_m < -0.2$ K   in Fig.~\ref{fig:deltameanplate}b.  There is a reduction in the slope of both $\Delta \bar T_{0,co}$ and $\bar\delta_{co}$ with $\Delta T_m$ for 0.2 K $ < \Delta T_m < 1$ K.  This change in slope is not predicted by the model.  In the range 0.2 K $ < \Delta T_m < 1$ K, using the measured slope  $\partial \Delta \bar T_{0,co}/\partial\Delta T_m = (0.019\pm0.007)$ the predicted slope $|\partial \bar\delta_{co}/\partial\Delta T_m| = 0.006\pm0.001$ is 5 times smaller than the measured slope $0.031\pm 0.002$.  
%page 48 notebook

%The model consistently underestimates the change in $\bar\delta$ in co-rotating states, both here and in $\delta_{co}$ at $\Delta T_m=0$.
%Decrease in $\tau_{\Delta \bar T_0}$ may account for 12\% increase in $\bar\delta$
%should the factor 0.55 in Eq.~ref{eqn:deltabarcolinear} be larger?

\subsubsection{Asymmetry $\delta_2-\delta_1$ in Counter-rotating states}

\begin{figure}
\includegraphics[width=0.4\textwidth]{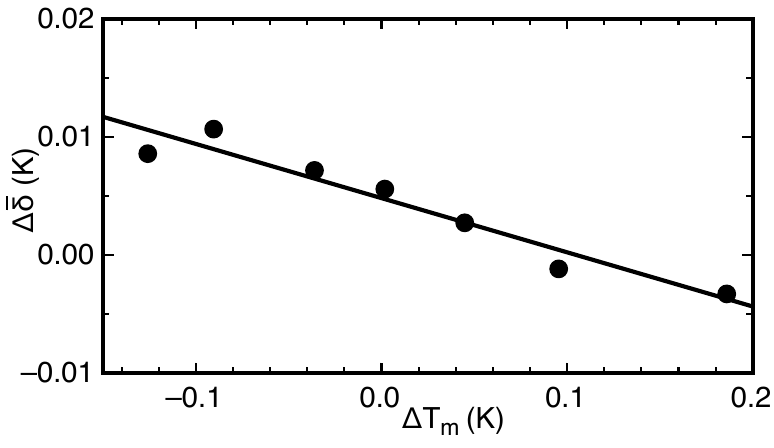} 
\caption{Difference in $\bar\delta$ between neighboring rolls in counter-rotating states.   The line is a linear fit based on Eq.~\ref{eqn:Deltadelta}.  The linearity and consistency with the magnitude of the prediction within 20\% confirm the validity of the detailed functional forms of the first and second terms of Eq.~\ref{eqn:deltaforcing}.
}
\label{fig:Deltadelta_DeltaTm_counter}
\end{figure}
 
%counter
When a small $\Delta T_m$ produces counter-offset states, there is a shift in $\bar\delta$ with increasing $\Delta T_m$ in opposite directions in the two cells.   This shift $\Delta\bar\delta \equiv(\bar\delta_2-\bar\delta_1)/2$ is shown in Fig.~\ref{fig:Deltadelta_DeltaTm_counter} for the counter-offset data from Fig.~\ref{fig:deltameanplate}.  Such a difference is predicted from Eq.~\ref{eqn:deltaforcing}.  $\Delta \bar\delta$ is obtained by inserting Eq.~\ref{eqn:deltaforcing} with a linear expansion in $\Delta\theta_p$ around a diagonal (e.g.~$\theta_p=\pi/4$ rad)  into a linear expansion around the stable fixed point of $\delta=\delta_0$ in Eq.~\ref{eqn:delta_model}, resulting in
\begin{equation}
\Delta \bar\delta = 2\tau_{\delta}\left(\frac{\dot\delta_{\kappa,2}-\dot\delta_{\kappa,1}}{2}\right)_{counter} = \frac{2\tau_\delta A\kappa_t}{\pi L^2}(\bar\delta\Delta\theta_p-\Delta \bar T_0 ) \ . 
\label{eqn:Deltadelta}
\end{equation}
 Using $\tau_{\delta} = 81$ s from measurements in a single cubic cell \cite{JB20a}, the fits of $\Delta \bar T_0=0.57\Delta T_m$ and $\Delta \theta_p = (3.7\pm0.4)\Delta T_m$ rad/K from Fig.~\ref{fig:deltameanplate}, and measurements $\kappa_t/\kappa=3200\pm400$ and $\bar\delta =0.12$ K, yields the prediction $\Delta\bar\delta = (-0.039\pm0.006)\Delta T_m$.    The data in Fig.~\ref{fig:Deltadelta_DeltaTm_counter} is fit by a linear function plus a constant, yielding  $\Delta\bar\delta = (-0.046\pm0.006)\Delta T_m +0.005$ K. The offset of 0.005 K is indicative of an asymmetry of the setup, comparable to the uncertainty on thermistor measurements.  The fit slope is consistent with the predicted value and within 20\%, which confirms the validity of the detailed form of the first and second terms of Eq.~\ref{eqn:deltaforcing}.

\subsubsection{Second order expansion of $\bar\delta(\Delta T_m)$}

Figure \ref{fig:deltameanplate}b showed an increase in $\bar\delta$ with $|\Delta T_m|$ in counter-offset states.  Since it increases with both positive and negative $\Delta T_m$, we fit a quadratic function, obtaining a curvature $\partial^2\bar\delta/\partial \Delta T_m^2 = 0.26\pm 0.11$/K.  
This quadratic trend can be predicted from Eq.~\ref{eqn:deltaforcing} using the prediction and observation from Fig.~\ref{fig:deltameanplate} that both $\Delta\theta_p$ and $\Delta \bar T_0$ are linear in $\Delta T_m$, so that $\cos\theta_{p,i} = \sqrt{2}(1\pm\Delta\theta_p)$.   Averaging over cells 1 and 2, the first-order expansions  in $\Delta \bar T_0$ or $\theta_p$ of the first two terms of Eq.~\ref{eqn:deltaforcing} have opposite signs for the two different cells at a diagonal (e.g.~$\theta_i=\pi/4$ rad), and cancel out in the average over the two cells (these are the terms that were responsible for $\Delta\bar\delta$).  The  expansion of the first and third terms of Eq.~\ref{eqn:deltaforcing} up to second-order in $\Delta \theta_p$ and $\Delta \bar T_0$ yields
%\begin{equation}
$\overline{\dot\delta}_{counter}  = (A\kappa_t/\pi L^2)(2\Delta \theta_p\Delta \bar T_0-\pi\delta_0\Delta\theta_p^2/8)$. 
%\end{equation}
 Plugging this into a first-order expansion of Eq.~\ref{eqn:delta_model}, using the fit $\Delta \bar T_0 = (0.57\pm 0.05)\Delta T_m$ from Fig.~\ref{fig:deltameanplate}c, the fit $\Delta\theta_p = (3.7\pm0.4)\Delta T_m$ rad/K from Fig.~\ref{fig:deltameanplate}a, and using the measured value of $\bar\delta$ at $\Delta T_m=0$ to approximate $\delta_0=0.12$ K yields the prediction 
\begin{equation}
\bar\delta_{counter} = \delta_0 +2\tau_{\delta}\overline{\dot\delta}_{counter} = \delta_0 + \frac{3.6A\kappa_t\tau_{\delta}}{\pi L^2}\Delta T_m^2
\end{equation}
Using $\tau_{\delta}=81$ s \cite{JB20a} and $\kappa_t/\kappa=3200\pm 400$, the predicted curvature in $\Delta T_m^2$ is $0.54\pm0.18$, %(adding errors in quadrature), 
consistent with the fit curvature $0.26\pm 0.11$/K.  
This confirms that Eq.~\ref{eqn:deltaforcing} is consistent with measurements even to 2nd order. 

%\subsection{Self-consistency of $\Delta T_m$-dependence of model}
The self-consistency of all of the model predictions of the dependencies of $\Delta\theta_p$, $\Delta \bar T_0$, $\bar\delta$, and $\Delta\bar\delta$  on $\Delta T_m$ was confirmed within a factor of 3 in Sec.~\ref{sec:thetap_counteroffset}, \ref{sec:DeltaT0_DeltaTm_co}, \ref{sec:DeltaT0_DeltaTm}, and \ref{sec:delta_DeltaTm}.  This confirms the validity of the dependence of Eqs.~\ref{eqn:deltaforcing}, \ref{eqn:thetaforcing},  and \ref{eqn:Tforcing} on $\Delta T_m$.

\subsection{Does the model quantitatively predict the conditions for stable co-rotating states?}
\label{sec:stabilityratio}
 
 \begin{figure}
\includegraphics[width=0.45\textwidth]{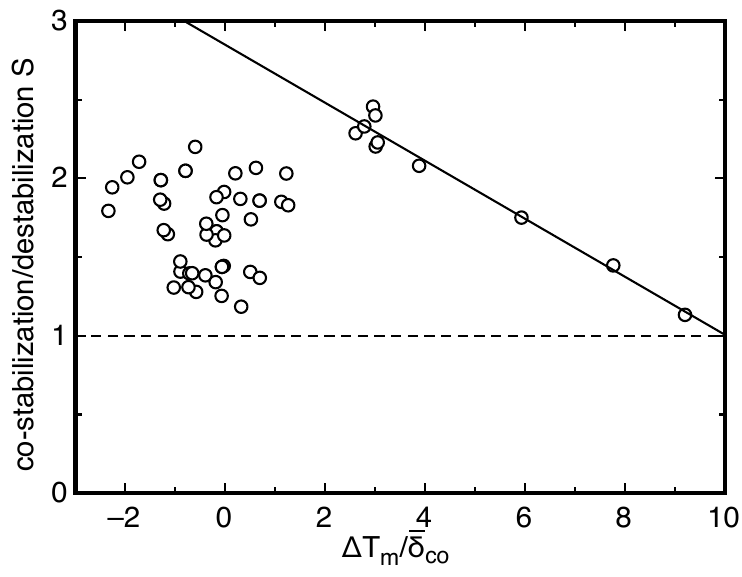} 
\caption{The ratio $S$ of stabilizing force from effective turbulent thermal diffusion at the orientations of co-rotating states to the destabilizing forcing from the geometry of the cell at these orientations.   Co-rotating states exist when this value is greater than $S > 1.13$, close to the predicted value of 1.  This confirms that the model prediction that co-rotating states are stable when the neighboring roll interaction due to turbulent thermal diffusion is strong enough to overcome the geometric forcing. Solid line: linear extrapolation of $S$ with increasing $\Delta T_m$.  The co-rotating state is predicted to become unstable when this drops below 1.
 }
\label{fig:asymmetricforcing_DeltaTm}
\end{figure}

%see deltacosDeltatheta_omegar2
%DeltaT0driving_omegar2

In this section, we check whether the model can quantitatively predict the conditions in which co-rotating states are stable.  To be stable at the alignment $\theta_p=0$ or $\pi$ rad,  there has to be a forcing to overcome the geometric forcing $-\nabla V_g$, which has unstable fixed points at $\theta_p=0$ and $\pi$ rad \cite{JB20a}. A first-order expansion in $\theta_i-\theta_p$ at fixed $\delta_i$ of the predicted unstable  geometric forcing is $-\nabla V_g \approx \pi\omega_r^2\bar\delta_{co}^2(\theta_1-\theta_p)/10\delta_0^2$ \cite{JB20a}.   This expansion for the geometric potential was untested due to limited data near the unstable fixed points of the potential $V_g$ in previous experiments \cite{JB20a}. We compare this predicted unstable forcing to the stabilizing forcing from the neighboring roll interaction, using a linear expansion in $\theta_i-\theta_p$ of Eq.~\ref{eqn:thetaforcing} for a fixed $\delta_i$  (this corresponds to the second term of Eq.~\ref{eqn:thetaforcinglinearco}).  We calculate the magnitude of the ratio $S$ of stabilizing forcing for co-rotating states to the destabilizing forcing from the expansion of the geometric forcing$-\nabla V_g$ to obtain the forcing ratio
\begin{equation}
S = \frac{10A\kappa_t\delta_0[\sqrt{2}|\Delta \bar T_{0,co}|/\bar\delta_{co}-2]}{\pi^2 L^2 \tau_{\dot\theta}\omega_r^2\bar\delta_{co}} \ .
\label{eqn:asymmetricforcing}
\end{equation}
To predict values of $S$ for different datasets, we use $\kappa_t/\kappa=0.8Ra^{0.41}$ from the fit in Fig.~\ref{fig:turbulentkappa},  $\tau_{\dot\theta} = 2.7\times10^4Ra^{-0.35}$ from a fit of data with a single LSC in a cubic cell data \cite{JB20a}, $\omega_r^2 = 6.3\times10^{-13}Ra^{0.98}$ from linear regression of data with a single LSC in a cubic cell \cite{JB20a}, $\bar\delta/\delta_0=0.57$ for co-rotating states from Fig.~\ref{fig:delta_ra},  and the measured $\Delta \bar T_{0,co}/\bar\delta_{co}$ for each dataset.  The ratio of stabilizing to destabilizing forces  is plotted Fig.~\ref{fig:asymmetricforcing_DeltaTm}a for each co-rotating state, including data at all measured Ra and $\Delta T_m$ for completeness.  Since counter-diagonal states are predicted to be stable to both forcings, the stabilizing ratio is not relevant  for them.  Co-rotating states are found to be concentrated in Fig.~\ref{fig:asymmetricforcing_DeltaTm} with $S>1.13$.   This concentration of many points just above this measured minimum of $1.13$  suggests a critical value slightly below this is required for co-rotating states.  This threshold is near the predicted threshold of 1, well within typical uncertainties of this model.  This confirms that the model successfully predicts the conditions for co-rotating states to be stable due to turbulent thermal diffusion, when it overcomes the geometric forcing that destabilizes co-rotating state orientations.  In co-rotating states, the stabilizing factor $b\sin(\theta_2-\theta_1)$ is typically small (on average, $b=0.36a$ in co-rotating states), thus while an understanding of the source of the stabilizing $b$ in co-rotating states might adjust the threshold value of $S$, the stability of co-rotating states can be explained independent of the unexpectedly stabilizing values of $b$.

\subsection{Unpredicted stable state at large $\Delta T_m$}

 \begin{figure}
\includegraphics[width=0.45\textwidth]{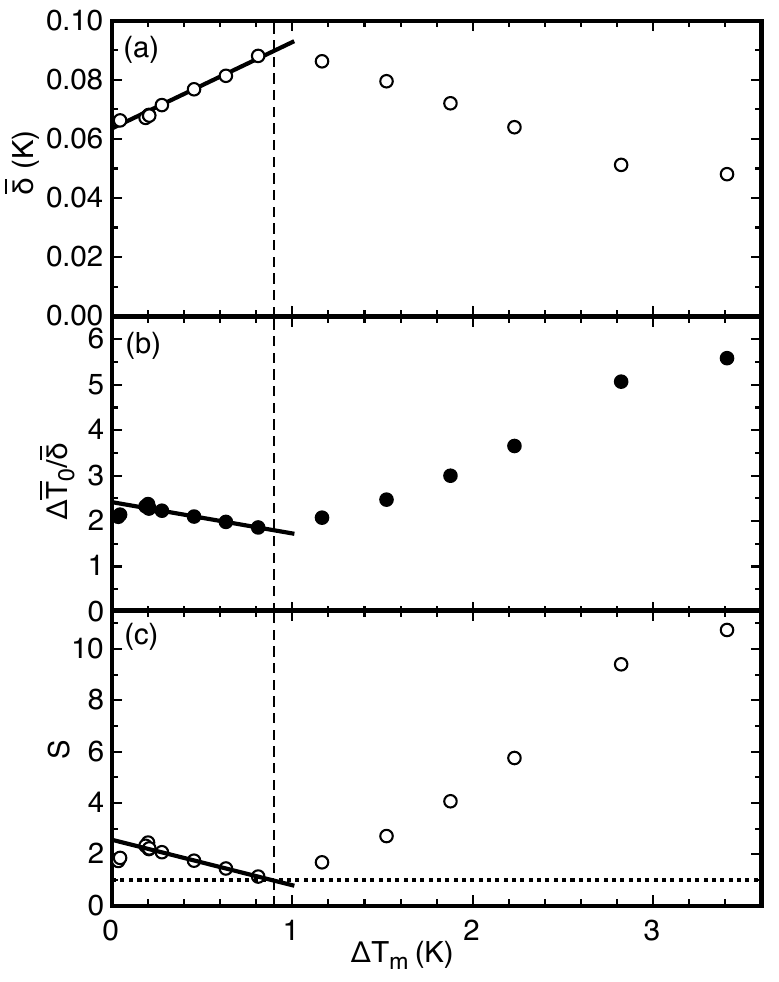} 
\caption{Measurements at large $\Delta T_m$  revealing a state with new scaling behavior for $\Delta T_m>0.9$ K when $\Delta T=3.8$ K. (a) mean LSC amplitude $\bar\delta$. (b) mean temperature difference between rolls $\Delta \bar T_0/\bar\delta$. (c) ratio of stabilizing to destabilizing forcing for co-rotating states.  Solid lines show fits of the predicted linear behavior of co-rotating states for small $\Delta T_m$. Vertical dashed line shows the point where the co-rotating state is predicted to become unstable at $S=1$.  The new state is oriented like a co-rotating state, but with a much larger $\Delta \bar T_0/\bar\delta$, suggesting it is no longer driven by turbulent thermal diffusion. 
%for $\Delta T=3.8$ K
 }
\label{fig:highDeltaTmstate}
\end{figure}

In Fig.~\ref{fig:asymmetricforcing_DeltaTm}, the ratio of stabilizing to destabilizing  forcings for co-rotating states $S$ is seen to drop towards the threshold of stability at $S=1$ for large $\Delta T_m$.  This linearly decreasing $S$ and loss of stability is predicted because the stabilizing forcing (which scales as $\Delta \bar T_0/\delta_0$) is growing linearly with $\Delta T_m$, but not as fast as the linearly increasing destabilizing term (which scales as $\bar\delta_{co}^2/\delta_0^2$) (see trends in Fig.~\ref{fig:deltameanplate}).  A linear fit of $S$ for a series of points at $\Delta T=3.8$ K for $\Delta T_m>0.14$ K is shown to intercept the predicted threshold of stability at $\Delta T_m/\bar\delta_{co}=10$ in Fig.~\ref{fig:asymmetricforcing_DeltaTm}.  At this point, both the co-rotating state and counter-rotating state are predicted to be unstable, so some new state should appear.
%however, The width of $p(\theta_i)$ is observed to become very narrow compared to co-rotating states, more like counter-rotating states, although the significance of this is unknown.  

To see if the co-rotating state destabilizes at larger $\Delta T_m$, we plot $\bar\delta$ and $\Delta \bar T_0/\bar\delta$ as a function of $\Delta T_m$ for fixed $\Delta T=3.8$ K (Ra $=5.5\times10^8)$ in Fig.~\ref{fig:highDeltaTmstate}, extending the range of experiments from Fig.~\ref{fig:deltameanplate}.    For $\Delta T_m<0.9$ K, the data is the same as Fig.~\ref{fig:deltameanplate}, with the predicted instability threshold of the co-rotating state at $\Delta T_m=0.9$ K.  When $\Delta T_m>0.9$ K, new scaling behaviors appear where $\bar\delta$ starts to decrease with $\Delta T_m$, and $\Delta \bar T_0/\bar\delta$ starts to increase significantly above its typical value of $2.22\pm0.13$ for co-rotating states. Values of $\theta_{p,i}$ are not shown because they remain constant as $\Delta T_m$ increases.   While the values of $\theta_{p,i}$ appear as if this is still a co-rotating state, the decrease in $\bar\delta$ and large increase in $\Delta \bar T_0/\bar\delta$ are inconsistent with Eqs.~\ref{eqn:deltaforcing} and \ref{eqn:Tforcing}.  In particular, values of $\Delta \bar T_0/\bar\delta \gg 2$ mean the temperature on the two sides of the interface between neighboring rolls is very different, and no longer dominated by a balance of turbulent thermal diffusion across the interface.   A likely candidate for this behavior is more coherent advection of heat between the cells, which can transport heat more coherently between the cells to balance the larger turbulent thermal diffusion for $\Delta \bar T_0/\bar\delta \gg 2$.  However, we cannot make specific predictions without more detailed knowledge of such advective flow fields.  
 The rapid increase of $\Delta \bar T_0/\bar\delta$ means $S$ increases again for $\Delta T_m>0.9$ K and does not drop below the threshold of instability for co-rotating orientations, so the stability of this  high-$\Delta T_m$ state is still consistent with Eq.~\ref{eqn:thetaforcing} for $\dot\theta_{\kappa,i}$, and perhaps only Eq.~\ref{eqn:Tforcing} for $\dot T_{\kappa,i}$ requires modification.  
%narrow $p(\theta_i)$ at intermediate $\Delta T_m$ is inconsistent  with Eq.~\ref{eqn:thetaforcing}, 

\subsection{Stability of counter-axis states}
\label{sec:counteraxis}

%counter axis - observations

Counter-axis states -- in which the preferred orientation is somewhere between a corner and axis --  were found for $\Delta T < 2$ K instead of counter-diagonal states. % in practice $\theta_1$ and $\theta_2$ are within 0.12 rad of each other.   
However, counter-axis states are not predicted to be stable from Eq.~\ref{eqn:thetaforcing}.  

%axis forcing and Delta T0
What might cause counter-axis states?  Equations of motion in $\ddot\theta_i$ could in principle produce stable counter-axis states if there is a force pushing both cells to be stable in $\pm\sin\theta_i$.  For example, if there was an asymmetry in the temperature profile of the plates such that the both plates are hotter near the interface.  It is notable that our counter axis states all have an unexpectedly large $\Delta \bar T_0 >\Delta T_m$, which is inconsistent with Eq.~\ref{eqn:DeltaT0} and thus Eq.~\ref{eqn:Tforcing} for turbulent thermal diffusion.  Thus, the large $\Delta \bar T_0$ must come from some other mechanism for these counter-axis states.  This unexpectedly large $\Delta \bar T_0$ might be an indication of asymmetric heating within the plates, although the forcing that would push both cells toward a counter-axis state would not be best represented by the parameter $\Delta \bar T_0$.  We found counter-axis states to be more likely when the flow direction in some of the cooling baths controlling the plate temperatures was switched (individual plate thermistor mean temperatures typically changed by $0.001\Delta T$), suggesting the plate temperature profile can produce such an asymmetry.  We also found counter-axis states to be more prominent in an early version of the apparatus with a different middle wall covering only the top half of the interface between the two rolls. %which produced counter-axis states instead of counter-diagonal states or co-rotating states at small $\Delta T_m$,
This suggests that there was some stabilizing force in $\pm\sin\theta_i$ with the same sign in both cells, presumably due to an interaction of the LSCs with the middle wall \cite{Dandanthesis}.  This asymmetry could have resulted from an interaction between a corner-roll and the middle wall, as corner-rolls are more prominent at the top only on the side of the up-flow (not down-flow), so the interaction could be different for a top-half middle wall depending on whether the flow in the middle is upward or downward.   Whatever the source, it seems likely that the observed counter-axis states are due to some unpredicted asymmetry of our setup, and the forcing can be surprisingly significant in systems with mild-seeming  asymmetries.

\subsection{Tilt}

Tilting the cell relative to gravity by an angle $\beta$ in the vertical plane going between the center of both cells results in a component of buoyant forcing aligned with  $\theta=0$ in one cell and $\theta=\pi$ rad in the other cell \cite{CRCC04, BA08b, JBB20}.  Since this is similar to the effect of a temperature difference $\Delta T_m$, it should be no surprise that a tilt of about $\beta=2$ degrees generally results in co-rotating states, with a mix of co-rotating and  counter-offset states at smaller $|\beta|$ (data is included in Fig.~\ref{fig:thetap1_thetap2}),  analogous to Fig.~\ref{fig:deltameanplate}a.  A detailed analysis of tilting a single cubic cell was presented previously \cite{JBB20}, and the same forcing terms are expected to apply here, in addition to the forcing in Eqs.~\ref{eqn:deltaforcing} and \ref{eqn:thetaforcing}.    Some quantitative data on the tilt-dependence is presented for the earlier version of our experiment with the middle wall only in the top half of the cell, which also included an extra forcing term that is responsible for the counter-axis states \cite{Dandanthesis}.  These observations serve as a confirmation that forcing terms from different physical mechanisms can be added in the low-dimensional model.

\section{Dynamics of switching between co-rotating and counter-rotating}
\label{sec:switching}

We observed numerous stochastic switching events between co-rotating and counter-rotating states such as in Fig.~\ref{fig:switchingexample}.  To obtain statistics to characterize these events, we searched through experiments with a total run time of 148 days with different Ra, $\Delta T_m$, and tilt angles $\beta$.  To intentionally produce more switching events, we also occasionally enforced a co-rotating state in one direction or the other by biasing $|\Delta T_m|/\delta_0\stackrel{>}{_\sim}1.7$ using Fig.~\ref{fig:deltameanplate} as a guide.

%algorithm
To systematically identify switching events between counter-rotating and co-rotating states, we define a transition based on the time-dependent difference in LSC orientations $|\theta_1-\theta_2|$.  We first smoothed values of $\theta_i$ over a duration of $1.5\tau_{\delta}$.  We define a transition from counter-rotating to co-rotating as starting when  $|\theta_1-\theta_2|$ last exceeds $\pi/2$ rad, and ending when $|\theta_1-\theta_2|$ first exceeds $3\pi/4$ rad, without returning below $\pi/4$ rad in between.  These first and last crossing times at different thresholds are used to avoid counting jitter around these threshold values.  We set the threshold $\pi/2$ rad further from the ideal counter-diagonal state value to include counter-axis states where $|\theta_1-\theta_2|$ is larger.   Likewise, a change from co-rotating to counter-rotating is defined to start when $|\theta_1-\theta_2|$ last drops below $3\pi/4$ rad, and end when $|\theta_1-\theta_2|$ first drops below $\pi/2$ rad, without returning above $3\pi/4$ rad in between.  To make sure the change in $|\theta_1-\theta_2|$ is not a false event due to a temporary fluctuation in one variable, we also calculated short-time averages of parameters  $|\theta_1-\theta_2|$, $(\delta_1+\delta_2)/2$, and $|T_{0,2}-T_{0,1}|$ in the steady states before and after the event --  in the final analysis we used the range of $14\tau_{\delta}$ to $56\tau_{\delta}$  before and after the event, excluding data where $\cos(\theta_2-\theta_1)$ has the wrong sign for the expected state, and excluding data with either $\delta_i<0.3\delta_0$. 
%keeping only events where at least 100 data points where included in these pre- and post- averages.   $\tau_{\delta}$ ranges from 32 to 130 s, depending on Ra \cite{JB20a}, 
%\tau_T$ is larger than $\tau_{\delta}$
%We used a slightly smaller range of $14\tau_{\delta}$ to $28\tau_{\delta}$ for identifying the stable states for few events that were short or near the start or end of the data set do to the array size cutting off some data 
We only considered a switching event to have occurred if the parameters $|\theta_1-\theta_2|$, $(\delta_1+\delta_2)/2$, and $|T_{0,2}-T_{0,1}|$ averaged over the specified time range changed from before to after the event by more than 60\% toward the expected counter- or co-rotating state: the expected differences going from counter- to co-rotating states are $|\theta_1-\theta_2|$ increased by $\pi$ rad (when values of $\theta_1-\theta_2$ are reduced to the range $-\pi$ rad to $\pi$ rad),  $|T_{0,2}-T_{0,1}|$ increased by  $2.22\bar\delta_{co}$ based on Fig.~\ref{fig:DeltaT0_ra}, and $(\delta_1+\delta_2)/2$ decreased by  $0.37\delta_0$  based on Fig.~\ref{fig:delta_ra}.    

%summary
We found 89 switching events between counter- and co-rotating states, corresponding to an average frequency of 0.60 per day, and only 0.48 per day if we do not count events that occurred shortly after we forced a switch to a co-rotating state by applying a large $|\Delta T_m|$.  We only observed a few direct switching events from one counter-diagonal state to another corner.  We never observed a direct switch between two co-rotating orientations.  Even in cases where a co-rotating state was stable, and we changed $\Delta T_m$ to drive the co-rotating state in the opposite direction, the system always switched to a counter rotating state and resided there for at least 1000s before switching to the new preferred co-rotating state.  Switching between two orientations of counter-axis states is relatively frequent, and statistics of switching between counter-axis orientations are reported elsewhere \cite{Dandanthesis}.  Mechanisms of switching between counter-axis states will not be  addressed here because they are driven by some uncharacterized asymmetry.

\subsection{Different driving mechanisms for switching events} 

In previous work, changes in LSC orientation between stable fixed point states were often found to be due to dramatic reorientations in $\theta_0$,  driven by extreme fluctuations in $\theta_0$ coming from the stochastic term $f_{\dot\theta}(t)$ \cite{BA08a, XX08, BJB16, Dandanthesis}.  Other events are driven by cessations -- a reduction in $\delta$ to near zero driven by stochastic fluctuations to overcome a potential barrier due to Eq.~\ref{eqn:delta_model}.  Cessations reduce the damping in Eq.~\ref{eqn:theta_model} so that the stochastic term $f_{\dot\theta}(t)$ allows $\theta_0$ to change more rapidly \cite{BA07a, BA08a, XX07, XX13}.  
For either mechanism the behavior could be quantified statistically as a rare event due to diffusive fluctuations crossing a large potential barrier in Eq.~\ref{eqn:delta_model} or \ref{eqn:theta_model}, respectively \cite{BA08a,  BJB16, JB20a}.  However, with only 89 switching events and a large number of variables, we do not have enough statistics to meaningfully test quantitative predictions of barrier crossing rates as a function of any parameter.  

\begin{table}
\begin{tabular}{ccrr}
 \hline\hline
parameter ~& change in & to counter-rotating  & ~ to co-rotating  \\ %& co  ($6\tau_{\delta}$) \\ 
 \hline
$d_{\theta}$ & $|\theta_1-\theta_2|$ & 0.12 & 0.37\\ %  &  0.16  \\ 
$d_T$ & $|T_{0,2}-T_{0,1}|$ & 0.01& 0.52\\ % & 0.37\\
$d_\delta$ & $(\delta_1+\delta_2)/2$  & -0.23 & 0.53\\ % & 0.26\\
$d_c$ & $\bar\delta_i-\delta_i$  &  0.36  &0.31\\ % & 0.22 \\
 \hline\hline
\end{tabular} 
\caption{Fractional changes of various parameters towards their new steady state values before switching events to counter- and co-rotating states.  Switching to counter-rotating states is mostly driven by cessations ($d_c$) during the event, in which a reduction in one of the $\delta_i$ reduces the damping on motion in $\theta_i$ which can be driven to large changes by random fluctuations.   Switching to co-rotating states is driven by a mix of diffusive fluctuations in $\theta_1-\theta_2$, $T_{0,2}-T_{0,1}$, and $\delta_i$ which have a positive feedback that destabilizes the counter-rotating state.
}
\label{tab:switchdriving}
\end{table}

%diffusion driving parameters
Instead, to identify the relevant mechanisms for switching between counter- and co-rotating states, we calculate correlations between changes in LSC parameters before an event, with the expected change towards the new stable value after switching.  These fractional changes $d_{\theta}$, $d_{\delta}$, and $d_T$ are calculated as the average fractional change of each parameter ($|\theta_1-\theta_2|$, $(\delta_1+\delta_2)/2$, and $|T_{0,2}-T_{0,1}|$, respectively) in the time period of $2\tau_{\delta}$ before the start of a switching event, relative to the expected change towards the stable state value after switching.   These fractions are averaged separately for events switching to either a counter- or co-rotating state, and summarized in Table \ref{tab:switchdriving}. We also include a row for a fractional change toward cessation driving $d_c$, calculated as the fractional change in $\delta$ toward zero ($\bar\delta_i-\delta_i$) relative to the stable value of $\bar\delta$ before switching .  We only count the maximum $d_c$ among the two cells, since a cessation of one of them is enough to cause a change in state. 

%counter
Table \ref{tab:switchdriving} suggests that  switching from the co-rotating to counter-rotating state  is most strongly correlated with cessation, with an average $d_c=0.36$, indicating a significant drop in $\delta$ from its previous stable fixed point in the co-rotating state, which weakens the damping of motion in $\theta_i$ (Eq.~\ref{eqn:theta_model}) \cite{BA08a}.    This strong correlation is especially notable because the new stable fixed point in the counter-rotating state has a larger $\delta_i$, meaning the cessation initially moves $\delta_i$ further from its new stable fixed point value at the start of the transition between states.
%would correspond to in $\delta$ in the opposite direction corresponding to $d_c=-0.65$.
All but one of the switching events to counter-rotating states has a reduction in one of the $\langle\delta_i\rangle$  either within the $2\tau_{\delta}$ before the event, or during the event,    The single event with negative $d_c$ is telling, because it occurred in the first 4000 s of one of the counter-offset states in Fig.~\ref{fig:deltameanplate}a, which we did not find to have a longer stable lifetime in a counter-rotating state, suggesting the  potential barrier that needed to be crossed was relatively low in that case.  
%smallest d_c 210421

\begin{figure}
\includegraphics[width=0.485\textwidth]{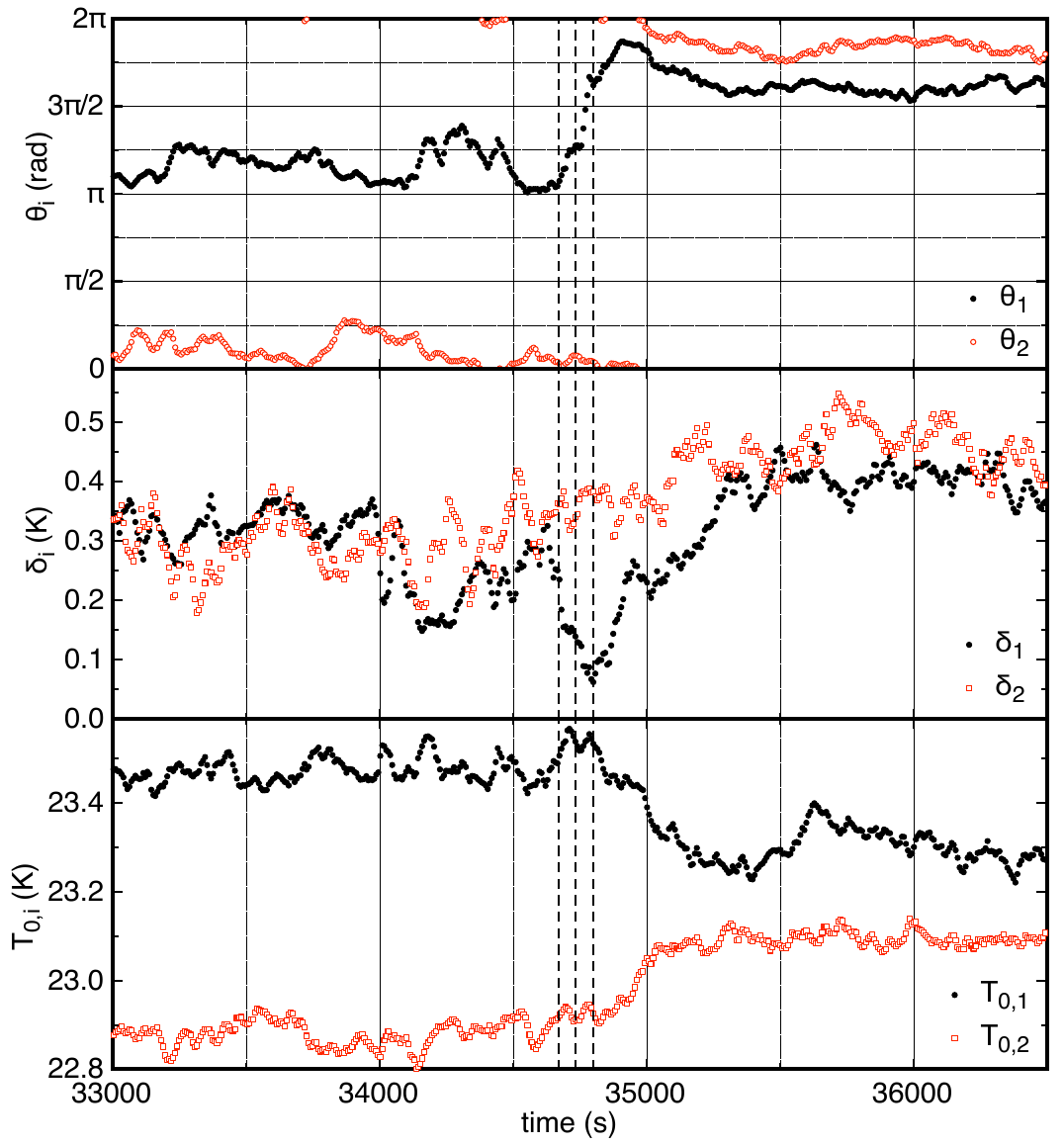} 
\caption{Example of a switching event from a co-rotating state to a counter-rotating state.  This transition is driven by a large drop in one of the $\delta_i$, which allows $\theta_i$ in the same cell (in this case in cell 1) to be rapidly driven by stochastic fluctuations, as are most transitions to counter-rotating states.  The vertical dashed lines indicate from left to right the time $2\tau_{\delta}$ before the start of the switching event, the start, and end of the switching event based on crossing thresholds in $|\theta_1-\theta_2|$.
 }
\label{fig:switchingexamplecessdriven}
\end{figure}

%median d_c lead and d_c during, 201205, 34733 s
An example of a switching to a counter-rotating state Ra $= 2.7\times10^9$ is shown in Fig.~\ref{fig:switchingexamplecessdriven}.  To represent a typical state, we chose the event with the median value of $d_c$ among switching events to counter-rotating states.  In this example, there is a large drop in $\delta_1$ in cell 1 that allows a rapid change of $\theta_1$ due to smaller damping on $\theta_1$ from Eq.~\ref{eqn:theta_model} \cite{BA08a}.   Once $\theta_1$ reaches near a new stable fixed point orientation for a counter-rotating state, the other parameters $\theta_2$, $T_{0,1}$, $T_{0,2}$, $\delta_1$ and $\delta_2$ follow to their new stable fixed point values within a few turnover times, or a few $\tau_{\delta}$.

%cessation mechanism counter
Generally, switching from co-rotating to counter-rotating states due to cessation (reduced $\delta_i$) can be understood to be due to reduced damping in Eq.~\ref{eqn:theta_model}, which allows $f_{\dot\theta}$ to drive large changes in $\theta_i$ in the same cell to cross the potential barrier that stabilizes co-rotating states in Eq.~\ref{eqn:thetaforcing}.  We generally find during switches to counter-rotating states that the drop in $\delta_i$ causes $\theta_i$ in the same cell to change further than the neighboring cell, consistent with this explanation.  Note that the reduction in $\delta_i$ only allows the motion of $\theta_i$ to be dominated by the stochastic term if the damping factor in Eq.~\ref{eqn:ddottheta} includes the factor of $\delta_i/\delta_0$ to cancel out the $1/\delta_i$ in Eq.~\ref{eqn:thetaforcingappendix}, so this observation is our best justification for using that factor $\delta_i/\delta_0$ in Eq.~\ref{eqn:ddottheta}.   In principle, Eq.~\ref{eqn:thetaforcing} also allows for the co-rotating state to become unstable due to large fluctuations that reduce $|T_{0,2}-T_{0,1}|$.  We see no evidence that this is a major driving factor in switching from co-rotating to counter-rotating states ($d_T=0.01$) without the aid of a cessation, perhaps because fluctuations in $T_0$ tend to be relatively small (see Fig.~\ref{fig:switchingexample}).

%co
In the other case of switching from a counter-rotating state to a co-rotating state, Table \ref{tab:switchdriving} shows that cessation ($d_c$) has the weakest correlations with these switching events, and is difficult to distinguish from a driving towards $\delta_{co}$, which also involves a reduction in $\delta$ in this case.  The correlations  with an increase in $|T_{0,2}-T_{0,1}|$, decrease in $(\delta_1+\delta_2)/2$ and increase in $|\theta_1-\theta_2|$ are all stronger than the driving due to cessation, and much stronger than when switching to a counter-rotating state. 

\begin{figure}
\includegraphics[width=0.485\textwidth]{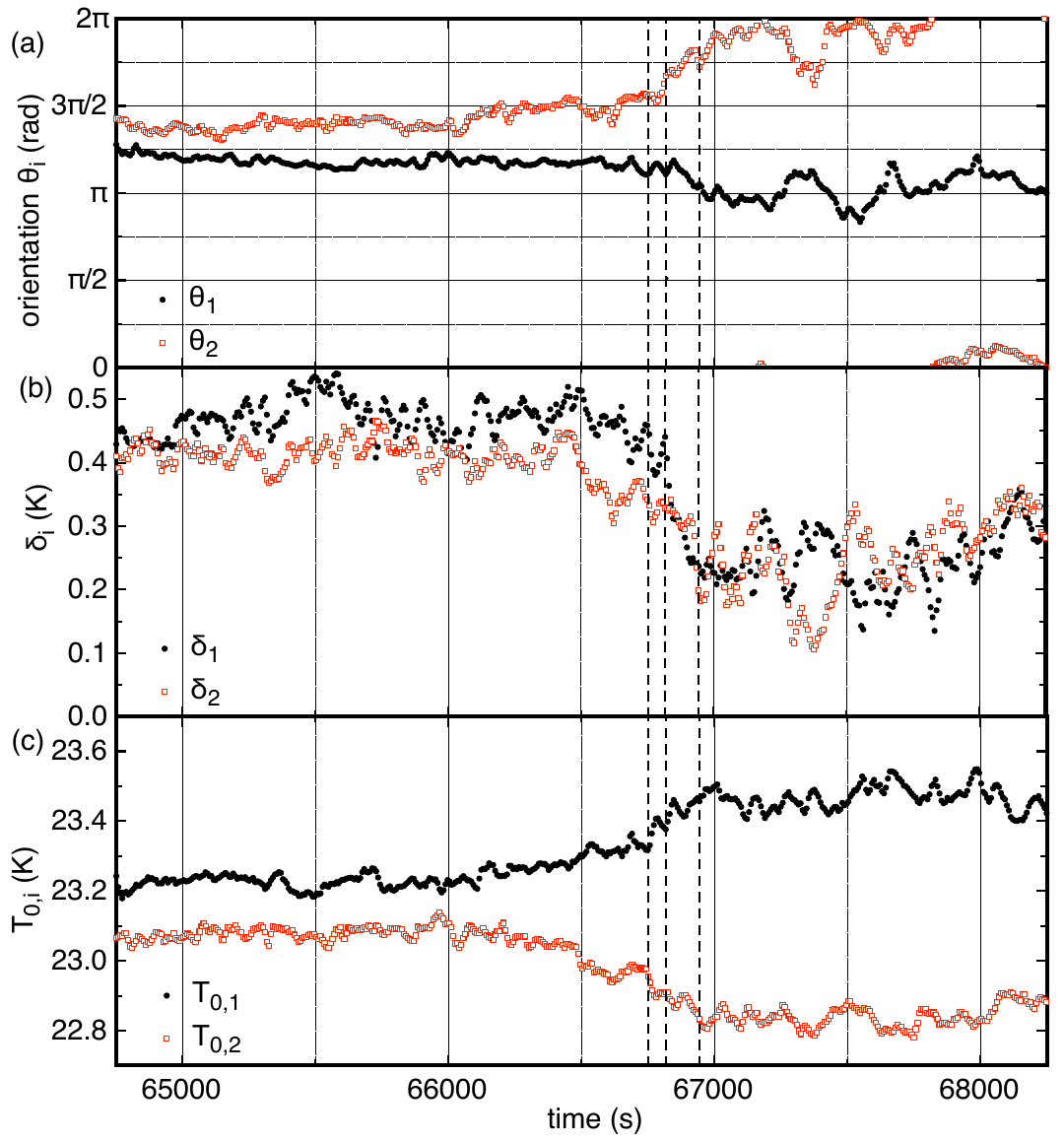} 
\caption{A switching from a counter-rotating state to a co-rotating state driven by a diffusion in $T_{0,2}-T_{0,1}$.    The vertical dashed lines indicate from left to right the time $2\tau_{\delta}$ before the start of the switching event, the start, and the end of the switching event.  Positive feedback between fluctuations in $T_{0,i}$, $\theta_i$, and $\delta_i$ away from the counter-rotating-state fixed point ultimately and a destabilizes the counter-rotating state, leading to a co-rotating state.
}
\label{fig:switchingexampleT0driven}
\end{figure}

%201210
 An example of a switching event from a counter-rotating state  to a co-rotating state led by random diffusive motion in $T_{0,2}-T_{0,1}$ at Ra $= 2.7\times10^9$ is shown in Fig.~\ref{fig:switchingexampleT0driven}.  To represent a typical switching to a co-rotating state, we chose the event with nearly median values in each of $d_\theta$, $d_T$, and $d_\delta$.  In this example, $T_{0,1}$ and $T_{0,2}$ start slowly drifting apart about 66,200 s, well before the change in $\theta_i$ and $\delta_i$ to their new stable values in a co-rotating state. Generally, as $|T_{0,2}-T_{0,1}|$ increases, it produces an increasing forcing on $\theta_i$ in the third term of Eq.~\ref{eqn:thetaforcing} to push $\theta_1$ and $\theta_2$ apart. Additionally, fluctuations in both $\theta_i$ and $\delta_i$ away from the counter-rotating state fixed point drive an increase in $|T_{0,2}-T_{0,1}|$ in Eq.~\ref{eqn:Tforcing}, providing a positive feedback that destabilizes a counter-rotating state to drive a transition to a co-rotating state.   This positive feedback can be seen in Fig.~\ref{fig:switchingexampleT0driven} as the changes in $\theta_i$, $\delta_i$, and  $T_{0,i}$ all start to increase more rapidly at about the same times. 
  In Eq.~\ref{eqn:thetaforcing}, when $\sin\theta\approx \sqrt{2}/2$ near the stable counter-rotating state, the third term that destabilizes the counter-rotating state becomes dominant over the first term that stabilizes the counter-rotating state when $T_{0,2}-T_{0,1} > \pi\delta/4$ (alternatively, the threshold of instability in Eq.~\ref{eqn:thetaforcing} can also be identified as $|\theta_1-\theta_2|>\pi/2$ as seen in Fig.~\ref{fig:deltameanplate}).  In the example in Fig.~\ref{fig:switchingexampleT0driven}, both of these thresholds are crossed roughly in between the first 2 timing lines, which is about when the rate of change in $\theta_i$ starts to increase.  Once the threshold of instability is crossed, the values of $\theta_i$, $\delta_i$, and $T_{0,2}-T_{0,1}$ quickly move towards their stable fixed point values for co-rotating states. 
  Since the positive feedback between increasing $|T_{0,2}-T_{0,1}|$, decreasing $(\delta_1+\delta_2)/2$, and increasing $|\theta_1-\theta_2|$ is what destabilizes counter-rotating states, this explains why $d_T$, $d_\delta$, and $d_\theta$ are all correlated with switching to the co-rotating state in Table \ref{tab:switchdriving}.   
  
  The behavior described above for switching from counter-rotating to co-rotating states is qualitatively similar in most events.  Notably, during switching from counter- to co-rotating states, we generally  observe that $T_{0,1}$ and $T_{0,2}$ are driven in opposite directions at nearly equal rates, as seen for example in Fig.~\ref{fig:switchingexampleT0driven}, and as expected from Eq.~\ref{eqn:Tforcing}.  The cell that starts slightly hotter also generally continues to get hotter, as seen for example in Fig.~\ref{fig:switchingexampleT0driven}c.  This determines which orientation is easier to switch to, as smaller fluctuations in $\theta_i$ and $T_{0,2}-T_{0,1}$ in this direction are required to destabilize the counter-offset state in Eq.~\ref{eqn:Tforcing}. 

%transient time
%After one parameter has reached its new stable state value during a switching event, the other parameters quickly move towards their new stable state values.  This is because the stable fixed point states of Eqs.~\ref{eqn:thetaforcing}, \ref{eqn:deltaforcing}, and \ref{eqn:Tforcing}  have incompatible values of $\theta_i$ and $\Delta T_{0,i}/\delta_i$ , so that once one of  $\theta_i$ or $\Delta T_{0,i}/\delta_i$ has changed, either the other values have to change to the new stable state or the system must return to its original state (the latter case is not counted towards switching events).  The correlation times of parameters, or transient relaxation timescales to a new steady state after a parameter value changes  are $\tau_{\dot\theta}$, $\tau_{\delta}$, and $\tau_T$ for equations \ref{eqn:thetaforcing}, \ref{eqn:deltaforcing}, and \ref{eqn:Tforcing}, respectively.  The longest of these timescales is $\tau_{\delta} =32$ s for the examples at Ra $= 2.7\times10^9$ in Fig.~\ref{fig:switchingexamplecessdriven} and \ref{fig:switchingexampleT0driven}.  
%there is no correlation between fluctuations in $T_0$ of the 2 cells

%prediction
If switching to counter-rotating states is due to cessation and other large drops in $\delta_i$, we can predict that switching to counter-rotating would occur in some fraction of cessations, which depends mainly on the mean value of $\delta_i$.  Any forcing that leads to a larger $\bar\delta_i$ is expected to decrease the likelihood of events \cite{BA08a, BJB16}, such as increasing $\Delta T_m$ or tilting the cell.  On the other hand, a prediction of switching to co-rotating states would instead depend on crossing barriers in a three-dimensional space of $|T_{0,2}-T_{0,1}|$,  $(\delta_1+\delta_2)/2$, and  $|\theta_1-\theta_2|$.  For either type of switching, due to the rarity of events, we do not have enough statistics to test such predictions.

\section{Discussion}

%model options
\subsection{Rejected model: modified drag at the interface between two rolls}
\label{sec:drag}
%diffusion analogy
 We originally considered a modification of viscous drag at the interface to model the interactions between neighboring rolls.  Since the effects of viscous drag at the interface between neighboring rolls could be modeled as a turbulent viscous diffusion of a vertical velocity  $U\cos(\theta_i-\theta_0)$ at the interface, some of its effects could be in part mathematically analogous to  turbulent thermal diffusion of temperature $T_i=T_{0,i}+\delta\cos(\theta_i-\theta_0)$.  
 
 %offset T_0
 One important difference between the velocity and temperature is the offset temperature $T_{0,2}-T_{0,1}$ between the neighboring cells, as there is no analogous offset for velocity.  The difference between the mean temperature of neighboring rolls $T_{0,2}-T_{0,1}$  is important because it is the term that stabilizes the observed co-rotating state in Eq.~\ref{eqn:thetaforcing}.    Without a mathematically analogous offset, viscous drag would only tend to align neighboring rolls toward a counter-rotating state.  $T_{0,2}-T_{0,1}$ is also the intermediary parameter for effects of the difference in mean plate temperature of the two cells $\Delta T_m$.   Thus, we rule out the hypothesis that the drag force is the most important interaction between neighboring convection rolls in our experiments, at least when it comes to determining the temperature profile and stable states of the flow structures.  

%If also include drag
On the other hand, viscous drag may play a smaller role in modifying parameter values.  Since the Prandtl number of the water is $Pr=\nu/\kappa = 6.41$, viscous diffusion should in principle be stronger than thermal diffusion by that factor.  Since we could represent both the temperature difference and velocity profiles as sine functions in the coordinate $\theta$  \cite{BA08a},  the diffusion equations and effective thermal diffusivity $\kappa_t$ measured may actually contain contributions from viscous diffusion.  Larger coefficients depending on $Pr$ might be expected for terms without $T_{0,2}-T_{0,1}$.    However, since the threshold for stability of co-rotating states is just barely met by the third term of Eq.~\ref{eqn:thetaforcing} overcoming the first and second terms, a significant destabilizing drag added to other terms would overwhelm the stabilizing term which depends on $T_{0,2}-T_{0,1}$, resulting in co-rotating states being unstable, in disagreement with observations.  Thus, any contribution of drag is probably small compared to the stabilizing term in our experiments.   It remains to be seen if systems with larger $Pr$, or those driven by forced convection rather than buoyancy, are dominated by turbulent viscous diffusion rather than turbulent thermal diffusion.

In Eq.~\ref{eqn:deltaforcing} for $\dot\delta_{\kappa,i}$, a modified drag term would be smaller in a counter-rotating state relative to a single LSC due to the flow in the same direction at the interface.  On the other hand, this modified drag may also change the time a fluid parcel spends near the top and bottom plates to lose or gain heat, respectively, and change the rate of turbulent mixing.  These effects may cancel each other out in the temperature profile, since there is little change in $\bar\delta$ [we observed a 6\%  reduction of $\bar\delta$ in the counter-rotating state compared to a single roll (Fig.~\ref{fig:delta_ra})].  The fact that there is not a larger change in $\bar\delta$ in the counter-rotating state suggests that any change in drag due to neighboring rolls has a small effect on $\bar\delta$.

\subsection{Discussion of previous work on co-rotating states}

Previous observations of co-rotating rolls in systems with horizontal forcing due to lateral heating \cite{PMELB04} or tilt \cite{MBAK19} are analogous to a one-dimensional version of our model with forcing due to $\Delta T_m$.  We found that motion in $\theta_i$ is important for determining the stability of states in three dimensions.  In particular, the stability of counter- and co-rotating states is determined by Eq.~\ref{eqn:thetaforcing}, which has terms that both stabilize and destabilize flow in terms of $\theta_i$, and stabilize in $\theta_1-\theta_2$ (Sec.~\ref{sec:thetastability}). In addition, in any non-circular cross-section,  the forcing $-\nabla V_g(\theta_i)$ due to the geometry of the cell can potentially add to the stabilizing or destabilizing terms  (Sec.~\ref{sec:stabilityratio}).

% Stacked rolls 
The behaviors of stacked co-rotating rolls \cite{XX13} are not explained by our model, as the forcing terms in the vertical direction would be different, and vertically stacked co-rotating states likely have more coherent advection between the two cells.  %Those states are still better explained by stochastic switching between viscous-coupled counter-rotating states and buoyancy-coupled co-rotating states \cite{PB03}. % In RBR89, thermal coupling did not result in co-rotating stacked rolls

\subsection{Preferred states in other systems}
 
%generalized model -- stability of co-rotating 
Our observations of stable co-rotating states leads to the question of why are co-rotating states not reported more often in other convective flows?  Equation \ref{eqn:thetaforcing} predicts stable co-rotating states when the stabilizing forcing in $\theta_1$ from the 3rd terms of Eq.~\ref{eqn:thetaforcing} are larger than the 1st and 2nd terms that stabilize counter-rotating states and any destabilizing forcing $-\nabla V_g(\theta_i)$ due to the geometry of the cell (Sec.~\ref{sec:stabilityratio}).  For co-rotating states to be stable with $\theta_2=\theta_1+\pi$ rad and assuming $\bar\delta_2=\bar\delta_1$, this requires $\Delta \bar T_{0,co}/\bar\delta_{co} > 1.62$, not including contributions from $V_g$ which can increase this threshold by an amount depending on the geometry.    Equation \ref{eqn:Tforcing} gave a prediction of $\Delta \bar T_{0,co}/\bar\delta_{co}=1.76$, which is only 9\% greater than the threshold of stability for co-rotating states, and we measured $T_{0,co}/\bar\delta_{co}=2.22$, which is 40\% larger than this threshold.   We do not yet know how $\Delta \bar T_{0,co}/\bar\delta_{co}$ varies in other systems.   Since the numerical coefficients in Eq.~\ref{eqn:thetaforcing} are expected to change with the shape of the interface and the shape of the convection patterns, it seems plausible that a version of Eqs.~\ref{eqn:thetaforcing} and \ref{eqn:Tforcing} for another system might not produce a large enough  $\Delta \bar T_{0,co}/\bar\delta_{co}$ to be above the threshold required for stability of co-rotating states.  These variables, might be part of the reason that co-rotating states are not often reported.  In other words, while this idealized laboratory system can reproduce many qualitative and quantitative features of natural convection rolls, the large number of model terms and the sensitivity of behavior to parameter values means that behavior may not be representative of all natural convection systems.

%other systems -- examples
Equation \ref{eqn:thetaforcing} by itself does not predict any preferred orientation for counter-rotating states.  Thus, counter-rotating rolls can in principle be aligned at any orientation.  Specific systems likely have a preference for certain orientations based on their geometric potential $V_g$ (i.e.~the longest diagonals of a closed cell), or other forces.  The counter-rotating predicted by Eq.~\ref{eqn:thetaforcing} corresponds to $\theta_1=\theta_2$ such that the LSC orientation vectors line up head-to-head,  and do not in general have to be parallel.  For example, a hexagonal convection pattern with 3 counter-rotating rolls at 120 degree angles relative to each other meeting at a vertex with adjacent upward-flowing regions meet this criteria for counter-rotating states -- in this case it is the tessellation of a hexagonal pattern that determines the preferred orientations of counter-rotating states \cite{BES10}.

\section{Conclusions}

%co and counter stability. qualitative model
We observed both counter-rotating and co-rotating states to be stable for adjacent convection rolls in cubic Raleigh B{\'e}nard convection cells with  $7.6\times10^7 < $ Ra $ < 3.5\times10^9$ and Prandtl number 6.41 (Figs.~\ref{fig:switchingexample}, \ref{fig:thetap1_thetap2}).   Notably co-rotating states were observed in nominally symmetric systems which are usually argued to support only counter-rotating states.  The stability of the observed counter-rotating states and co-rotating states can be explained by a low-dimensional model consisting of three stochastic ordinary differential equations for each roll.  The interactions between neighboring rolls are predicted from the scaling of heat transport by turbulent thermal diffusion, and added to existing model equations \cite{BA08a, BA08b, JB20a} for the LSC amplitude $\delta_i$ (Eqs.~\ref{eqn:delta_model}, \ref{eqn:deltaforcing}), orientation $\theta_i$ (Eqs.~\ref{eqn:theta_model}, \ref{eqn:thetaforcing}), and a new equation for mean temperature $T_{0,i}$ (Eq.~\ref{eqn:Tforcing}).   Despite there only being one new mechanism in the model, projecting turbulent thermal diffusion for two neighboring rolls with just simple sinusoidal temperature profiles into the parameters $\theta_i$, $\delta_i$ and $T_{0,i}$ resulted in eleven new model terms.

%quantitative predictions of model and kappa_t
To make quantitative predictions, the model required a parameter as experimental input: a turbulent thermal diffusivity $\kappa_t$ in the horizontal direction across the interface between neighboring rolls.  Once $\kappa_t$ was measured based on a fit of one model term, it could be used to predict the magnitudes of all eight model terms in which $\kappa_t$ affects steady-state solutions  for $5.5\times10^8 < $ Ra $ < 3.5\times10^9$.  Predictions were not tested in the lower measured range of Ra due to an unknown additional forcing from some asymmetry that was unquantified.     One term of the thermal diffusivity model depends on the Nusselt number for vertical heat transport, and the remaining two predicted terms of the model affect transients only, which were not tested.  Specifically, the turbulent thermal diffusion model predicts the stabilizing forcing in $\sin(\theta_1-\theta_2)$ for counter-rotating states and the stabilizing forcing in $\theta_i$ for parallel co-rotating states (Figs.~\ref{fig:probtheta}, \ref{fig:forcing_theta_d2theta}), given the value of $\kappa_t$ obtained from an independent measurement of correlation time of $T_{0,i}$ (Figs.~\ref{fig:T0diffusivity}, \ref{fig:turbulentkappa}).  The co-rotating states are stable if the difference $\Delta \bar T_{0,co}$ in the mean temperature of the two rolls is large enough that the heat flux between the rolls stabilizes the temperature profile of aligned co-rotating states.  It was confirmed that the model correctly predicts the value $\Delta \bar T_{0,co}$ is large enough to produce a stabilizing forcing (Figs.~\ref{fig:delta_ra}, \ref{fig:DeltaT0_ra}), and that we find stable co-rotating states in our cubic cell if and only if this forcing is large enough to overcome the forcing due to the cell geometry (Fig.~\ref{fig:asymmetricforcing_DeltaTm}).  This large $\Delta \bar T_{0,co}$ could be driven by a difference $\Delta T_m$ between the mean temperatures of the plates of the two cells (Fig.~\ref{fig:deltameanplate}), tilt of the cell  (Fig.~\ref{fig:thetap1_thetap2}), or even by stochastic fluctuations from a stable counter-rotating state in a nominally symmetric setup where both states are stable (Fig.~\ref{fig:switchingexample}).
These predictions  are all consistent within a factor of 3 of measurements when using the same value of $\kappa_t$ for each term, a typical uncertainty of this modeling approach \cite{BA08a, BA08b, JB20a}.
%although in our experiments the preferred orientation of counter-rotating states is determined by a pressure due to the shape of the sidewall \cite{BA08b, JB20a}.  
The model can predict the decrease in the LSC amplitude $\bar\delta_{co}$ in co-rotating states if the measured $\Delta \bar T_{0,co}$ is used as an input (Fig.~\ref{fig:delta_ra}), however, the model is not accurate enough to predict the sign or lower bound of this change without experimental input due to the predictions depending on differences of terms with comparable magnitude and having large uncertainties.
  
  %Delta Tm
When an asymmetry between neighboring cells is introduced through an imposed temperature difference $\Delta T_m$ between the mean plate temperatures of the two cells, the model for  effective turbulent thermal diffusion explains an observed increase in $\Delta \bar T_0$ with increasing $\Delta T_m$, which shifts the preferred orientations of  counter-rotating rolls in opposite directions by $\Delta \theta_p$, until the shift is large enough that the state becomes unstable and switches to a co-rotating state (Fig.~\ref{fig:deltameanplate}).   The model self-consistently predicted the trends $\Delta \theta_p$, $\bar\delta$, and $\Delta \bar T_0$ in $\Delta T_m$ for both counter- and co-rotating states, and a difference $\Delta\bar\delta$ between the two cells in counter-rotating states, when using the other trends as input (Figs.~\ref{fig:deltameanplate}, \ref{fig:DeltaT0_Ddeltacostheta}, \ref{fig:Deltadelta_DeltaTm_counter}) all within a factor of 3. However, the sensitivity of the model to differences in estimated terms makes it unable to predict these trends without any experimental input.    

%switching
We also observed spontaneous switching events of the LSC between counter-rotating and co-rotating states.  Events switching from co-rotating to counter-rotating states were almost all driven by cessations of the LSC -- or at least a significant reduction in $\delta$ (Fig.~\ref{fig:switchingexamplecessdriven})  -- which reduces the damping term in Eq.~\ref{eqn:theta_model}, so that the orientation of the LSC can be driven easily by random turbulent fluctuations \cite{BA08a, BJB16}.  We found that switching from counter-rotating states to co-rotating states is generally driven by a combination of fluctuations that cause increases in $|T_{0,2}-T_{0,1}|$, decreases in $\delta_i$, and increases in $|\theta_1-\theta_2|$, which result in a positive feedback in Eqns.~\ref{eqn:thetaforcing} and \ref{eqn:Tforcing} that destabilizes the counter-rotating state (Fig.~\ref{fig:switchingexampleT0driven}).  %The role of $T_{0,2}-T_{0,1}$ in changes of state is new, since $T_{0,2}-T_{0,1}$ did not play any role in forcing in previous experiments without neighboring rolls.

%kappa_t significance
The observations above tested each of the model terms that affect stable fixed point values (including the first 3 terms of Eq.~\ref{eqn:thetaforcing}, the first 3 terms of Eq.~\ref{eqn:deltaforcing}, and all 3 deterministic terms of Eq.~\ref{eqn:Tforcing}), confirming the validity of the turbulent thermal diffusion model for the interaction between neighboring rolls.   This level of detail suggests that the hypothesized mechanism of turbulent thermal diffusion is dominant, and the $\kappa_t$ that controls the interactions between neighboring convection rolls is the same turbulent thermal diffusivity that describes macroscopically averaged heat transport.  It remains to be confirmed whether the $\kappa_t$ measured here based on the interaction between convection rolls has the same magnitude as an effective thermal diffusivity for macroscopically averaged heat transport, or if the value of $\kappa_t$ can be predicted from a general scaling model.    It also remains to be confirmed if the prediction for the term of Eq.~\ref{eqn:Tforcing} corresponding to vertical heat transport is determined by the Nusselt number, as our data had very limited resolution to test this (Sec.~\ref{sec:DeltaT0_DeltaTm_co}). If these can be confirmed, then it would suggest that low-dimensional models can include both coherent flow structures and transport in the same model, using equations such as Eqs.~\ref{eqn:delta_model}, \ref{eqn:theta_model}, \ref{eqn:deltaforcing},  \ref{eqn:thetaforcing}, and \ref{eqn:Tforcing} for flow structure interactions (including interactions with boundaries through predictable terms such as $V_g$), and relating them to equations such as Eq.~\ref{eqn:turbulentdiffusion} for heat transport.

%sinusoid assumption - significance?
%Another key assumption of our turbulent thermal diffusion model was to characterize the LSC in terms of an amplitude $\delta_i$, orientation $\theta_i$, and mean temperature $T_{0,i}$.  While this was an assumption in the base model as equations were only introduced with these parameters \cite{BA08a}, here the assumption of a sinusoidal temperature profile was also used in Eq.~\ref{eqn:Tdot} so only the lowest order Fourier terms appear.  Even with this simplifying assumption, the Fourier transforms added ten new terms to the model in Eqs.~\ref{eqn:thetaforcing}, \ref{eqn:deltaforcing}, and \ref{eqn:Tforcing}.  Higher-order moments in temperature profiles have been observed to be relevant parts of oscillation modes, including  the jumprope oscillation \cite{VHGA18, AYSSHVE22, HSA22}, an advected oscillation that appears as twisting and sloshing modes  \cite{FA04, XZZCX09, ZXZSX09, BA09}, and an oscillation in the shape of the temperature profile induced by the shape of the cell \cite{JB20b}.  While our model could in principle be expanded further to include higher-order Fourier modes in the temperature profile, it would add several model terms per Fourier mode.
  %Higher-order terms and their interactions were not required to characterize any of the features and dynamics modeled here.  More complex models would of course be needed for details that are too fine to be characterized by the model parameters $\theta_i$, $\delta_i$, and $T_{0,i}$.   

%failures of model -co-rotating state stability, new state, advection
While most of the features of the interaction between neighboring LSCs could be explained by an effective turbulent thermal diffusion between the cells, some stabilizing features  of the co-rotating state could not be explained.   This includes the stabilizing forcing on the orientation of co-rotating states in $\theta_1-\theta_2$, and the value of  $\Delta \bar T_{0,co}/\bar\delta_{co}>2$, which increases to much higher values at large $\Delta T_m$ (Fig.~\ref{fig:highDeltaTmstate}).  These observations are not consistent with turbulent thermal diffusion.  It is likely that these discrepancies are due to an additional coherent advection of heat  from one cell to the other, which was not included in the model.

%generalizations
The interaction between two neighboring rolls should be straightforwardly generalizable to systems of more rolls with the addition of similar interaction terms to Eqs.~\ref{eqn:deltaforcing}, \ref{eqn:thetaforcing},  and \ref{eqn:Tforcing} for each neighboring roll, assuming that there are no higher-order interaction terms that involve 3 or more rolls.  In other words, systems with multiple rolls -- such as the cases with large lateral extend with a series of adjacent rolls --  would then have a set of equations of motion in $\theta_i$, $\delta_i$, and $T_{0,i}$ for each roll (the number of rolls would have to be known), and each equation of motion would have a  set of interaction terms for each neighbor.  When considering flows in different geometries, the geometry-dependence is expected to be mostly addressed by the separate term $V_g(\theta_i)$ characterizing the interaction with the boundary, which was predicted  in earlier papers \cite{BA08b, JB20a}, while the neighboring-roll interaction terms are expected to be separate from boundary geometry effects.  The model terms were derived here based on a sinusoidal temperature profile, which could in principle have higher order Fourier terms in other flows, adding even more terms to the model, although the sinusoidal temperature profile has been found to be a good approximation (within a few percent) for flows in circular cross-sections of different aspect ratios as well \cite{BA07b, VHGA18}.  The bounds of the integrals when evaluating terms of Eqs.~\ref{eqn:fouriertransform}, \ref{eqn:fouriertransformdelta}, and \ref{eqn:dotTintegral} depend on the length of the interaction region, but would only affect the coefficients of each model term, and not the form of the terms.  These coefficients could be re-derived for an interaction region of different length following the procedure in the Appendix.

\section{Acknowledgements}

Preliminary experimental investigations and model analysis were done by D. Ji, which was essential to develop our understanding of the system.  Experimental investigations for published data, analysis, and writing of the manuscript was done by E. Brown.  We thank the University of California, Santa Barbara machine shop and K. Faysal for helping with construction of the experimental apparatus.  This work was supported by Grant CBET-1255541 of the U.S. National Science Foundation.

\section*{Appendix: model derivations}
\label{sec:modelappendix}

\subsection{Forcing terms $\dot\delta_{\kappa,i}$ and $\ddot\theta_{\kappa,i}$ due to neighboring roll interactions}

%model summary
We start with the assumption that Eqs.~\ref{eqn:delta_model} and \ref{eqn:theta_model} describe a baseline model for each LSC, and that we only need to derive terms for additional forcing due to neighboring-roll interactions.  The forcing on the temperature profile along the interface between neighboring convection rolls is assumed to come from the turbulent diffusion of heat in across the interface with effective turbulent thermal diffusivity $\kappa_t$, which is assumed to be uniform and constant. The turbulent thermal diffusion represents the enhancement of heat transport by eddies relative to the thermal diffusivity $\kappa$ due to advection, analogous to a turbulent viscosity or eddy viscosity.  This thermal diffusion is approximated from a heat equation for heat flux only in the direction perpendicular to the interface between neighboring cells, as a function of the coordinate $\theta$ along the interface, as heat transport in other directions should already be accounted for in the existing Eqs.~\ref{eqn:delta_model} and \ref{eqn:theta_model}.   We estimate the thermal diffusion term as $\kappa_t \nabla^2 T \approx \kappa_t[T_1(\theta, \theta_1) -T_2(\theta,\theta_2)]/L^2$, where $T_1(\theta, \theta_1) -T_2(\theta,\theta_2)$ is the temperature difference between the neighboring cells over an interfacial mixing layer of unknown thickness.  We include the unknown mixing layer thickness in the value of $\kappa_t$, which will be a fit parameter, so we can use the known cell size $L$ as the lengthscale in the thermal diffusion equation.  This formulation is mathematically analogous to boundary layer approximations with thermal diffusion in an interfacial mixing layer of thickness $L\kappa/\kappa_t$ between the two cells,  although it seems unlikely that such a thin mixing layer exists in this case.  Because of the middle insulating wall blocking half of the interface, the heat transport acts over the exposed fractional area $A=0.50$ of the interface between the two cells.   An additional vertical heat transport from each of the top and bottom plates is calculated using the standard boundary layer approximation that the heat transport from each plate at temperature $T_{m,i}\pm \Delta T/2$ to the bulk at temperature $T_{0,i}$ is diffusive across thermal boundary layers of thickness $H/2 \mbox{Nu}$, where the Nusselt number Nu is the dimensionless turbulent heat transport in the vertical direction.   In the vertical heat transport term we ignore the variation in bulk temperature with $\theta$, which is a small correction since  $\delta \ll \Delta T$.    The resulting simplification of the heat equation for cell 1 is
\begin{equation}
\dot T_1(\theta) = \frac{A\kappa_t}{L^2} \left[T_2(\theta, \theta_2) -T_1(\theta,\theta_1)\right]+\frac{4\mbox{Nu}\kappa}{H^2}[T_{m,1}-T_{0,1}] \ .
\label{eqn:turbulentdiffusion}
\end{equation}
  For brevity, we write all equations in this section for the forcing on cell 1 only, as the equations for cell 2 are identical other than an exchange of the subscripts 1 and 2 in each equation. 

%projection to delta and theta
Equation \ref{eqn:turbulentdiffusion} is a function of the coordinate $\theta$ along the interface between cells.   To instead put it in terms of parameters $\theta_i$, $\delta_i$, and $T_{0,i}$ of the LSC, we need to transform the equation into that parameter space.  To do this, we make use of the time derivative of the temperature profile of Eq.~\ref{eqn:tempprofile}: 
\begin{equation}
\dot T_1(\theta) = \dot T_{0,1} + \delta_1 \dot\theta_1\sin(\theta-\theta_1)  + \dot\delta_1\cos(\theta-\theta_1) \ . 
\label{eqn:Tdot}
\end{equation}
In this equation, the lowest order terms of a Fourier series of $\dot T_1(\theta)$ relate directly to the forcing terms $\dot\theta_1$ and $\dot\delta_1$.  We assume for simplicity that these lowest order Fourier series terms are a good approximation of the LSC -- i.e.~the only parameters needed to describe the LSC are $\theta_i$, $\delta_i$, and $T_{0,i}$ from the sinusoidal temperature profile of Eq.~\ref{eqn:tempprofile}, and ignore the small higher order terms in the Fourier series \cite{BA07b, VHGA18, JB20b}.   Thus, we perform a Fourier transform of Eq.~\ref{eqn:Tdot}  in terms of sines and cosines to obtain expressions for forcing terms on the LSC parameters
\begin{equation}
\dot\theta_{\kappa,1} = \frac{1}{\pi\delta_1}\int_0^{2\pi}[\dot T_1(\theta)-\dot T_{0,1}]\sin(\theta-\theta_1)d\theta
\label{eqn:fouriertransform}
\end{equation}
and
\begin{equation}
\dot\delta_{\kappa,1} = \frac{1}{\pi}\int_0^{2\pi}[\dot T_1(\theta)-\dot T_{0,1}]\cos(\theta-\theta_1)d\theta \ .
\label{eqn:fouriertransformdelta}
\end{equation}

Equations \ref{eqn:fouriertransform} and \ref{eqn:fouriertransformdelta} are evaluated by inserting Eq.~\ref{eqn:turbulentdiffusion} for $\dot T_1(\theta)$, which uses as input Eq.~\ref{eqn:tempprofile} for $T_1(\theta,\theta_1)$ and $T_2(\theta,\theta_2)$.  Expanding the term $\cos(\theta-\theta_2) = \cos(\theta-\theta_1)\cos(\theta_1-\theta_2) - \sin(\theta-\theta_1)\sin(\theta_1-\theta_2)$ from Eq.~\ref{eqn:turbulentdiffusion} makes the integrands of Eqs.~\ref{eqn:fouriertransform} and \ref{eqn:fouriertransformdelta} functions only of $\theta-\theta_1$. The neighboring roll interaction terms along the interface only need to be integrated over the range  $-\pi/4 < \theta < \pi/4$ rad, while the second term of Eq.~\ref{eqn:turbulentdiffusion} due to vertical heat transport drops out when integrated over $\theta$ from 0 to $2\pi$ rad.  Integrating Eqs.~\ref{eqn:fouriertransform} and \ref{eqn:fouriertransformdelta} over $\theta$ results in:
\begin{multline}
\dot\theta_{\kappa,1} = - \frac{A\kappa_t}{\pi L^2 \delta_1}\bigg[ \delta_2\sin(\theta_1-\theta_2)[\pi/4-\cos(2\theta_1)/2]\\
+[\delta_2\cos(\theta_1-\theta_2) - \delta_1]\sin(2\theta_1)/2
+\sqrt{2}(T_{0,2}-T_{0,1})\sin\theta_1 \bigg]\\
+ \frac{\sqrt{2}\dot T_{0,1}}{\pi\delta_1}\sin\theta_1 
\label{eqn:thetaforcingappendix}
\end{multline}
and
\begin{multline}
\dot\delta_{\kappa,1} =  \frac{A\kappa_t}{\pi L^2}\bigg[ \sqrt{2}(T_{0,2}-T_{0,1})\cos\theta_1 \\
+\delta_2\sin(\theta_1-\theta_2)\sin(2\theta_1)/2\\
+[\delta_2\cos(\theta_1-\theta_2) - \delta_1][\pi/4+\cos(2\theta_1)/2] \bigg] \\
- \frac{\sqrt{2}\dot T_{0,1}}{\pi}\cos\theta_1\ .
\label{eqn:deltaforcingappendix}
\end{multline}

%Combining thermal diffusion forcing with the previous model
This forcing $\dot \delta_{\kappa,1}$ can be inserted directly to the existing  equation of motion (Eq.~\ref{eqn:delta_model}).  Since Eq.~\ref{eqn:theta_model} is in terms of $\ddot\theta_1$, but we derived a forcing in $\dot\theta_1$ (Eq.~\ref{eqn:thetaforcingappendix}), we assume the forcing in $\dot\theta_1$ can be treated as the forcing in the overdamped limit, corresponding to an offset in the $\dot\theta_1$-term of Eq.~\ref{eqn:theta_model}, so that 
\begin{equation}
\ddot\theta_{\kappa,1} = \frac{\dot\theta_{\kappa,1}\delta_1}{\tau_{\dot\theta}\delta_0} \ .
\label{eqn:ddottheta}
\end{equation}

%effect of including $\bar\delta/\delta_0$:
%forcing weaker in co, but for all terms equally in theta equation
%co-stabilizing forcing weaker (less stable), but more stable on axis when also including $\bar\delta^2$ on $\omega_r^2$
%during cessation, forcing to fixed points also weakens if delta-factor included
%correlation time in theta larger in co-rotating state -- only for destabilizing term and term with unknown sign

\subsection{Equation of motion for $T_{0,1}$}
%motivation
%The model for a single LSC did not include an equation of motion for the mean temperature $T_{0,i}$ because it was observed to have trivial behavior. With neighboring convection cells, we observe a systematic difference between  $T_{0,2}$ and $T_{0,1}$ (Fig.~\ref{fig:switchingexample}c), and this difference may drive dynamics in $\theta_i$ and $\delta_i$ according to Eqs.~\ref{eqn:thetaforcingappendix} and \ref{eqn:deltaforcingappendix}.  Thus, we need an equation of motion for $T_{0,i}$.  

In an equation of motion for $T_{0,i}$, we should also expect to have some diffusive fluctuations driven by turbulence, analogous to Eqs.~\ref{eqn:delta_model} and \ref{eqn:theta_model}, so we include a fluctuation term $f_T(t)$ with diffusivity $D_T$.   The deterministic part of the equation for $T_{0,1}$ is assumed to be due to the net heat flux between the cells and from the top and bottom plates to the cells, calculated by integrating both sides of Eq.~\ref{eqn:turbulentdiffusion} over $d\theta/\cos^2\theta$.  The denominator of $\cos^2\theta$ accounts for the relative interface area at each infinitesimal $d\theta$ to satisfy conservation of energy. %\footnote{Such a factor is not included in Eqs.~\ref{eqn:fouriertransform} and \ref{eqn:fouriertransformdelta}, because the quantities $\delta_0_{\kappa,i}$ and $\theta_{\kappa,i}$ do not scale with thermal energy.  In any case, the factor of $\cos^2 \theta$ in the integral only produces a 2% correction on the relative size of the 2 terms of Eq.~\ref{eqn:Tforcing}.   
The term of Eq.~\ref{eqn:turbulentdiffusion} with $A$ due to the neighboring roll interaction only contributes in the interaction region $|\theta| < \pi/4$ rad so is only integrated over that range.  
% and define $B\equiv \mbox{Nu} \kappa/\kappa_t$ as the ratio of vertical to horizontal turbulent thermal diffusivities at aspect ratio 1.  
This results in:  
\begin{multline}
\dot T_{0,1} = f_T(t) + \frac{4\mbox{Nu}\kappa}{H^2}(T_{m,1}-T_{0,1}) + \frac{A\kappa_t}{2 L^2}\times\\
\int_{-\pi/4}^{\pi/4} \frac{T_{0,2}-T_{0,1} + \delta_2\cos(\theta-\theta_2) - \delta_1\cos(\theta-\theta_1)}{\cos^2\theta}d\theta \ .
\label{eqn:dotTintegral}
 \end{multline}
  Using the same substitution as before for $\cos(\theta-\theta_2)$, and integrating, results in
 \begin{multline}
\dot T_{0,1} = f_T(t)+\frac{4\mbox{Nu}\kappa}{H^2}(T_{m,1}-T_{0,1})+\\
\frac{A\kappa_t}{L^2}\left[T_{0,2}-T_{0,1}+ 0.88(\delta_2\cos\theta_2 - \delta_1\cos\theta_1)\right] \ .
\label{eqn:Tforcingappendix}
 \end{multline}
  For simplicity of presenting equations for our cubic cells, we present equations in the main paper for aspect ratio 1 ($H=L$).
%page 64
%While this equation could be plugged into Eqs.~\ref{eqn:thetaforcing} and \ref{eqn:deltaforcing}, we find it more insightful to leave Eqs.~\ref{eqn:thetaforcing} and \ref{eqn:deltaforcing} as is for a more direct physical interpretation of their dependence on $T_{0,2}-T_{0,1}$ at the stable fixed points where $\dot T_{0,i}=0$.

%\subsection{Testing the model for symmetric cases ($\Delta T_m=0$)}
%\label{sec:interactionmodeltest}

%\subsection{Forcing terms on $\ddot\theta_i$ in $\theta_1-\theta_2$ and $\theta_i-\theta_p$ for $\Delta T_m=0$}
%\label{sec:thetastability}

%In Sec.~\ref{sec:preferredstates}, the observed counter- and co-rotating states suggest forcing terms on $\ddot\theta_i$ that are stable in $\theta_1-\theta_2$ and $\theta_i-\theta_{i,p}$, respectively.  In this section, we obtain predictions for these forcing terms  from Eq.~\ref{eqn:thetaforcing} for counter- and co-rotating states. % and test their functional forms, starting with the simpler special case of $\Delta T_m=0$.

%stability counter-rotating theta
\subsection{Stable fixed points for counter-rotating states for $\Delta T_m=0$} 
\label{sec:counterfixedpoints}

In a counter-rotating state where $\theta_1=\theta_2$, the equations for the two cells are symmetric in cell 1 and cell 2, so we expect a fixed point when  $\delta_1=\delta_2$ and $T_{0,1}=T_{0,2}=T_{m,1}=T_{m,2}$  in the Boussinesq approximation.  Inspection of Eqs.~\ref{eqn:thetaforcingappendix}, \ref{eqn:deltaforcingappendix} and \ref{eqn:Tforcingappendix} shows that each term is zero in this case, so $\dot\theta_i=0$, $\dot\delta_i=0$ and $\dot T_i=0$, corresponding to a fixed point for counter-rotating states.   Since Eq.~\ref{eqn:deltaforcing} is a forcing on $\delta_i$ in addition to Eq.~\ref{eqn:delta_model}, the zero forcing means that  $\delta_i$ has the same stable fixed point $\delta_0$ as Eq.~\ref{eqn:delta_model}.  

Is this fixed point stable? The first term of Eq.~\ref{eqn:thetaforcingappendix} has a  restoring force proportional to $-\sin(\theta_1-\theta_2)$, where the proportionality remains negative for all possible values of $\pi/4-\cos(2\theta_1)/2$, which is stable when $\theta_1=\theta_2$.  A Taylor expansion of the second term of Eq.~\ref{eqn:thetaforcingappendix} in $\theta_1-\theta_2$ has second-order terms, but no first order terms, so is linearly stable in $\theta_1-\theta_2$.  The third term of Eq.~\ref{eqn:thetaforcingappendix} is zero in the symmetric case when $T_{0,1}=T_{0,2}$, and the fourth term is zero at a fixed point where $\dot T_{0,i}=0$.  Thus, Eq.~\ref{eqn:thetaforcingappendix} is linearly stable in $\theta_1-\theta_2$, and counter-rotating states are a stable fixed point solution of this equation.  The first term of Eq.~\ref{eqn:thetaforcingappendix} which is responsible for this stability represents the heat transfer from effective turbulent diffusion from the LSC of roll 2 projected onto an orientation perpendicular to roll 1 to drive changes in orientation.   

%A linear expansion of Eq.~\ref{eqn:thetaforcingappendix} around the stable fixed point for counter-rotating states plugged into Eq.~\ref{eqn:ddottheta} at a preferred orientation of the cubic cell (e.g.~$\theta_1= \pi/4$ rad) and assuming $\bar\delta_i=\delta_0$ is  
%\begin{equation}
%\ddot\theta_{1,counter} \approx -0.79\frac{A\kappa_t}{\pi L^2\tau_{\dot\theta}} (\theta_1-\theta_2) \ .
%\label{eqn:thetaforcinglinearcounterappendix}
%\end{equation}
%note: checked linear expansion of 0.5\cos(2\theta_1) = \theta_1-\theta_p (small)
%This functional form will be tested in the later subsections of this section.

\subsection{Stable fixed points for co-rotating states for $\Delta T_m=0$}  

Observed stable fixed points for co-rotating states correspond to $\theta_1 =0$ and $\theta_2=\pi$ rad or $\theta_1=\pi$ rad and $\theta_2=0$,  $\delta_1=\delta_2$, and $\dot T_{0,i}=0$ in steady state.  All four terms of Eq.~\ref{eqn:thetaforcingappendix} are zero in in this case, confirming these are fixed points of Eq.~\ref{eqn:thetaforcingappendix} as well.  There are forcings on $\dot\delta_i$ in Eq.~\ref{eqn:deltaforcing} and $\dot T_{0,i}$ in Eq.~\ref{eqn:Tforcing} that shift their stable fixed point values, which are addressed in Sec.~\ref{sec:bardelta}.  Assuming constant $\Delta \bar T_{0,co} \equiv \bar T_{0,2}-\bar T_{0,1}$ and $\bar\delta_{co}\equiv\bar\delta_1=\bar\delta_2$ in the co-rotating state, a linear approximation  around $\theta_{p,1} = 0$ or $\pi$ rad for co-rotating states of Eq.~\ref{eqn:thetaforcingappendix} plugged into Eq.~\ref{eqn:ddottheta} is 
\begin{multline}
\ddot\theta_{1,co} \approx \frac{A\kappa_t\bar\delta_{co}}{\pi L^2\delta_0\tau_{\dot\theta}} \bigg[+0.28(\theta_1-\theta_2-\pi)\\
-\left(\frac{1.41\Delta \bar T_{0,co}}{\bar\delta_{co}}-2\right)(\theta_1-\theta_{p,1}) \ . \bigg]
\label{eqn:thetaforcinglinearcoappendix}
\end{multline}
  While the first term of Eq.~\ref{eqn:thetaforcingappendix}  is what stabilized the counter-rotating state, for the co-rotating state where $\theta_1 = \theta_2 + \pi$ rad, this forcing instead is predicted to be destabilizing in the first term of Eq.~\ref{eqn:thetaforcinglinearcoappendix}.  The second term of Eq.~\ref{eqn:thetaforcingappendix} results in a restoring force proportional to $+\sin(2\theta_1)$ which also destabilizes co-rotating states where $\theta_1 =0$ or $\pi$ rad.  The third term of  Eq.~\ref{eqn:thetaforcingappendix} results in a restoring force proportional to $\Delta \bar T_{0,co}\sin\theta_1$, which is stabilizing for $\theta_1=0$ and $\Delta \bar T_{0,co}>0$, or $\theta_1=\pi$ rad and $\Delta \bar T_{0,co}<0$.   These combinations correspond to same combinations as observed co-rotating states.
 This stabilizing term represents the heat transfer by effective turbulent thermal diffusion from the difference in mean temperatures $\Delta \bar T_{0,co}$, which provides a uniform heat flux across the interface, which when projected onto the LSC temperature profile, tends to align the LSC of cell 1 with $\theta_1=0$ if cell 2 is hotter ($\Delta \bar T_{0,co}>0$) or $\theta_1=\pi$ rad if cell 2 is colder ($\Delta \bar T_{0,co}<0$).  A stable co-rotating state requires the third term of Eq.~\ref{eqn:thetaforcingappendix} to be larger than the second term (which are combined in the second term of Eq.~\ref{eqn:thetaforcinglinearco}), and the first term of Eq.~\ref{eqn:thetaforcingappendix}, as well as the forcing from the geometric potential $-\nabla V_g(\theta_i)$ which is destabilizing at $\theta_i=0$ or $\pi$ rad \cite{JB20a}.  Whether $\Delta \bar T_{0,co}/\bar\delta_{co}$ is large enough to make the third term of Eq.~\ref{eqn:thetaforcingappendix} dominant over all other terms so the model has stable fixed points for co-rotating states is tested in Sec.~\ref{sec:stabilityratio}.

%\bibliography{../rbc}

\begin{thebibliography}{69}%
\makeatletter
\providecommand \@ifxundefined [1]{%
 \@ifx{#1\undefined}
}%
\providecommand \@ifnum [1]{%
 \ifnum #1\expandafter \@firstoftwo
 \else \expandafter \@secondoftwo
 \fi
}%
\providecommand \@ifx [1]{%
 \ifx #1\expandafter \@firstoftwo
 \else \expandafter \@secondoftwo
 \fi
}%
\providecommand \natexlab [1]{#1}%
\providecommand \enquote  [1]{``#1''}%
\providecommand \bibnamefont  [1]{#1}%
\providecommand \bibfnamefont [1]{#1}%
\providecommand \citenamefont [1]{#1}%
\providecommand \href@noop [0]{\@secondoftwo}%
\providecommand \href [0]{\begingroup \@sanitize@url \@href}%
\providecommand \@href[1]{\@@startlink{#1}\@@href}%
\providecommand \@@href[1]{\endgroup#1\@@endlink}%
\providecommand \@sanitize@url [0]{\catcode `\\12\catcode `\$12\catcode
  `\&12\catcode `\#12\catcode `\^12\catcode `\_12\catcode `\%12\relax}%
\providecommand \@@startlink[1]{}%
\providecommand \@@endlink[0]{}%
\providecommand \url  [0]{\begingroup\@sanitize@url \@url }%
\providecommand \@url [1]{\endgroup\@href {#1}{\urlprefix }}%
\providecommand \urlprefix  [0]{URL }%
\providecommand \Eprint [0]{\href }%
\providecommand \doibase [0]{http://dx.doi.org/}%
\providecommand \selectlanguage [0]{\@gobble}%
\providecommand \bibinfo  [0]{\@secondoftwo}%
\providecommand \bibfield  [0]{\@secondoftwo}%
\providecommand \translation [1]{[#1]}%
\providecommand \BibitemOpen [0]{}%
\providecommand \bibitemStop [0]{}%
\providecommand \bibitemNoStop [0]{.\EOS\space}%
\providecommand \EOS [0]{\spacefactor3000\relax}%
\providecommand \BibitemShut  [1]{\csname bibitem#1\endcsname}%
\let\auto@bib@innerbib\@empty
%</preamble>
\bibitem [{\citenamefont {Ahlers}\ \emph {et~al.}(2009)\citenamefont {Ahlers},
  \citenamefont {Grossmann},\ and\ \citenamefont {Lohse}}]{AGL09}%
  \BibitemOpen
  \bibfield  {author} {\bibinfo {author} {\bibfnamefont {G.}~\bibnamefont
  {Ahlers}}, \bibinfo {author} {\bibfnamefont {S.}~\bibnamefont {Grossmann}}, \
  and\ \bibinfo {author} {\bibfnamefont {D.}~\bibnamefont {Lohse}},\ }\bibfield
   {title} {\enquote {\bibinfo {title} {Heat transfer and large-scale dynamics
  in turbulent {{Rayleigh-B\'enard}} convection},}\ }\href@noop {} {\bibfield
  {journal} {\bibinfo  {journal} {Reviews of Modern Physics}\ }\textbf
  {\bibinfo {volume} {81}},\ \bibinfo {pages} {503--537} (\bibinfo {year}
  {2009})}\BibitemShut {NoStop}%
\bibitem [{\citenamefont {Lohse}\ and\ \citenamefont {Xia}(2010)}]{LX10}%
  \BibitemOpen
  \bibfield  {author} {\bibinfo {author} {\bibfnamefont {D.}~\bibnamefont
  {Lohse}}\ and\ \bibinfo {author} {\bibfnamefont {K.-Q.}\ \bibnamefont
  {Xia}},\ }\bibfield  {title} {\enquote {\bibinfo {title} {Small-scale
  properties of turbulent {{Rayleigh-B\'enard}} convection},}\ }\href@noop {}
  {\bibfield  {journal} {\bibinfo  {journal} {Annual Reviews of Fluid
  Mechanics}\ }\textbf {\bibinfo {volume} {42}},\ \bibinfo {pages} {335--364}
  (\bibinfo {year} {2010})}\BibitemShut {NoStop}%
\bibitem [{\citenamefont {Krishnamurti}\ and\ \citenamefont
  {Howard}(1981)}]{KH81}%
  \BibitemOpen
  \bibfield  {author} {\bibinfo {author} {\bibfnamefont {R.}~\bibnamefont
  {Krishnamurti}}\ and\ \bibinfo {author} {\bibfnamefont {L.~N.}\ \bibnamefont
  {Howard}},\ }\bibfield  {title} {\enquote {\bibinfo {title} {Large scale flow
  generation in turbulent convection},}\ }\href@noop {} {\bibfield  {journal}
  {\bibinfo  {journal} {Proc. Natl. Acad. Sci.}\ }\textbf {\bibinfo {volume}
  {78}},\ \bibinfo {pages} {1981--1985} (\bibinfo {year} {1981})}\BibitemShut
  {NoStop}%
\bibitem [{\citenamefont {Brown}\ and\ \citenamefont
  {Ahlers}(2006{\natexlab{a}})}]{BA06a}%
  \BibitemOpen
  \bibfield  {author} {\bibinfo {author} {\bibfnamefont {E.}~\bibnamefont
  {Brown}}\ and\ \bibinfo {author} {\bibfnamefont {G.}~\bibnamefont {Ahlers}},\
  }\bibfield  {title} {\enquote {\bibinfo {title} {Rotations and cessations of
  the large-scale circulation in turbulent {{Rayleigh-B\'enard}} convection},}\
  }\href@noop {} {\bibfield  {journal} {\bibinfo  {journal} {J. Fluid Mech.}\
  }\textbf {\bibinfo {volume} {568}},\ \bibinfo {pages} {351} (\bibinfo {year}
  {2006}{\natexlab{a}})}\BibitemShut {NoStop}%
\bibitem [{\citenamefont {Xi}\ and\ \citenamefont {Xia}(2007)}]{XX07}%
  \BibitemOpen
  \bibfield  {author} {\bibinfo {author} {\bibfnamefont {H.-D.}\ \bibnamefont
  {Xi}}\ and\ \bibinfo {author} {\bibfnamefont {K.-Q.}\ \bibnamefont {Xia}},\
  }\bibfield  {title} {\enquote {\bibinfo {title} {Cessations and reversals of
  the large-scale circulation in turbulent thermal convection},}\ }\href@noop
  {} {\bibfield  {journal} {\bibinfo  {journal} {Phys. Rev. E}\ }\textbf
  {\bibinfo {volume} {75}},\ \bibinfo {pages} {066307--1--5} (\bibinfo {year}
  {2007})}\BibitemShut {NoStop}%
\bibitem [{\citenamefont {Heslot}\ \emph {et~al.}(1987)\citenamefont {Heslot},
  \citenamefont {Castaing},\ and\ \citenamefont {Libchaber}}]{HCL87}%
  \BibitemOpen
  \bibfield  {author} {\bibinfo {author} {\bibfnamefont {F.}~\bibnamefont
  {Heslot}}, \bibinfo {author} {\bibfnamefont {B.}~\bibnamefont {Castaing}}, \
  and\ \bibinfo {author} {\bibfnamefont {A.}~\bibnamefont {Libchaber}},\
  }\bibfield  {title} {\enquote {\bibinfo {title} {Transition to turbulence in
  helium gas},}\ }\href@noop {} {\bibfield  {journal} {\bibinfo  {journal}
  {Phys. Rev. A}\ }\textbf {\bibinfo {volume} {36}},\ \bibinfo {pages}
  {5870--5873} (\bibinfo {year} {1987})}\BibitemShut {NoStop}%
\bibitem [{\citenamefont {Sano}\ \emph {et~al.}(1989)\citenamefont {Sano},
  \citenamefont {Wu},\ and\ \citenamefont {Libchaber}}]{SWL89}%
  \BibitemOpen
  \bibfield  {author} {\bibinfo {author} {\bibfnamefont {M.}~\bibnamefont
  {Sano}}, \bibinfo {author} {\bibfnamefont {X.~Z.}\ \bibnamefont {Wu}}, \ and\
  \bibinfo {author} {\bibfnamefont {A.}~\bibnamefont {Libchaber}},\ }\bibfield
  {title} {\enquote {\bibinfo {title} {Turbulence in helium-gas free
  convection},}\ }\href@noop {} {\bibfield  {journal} {\bibinfo  {journal}
  {Phys. Rev. A}\ }\textbf {\bibinfo {volume} {40}},\ \bibinfo {pages} {6421}
  (\bibinfo {year} {1989})}\BibitemShut {NoStop}%
\bibitem [{\citenamefont {Castaing}\ \emph {et~al.}(1989)\citenamefont
  {Castaing}, \citenamefont {Gunaratne}, \citenamefont {Heslot}, \citenamefont
  {Kadanoff}, \citenamefont {Libchaber}, \citenamefont {Thomae}, \citenamefont
  {Wu}, \citenamefont {Zaleski},\ and\ \citenamefont {Zanetti}}]{CGHKLTWZZ89}%
  \BibitemOpen
  \bibfield  {author} {\bibinfo {author} {\bibfnamefont {B.}~\bibnamefont
  {Castaing}}, \bibinfo {author} {\bibfnamefont {G.}~\bibnamefont {Gunaratne}},
  \bibinfo {author} {\bibfnamefont {F.}~\bibnamefont {Heslot}}, \bibinfo
  {author} {\bibfnamefont {L.}~\bibnamefont {Kadanoff}}, \bibinfo {author}
  {\bibfnamefont {A.}~\bibnamefont {Libchaber}}, \bibinfo {author}
  {\bibfnamefont {S.}~\bibnamefont {Thomae}}, \bibinfo {author} {\bibfnamefont
  {X.~Z.}\ \bibnamefont {Wu}}, \bibinfo {author} {\bibfnamefont
  {S.}~\bibnamefont {Zaleski}}, \ and\ \bibinfo {author} {\bibfnamefont
  {G.}~\bibnamefont {Zanetti}},\ }\bibfield  {title} {\enquote {\bibinfo
  {title} {Scaling of hard thermal turbulence in {{Rayleigh-B\'enard}}
  convection},}\ }\href@noop {} {\bibfield  {journal} {\bibinfo  {journal} {J.
  Fluid Mech.}\ }\textbf {\bibinfo {volume} {204}},\ \bibinfo {pages} {1--30}
  (\bibinfo {year} {1989})}\BibitemShut {NoStop}%
\bibitem [{\citenamefont {Ciliberto}\ \emph {et~al.}(1996)\citenamefont
  {Ciliberto}, \citenamefont {Cioni},\ and\ \citenamefont {Laroche}}]{CCL96}%
  \BibitemOpen
  \bibfield  {author} {\bibinfo {author} {\bibfnamefont {S.}~\bibnamefont
  {Ciliberto}}, \bibinfo {author} {\bibfnamefont {S.}~\bibnamefont {Cioni}}, \
  and\ \bibinfo {author} {\bibfnamefont {C.}~\bibnamefont {Laroche}},\
  }\bibfield  {title} {\enquote {\bibinfo {title} {Large-scale flow properties
  of turbulent thermal convection},}\ }\href@noop {} {\bibfield  {journal}
  {\bibinfo  {journal} {Phys. Rev. E}\ }\textbf {\bibinfo {volume} {54}},\
  \bibinfo {pages} {R5901--R5904} (\bibinfo {year} {1996})}\BibitemShut
  {NoStop}%
\bibitem [{\citenamefont {Takeshita}\ \emph {et~al.}(1996)\citenamefont
  {Takeshita}, \citenamefont {Segawa}, \citenamefont {Glazier},\ and\
  \citenamefont {Sano}}]{TSGS96}%
  \BibitemOpen
  \bibfield  {author} {\bibinfo {author} {\bibfnamefont {T.}~\bibnamefont
  {Takeshita}}, \bibinfo {author} {\bibfnamefont {T.}~\bibnamefont {Segawa}},
  \bibinfo {author} {\bibfnamefont {J.~A.}\ \bibnamefont {Glazier}}, \ and\
  \bibinfo {author} {\bibfnamefont {M.}~\bibnamefont {Sano}},\ }\bibfield
  {title} {\enquote {\bibinfo {title} {Thermal turbulence in mercury},}\
  }\href@noop {} {\bibfield  {journal} {\bibinfo  {journal} {Phys. Rev. Lett.}\
  }\textbf {\bibinfo {volume} {76}},\ \bibinfo {pages} {1465--1468} (\bibinfo
  {year} {1996})}\BibitemShut {NoStop}%
\bibitem [{\citenamefont {Cioni}\ \emph {et~al.}(1997)\citenamefont {Cioni},
  \citenamefont {Ciliberto},\ and\ \citenamefont {Sommeria}}]{CCS97}%
  \BibitemOpen
  \bibfield  {author} {\bibinfo {author} {\bibfnamefont {S.}~\bibnamefont
  {Cioni}}, \bibinfo {author} {\bibfnamefont {S.}~\bibnamefont {Ciliberto}}, \
  and\ \bibinfo {author} {\bibfnamefont {J.}~\bibnamefont {Sommeria}},\
  }\bibfield  {title} {\enquote {\bibinfo {title} {Strongly turbulent
  {{Rayleigh-B\'enard}} convection in mercury: comparison with results at
  moderate {{Prandtl}} number},}\ }\href@noop {} {\bibfield  {journal}
  {\bibinfo  {journal} {J. Fluid Mech.}\ }\textbf {\bibinfo {volume} {335}},\
  \bibinfo {pages} {111--140} (\bibinfo {year} {1997})}\BibitemShut {NoStop}%
\bibitem [{\citenamefont {Qiu}\ \emph {et~al.}(2000)\citenamefont {Qiu},
  \citenamefont {Yao},\ and\ \citenamefont {Tong}}]{QYT00}%
  \BibitemOpen
  \bibfield  {author} {\bibinfo {author} {\bibfnamefont {X.~L.}\ \bibnamefont
  {Qiu}}, \bibinfo {author} {\bibfnamefont {S.~H.}\ \bibnamefont {Yao}}, \ and\
  \bibinfo {author} {\bibfnamefont {P.}~\bibnamefont {Tong}},\ }\bibfield
  {title} {\enquote {\bibinfo {title} {Large-scale coherent rotation and
  oscillation in turbulent thermal convection},}\ }\href@noop {} {\bibfield
  {journal} {\bibinfo  {journal} {Phys. Rev. E}\ }\textbf {\bibinfo {volume}
  {61}},\ \bibinfo {pages} {R6075} (\bibinfo {year} {2000})}\BibitemShut
  {NoStop}%
\bibitem [{\citenamefont {Qiu}\ and\ \citenamefont {Tong}(2001)}]{QT01b}%
  \BibitemOpen
  \bibfield  {author} {\bibinfo {author} {\bibfnamefont {X.~L.}\ \bibnamefont
  {Qiu}}\ and\ \bibinfo {author} {\bibfnamefont {P.}~\bibnamefont {Tong}},\
  }\bibfield  {title} {\enquote {\bibinfo {title} {Onset of coherent
  oscillations in turbulent {{Rayleigh-B\'enard}} convection},}\ }\href@noop {}
  {\bibfield  {journal} {\bibinfo  {journal} {Phys. Rev. Lett}\ }\textbf
  {\bibinfo {volume} {87}},\ \bibinfo {pages} {094501} (\bibinfo {year}
  {2001})}\BibitemShut {NoStop}%
\bibitem [{\citenamefont {Niemela}\ \emph {et~al.}(2001)\citenamefont
  {Niemela}, \citenamefont {Skrbek}, \citenamefont {Sreenivasan},\ and\
  \citenamefont {Donnelly}}]{NSSD01}%
  \BibitemOpen
  \bibfield  {author} {\bibinfo {author} {\bibfnamefont {J.~J.}\ \bibnamefont
  {Niemela}}, \bibinfo {author} {\bibfnamefont {L.}~\bibnamefont {Skrbek}},
  \bibinfo {author} {\bibfnamefont {K.~R.}\ \bibnamefont {Sreenivasan}}, \ and\
  \bibinfo {author} {\bibfnamefont {R.~J.}\ \bibnamefont {Donnelly}},\
  }\bibfield  {title} {\enquote {\bibinfo {title} {The wind in confined thermal
  turbulence},}\ }\href@noop {} {\bibfield  {journal} {\bibinfo  {journal} {J.
  Fluid Mech.}\ }\textbf {\bibinfo {volume} {449}},\ \bibinfo {pages}
  {169--178} (\bibinfo {year} {2001})}\BibitemShut {NoStop}%
\bibitem [{\citenamefont {Qiu}\ and\ \citenamefont {Tong}(2002)}]{QT02}%
  \BibitemOpen
  \bibfield  {author} {\bibinfo {author} {\bibfnamefont {X.~L.}\ \bibnamefont
  {Qiu}}\ and\ \bibinfo {author} {\bibfnamefont {P.}~\bibnamefont {Tong}},\
  }\bibfield  {title} {\enquote {\bibinfo {title} {Temperature oscillations in
  turbulent rayleigh-b{\'e}nard convection},}\ }\href@noop {} {\bibfield
  {journal} {\bibinfo  {journal} {Phys. Rev. E}\ }\textbf {\bibinfo {volume}
  {66}},\ \bibinfo {pages} {026308} (\bibinfo {year} {2002})}\BibitemShut
  {NoStop}%
\bibitem [{\citenamefont {Qiu}\ \emph {et~al.}(2004)\citenamefont {Qiu},
  \citenamefont {Shang}, \citenamefont {Tong},\ and\ \citenamefont
  {Xia}}]{QSTX04}%
  \BibitemOpen
  \bibfield  {author} {\bibinfo {author} {\bibfnamefont {X.~L.}\ \bibnamefont
  {Qiu}}, \bibinfo {author} {\bibfnamefont {X.~D.}\ \bibnamefont {Shang}},
  \bibinfo {author} {\bibfnamefont {P.}~\bibnamefont {Tong}}, \ and\ \bibinfo
  {author} {\bibfnamefont {K.-Q.}\ \bibnamefont {Xia}},\ }\bibfield  {title}
  {\enquote {\bibinfo {title} {Velocity oscillations in turbulent
  {{Rayleigh-B\'enard}} convection},}\ }\href@noop {} {\bibfield  {journal}
  {\bibinfo  {journal} {Phys. Fluids.}\ }\textbf {\bibinfo {volume} {16}},\
  \bibinfo {pages} {412--423} (\bibinfo {year} {2004})}\BibitemShut {NoStop}%
\bibitem [{\citenamefont {Sun}\ \emph {et~al.}(2005)\citenamefont {Sun},
  \citenamefont {Xia},\ and\ \citenamefont {Tong}}]{SXT05}%
  \BibitemOpen
  \bibfield  {author} {\bibinfo {author} {\bibfnamefont {C.}~\bibnamefont
  {Sun}}, \bibinfo {author} {\bibfnamefont {K.~Q.}\ \bibnamefont {Xia}}, \ and\
  \bibinfo {author} {\bibfnamefont {P.}~\bibnamefont {Tong}},\ }\bibfield
  {title} {\enquote {\bibinfo {title} {Three-dimensional flow structures and
  dynamics of turbulent thermal convection in a cylindrical cell},}\
  }\href@noop {} {\bibfield  {journal} {\bibinfo  {journal} {Phys. Rev. E}\
  }\textbf {\bibinfo {volume} {72}},\ \bibinfo {pages} {026302} (\bibinfo
  {year} {2005})}\BibitemShut {NoStop}%
\bibitem [{\citenamefont {Tsuji}\ \emph {et~al.}(2005)\citenamefont {Tsuji},
  \citenamefont {Mizuno}, \citenamefont {Mashiko},\ and\ \citenamefont
  {Sano}}]{TMMS05}%
  \BibitemOpen
  \bibfield  {author} {\bibinfo {author} {\bibfnamefont {Y.}~\bibnamefont
  {Tsuji}}, \bibinfo {author} {\bibfnamefont {T.}~\bibnamefont {Mizuno}},
  \bibinfo {author} {\bibfnamefont {T.}~\bibnamefont {Mashiko}}, \ and\
  \bibinfo {author} {\bibfnamefont {M.}~\bibnamefont {Sano}},\ }\bibfield
  {title} {\enquote {\bibinfo {title} {Mean wind in convective turbulence of
  mercury},}\ }\href@noop {} {\bibfield  {journal} {\bibinfo  {journal} {Phys.
  Rev. Lett.}\ }\textbf {\bibinfo {volume} {94}},\ \bibinfo {pages} {034501}
  (\bibinfo {year} {2005})}\BibitemShut {NoStop}%
\bibitem [{\citenamefont {Funfschilling}\ and\ \citenamefont
  {Ahlers}(2004)}]{FA04}%
  \BibitemOpen
  \bibfield  {author} {\bibinfo {author} {\bibfnamefont {D.}~\bibnamefont
  {Funfschilling}}\ and\ \bibinfo {author} {\bibfnamefont {G.}~\bibnamefont
  {Ahlers}},\ }\bibfield  {title} {\enquote {\bibinfo {title} {Plume motion and
  large scale circulation in a cylindrical {{Rayleigh-B\'enard}} cell},}\
  }\href@noop {} {\bibfield  {journal} {\bibinfo  {journal} {Phys. Rev. Lett.}\
  }\textbf {\bibinfo {volume} {92}},\ \bibinfo {pages} {194502} (\bibinfo
  {year} {2004})}\BibitemShut {NoStop}%
\bibitem [{\citenamefont {Xi}\ \emph {et~al.}(2009)\citenamefont {Xi},
  \citenamefont {Zhou}, \citenamefont {Zhou}, \citenamefont {Chan},\ and\
  \citenamefont {Xia}}]{XZZCX09}%
  \BibitemOpen
  \bibfield  {author} {\bibinfo {author} {\bibfnamefont {H.-D.}\ \bibnamefont
  {Xi}}, \bibinfo {author} {\bibfnamefont {S.-Q.}\ \bibnamefont {Zhou}},
  \bibinfo {author} {\bibfnamefont {Q.}~\bibnamefont {Zhou}}, \bibinfo {author}
  {\bibfnamefont {T.-S.}\ \bibnamefont {Chan}}, \ and\ \bibinfo {author}
  {\bibfnamefont {K.-Q.}\ \bibnamefont {Xia}},\ }\bibfield  {title} {\enquote
  {\bibinfo {title} {Origin of the temperature oscillation in turbulent thermal
  convection},}\ }\href@noop {} {\bibfield  {journal} {\bibinfo  {journal}
  {Phys. Rev. Lett.}\ }\textbf {\bibinfo {volume} {102}},\ \bibinfo {pages}
  {044503--1--4} (\bibinfo {year} {2009})}\BibitemShut {NoStop}%
\bibitem [{\citenamefont {Zhou}\ \emph {et~al.}(2009)\citenamefont {Zhou},
  \citenamefont {Xi}, \citenamefont {Zhou}, \citenamefont {Sun},\ and\
  \citenamefont {Xia}}]{ZXZSX09}%
  \BibitemOpen
  \bibfield  {author} {\bibinfo {author} {\bibfnamefont {Q.}~\bibnamefont
  {Zhou}}, \bibinfo {author} {\bibfnamefont {H.-D.}\ \bibnamefont {Xi}},
  \bibinfo {author} {\bibfnamefont {S.-Q.}\ \bibnamefont {Zhou}}, \bibinfo
  {author} {\bibfnamefont {C.}~\bibnamefont {Sun}}, \ and\ \bibinfo {author}
  {\bibfnamefont {K.-Q.}\ \bibnamefont {Xia}},\ }\bibfield  {title} {\enquote
  {\bibinfo {title} {Oscillations of the large-scale circulation in turbulent
  {{Rayleigh-B\'enard}} convection: the sloshing mode and its relationship with
  the torsional mode},}\ }\href@noop {} {\bibfield  {journal} {\bibinfo
  {journal} {J. Fluid Mech.}\ }\textbf {\bibinfo {volume} {630}},\ \bibinfo
  {pages} {367--390} (\bibinfo {year} {2009})}\BibitemShut {NoStop}%
\bibitem [{\citenamefont {Brown}\ and\ \citenamefont {Ahlers}(2009)}]{BA09}%
  \BibitemOpen
  \bibfield  {author} {\bibinfo {author} {\bibfnamefont {E.}~\bibnamefont
  {Brown}}\ and\ \bibinfo {author} {\bibfnamefont {G.}~\bibnamefont {Ahlers}},\
  }\bibfield  {title} {\enquote {\bibinfo {title} {The origin of oscillations
  of the large-scale circulation of turbulent {{Rayleigh-B\'enard}}
  convection},}\ }\href@noop {} {\bibfield  {journal} {\bibinfo  {journal} {J.
  Fluid Mech.}\ }\textbf {\bibinfo {volume} {638}},\ \bibinfo {pages}
  {383--400} (\bibinfo {year} {2009})}\BibitemShut {NoStop}%
\bibitem [{\citenamefont {Vogt}\ \emph {et~al.}(2018)\citenamefont {Vogt},
  \citenamefont {Horn}, \citenamefont {Grannan},\ and\ \citenamefont
  {Aurnou}}]{VHGA18}%
  \BibitemOpen
  \bibfield  {author} {\bibinfo {author} {\bibfnamefont {T.}~\bibnamefont
  {Vogt}}, \bibinfo {author} {\bibfnamefont {S.}~\bibnamefont {Horn}}, \bibinfo
  {author} {\bibfnamefont {A.~M.}\ \bibnamefont {Grannan}}, \ and\ \bibinfo
  {author} {\bibfnamefont {J.~M.}\ \bibnamefont {Aurnou}},\ }\bibfield  {title}
  {\enquote {\bibinfo {title} {Jump rope vortex in liquid metal convection},}\
  }\href@noop {} {\bibfield  {journal} {\bibinfo  {journal} {Proc. Nat. Acad.
  Sciences}\ }\textbf {\bibinfo {volume} {115}},\ \bibinfo {pages}
  {12674--12679} (\bibinfo {year} {2018})}\BibitemShut {NoStop}%
\bibitem [{\citenamefont {Horn}\ \emph {et~al.}(2022)\citenamefont {Horn},
  \citenamefont {Schmid},\ and\ \citenamefont {arnou}}]{HSA22}%
  \BibitemOpen
  \bibfield  {author} {\bibinfo {author} {\bibfnamefont {S.}~\bibnamefont
  {Horn}}, \bibinfo {author} {\bibfnamefont {P.J.}\ \bibnamefont {Schmid}}, \
  and\ \bibinfo {author} {\bibfnamefont {J.M.}\ \bibnamefont {arnou}},\
  }\bibfield  {title} {\enquote {\bibinfo {title} {Unravelling the large-scale
  circulation modes in turbulent rayleigh-bénard convection},}\ }\href@noop {}
  {\bibfield  {journal} {\bibinfo  {journal} {Europhysics Letters}\ }\textbf
  {\bibinfo {volume} {136}},\ \bibinfo {pages} {14003} (\bibinfo {year}
  {2022})}\BibitemShut {NoStop}%
\bibitem [{\citenamefont {Liu}\ and\ \citenamefont {Ecke}(2009)}]{LE09}%
  \BibitemOpen
  \bibfield  {author} {\bibinfo {author} {\bibfnamefont {Y.}~\bibnamefont
  {Liu}}\ and\ \bibinfo {author} {\bibfnamefont {R.~E.}\ \bibnamefont {Ecke}},\
  }\bibfield  {title} {\enquote {\bibinfo {title} {Heat transport measurements
  in turbulent rotating {{Rayleigh-B\'enard}} convection},}\ }\href {\doibase
  10.1103/PhysRevE.80.036314} {\bibfield  {journal} {\bibinfo  {journal} {Phys.
  Rev. E}\ }\textbf {\bibinfo {volume} {80}},\ \bibinfo {pages} {036314}
  (\bibinfo {year} {2009})}\BibitemShut {NoStop}%
\bibitem [{\citenamefont {Song}\ \emph {et~al.}(2014)\citenamefont {Song},
  \citenamefont {Brown}, \citenamefont {Hawkins},\ and\ \citenamefont
  {Tong}}]{SBHT14}%
  \BibitemOpen
  \bibfield  {author} {\bibinfo {author} {\bibfnamefont {H.}~\bibnamefont
  {Song}}, \bibinfo {author} {\bibfnamefont {E.}~\bibnamefont {Brown}},
  \bibinfo {author} {\bibfnamefont {R.}~\bibnamefont {Hawkins}}, \ and\
  \bibinfo {author} {\bibfnamefont {P}~\bibnamefont {Tong}},\ }\bibfield
  {title} {\enquote {\bibinfo {title} {Dynamics of large-scale circulation of
  turbulent thermal convection in a horizontal cylinder},}\ }\href@noop {}
  {\bibfield  {journal} {\bibinfo  {journal} {J. Fluid Mech}\ }\textbf
  {\bibinfo {volume} {740}},\ \bibinfo {pages} {136--167} (\bibinfo {year}
  {2014})}\BibitemShut {NoStop}%
\bibitem [{\citenamefont {Bai}\ \emph {et~al.}(2016)\citenamefont {Bai},
  \citenamefont {Ji},\ and\ \citenamefont {Brown}}]{BJB16}%
  \BibitemOpen
  \bibfield  {author} {\bibinfo {author} {\bibfnamefont {K.}~\bibnamefont
  {Bai}}, \bibinfo {author} {\bibfnamefont {D.}~\bibnamefont {Ji}}, \ and\
  \bibinfo {author} {\bibfnamefont {E.}~\bibnamefont {Brown}},\ }\bibfield
  {title} {\enquote {\bibinfo {title} {Ability of a low-dimensional model to
  predict geometry-dependent dynamics of large-scale coherent structures in
  turbulence},}\ }\href@noop {} {\bibfield  {journal} {\bibinfo  {journal}
  {Phys. Rev. E}\ }\textbf {\bibinfo {volume} {93}},\ \bibinfo {pages}
  {023117--1--5} (\bibinfo {year} {2016})}\BibitemShut {NoStop}%
\bibitem [{\citenamefont {Foroozani}\ \emph {et~al.}(2017)\citenamefont
  {Foroozani}, \citenamefont {Niemela}, \citenamefont {Armenio},\ and\
  \citenamefont {Sreenivasan}}]{FNAS17}%
  \BibitemOpen
  \bibfield  {author} {\bibinfo {author} {\bibfnamefont {N.}~\bibnamefont
  {Foroozani}}, \bibinfo {author} {\bibfnamefont {J.J.}\ \bibnamefont
  {Niemela}}, \bibinfo {author} {\bibfnamefont {V.}~\bibnamefont {Armenio}}, \
  and\ \bibinfo {author} {\bibfnamefont {K.R.}\ \bibnamefont {Sreenivasan}},\
  }\bibfield  {title} {\enquote {\bibinfo {title} {Reorientations of the
  large-scale flow in turbulent convection in a cube},}\ }\href@noop {}
  {\bibfield  {journal} {\bibinfo  {journal} {Phys. Rev. E}\ }\textbf {\bibinfo
  {volume} {95}},\ \bibinfo {pages} {033107} (\bibinfo {year}
  {2017})}\BibitemShut {NoStop}%
\bibitem [{\citenamefont {Giannakis}\ \emph {et~al.}(2018)\citenamefont
  {Giannakis}, \citenamefont {Kolchinskaya}, \citenamefont {Krasnov},\ and\
  \citenamefont {Schumacher}}]{GKKS18}%
  \BibitemOpen
  \bibfield  {author} {\bibinfo {author} {\bibfnamefont {D.}~\bibnamefont
  {Giannakis}}, \bibinfo {author} {\bibfnamefont {A.}~\bibnamefont
  {Kolchinskaya}}, \bibinfo {author} {\bibfnamefont {D.}~\bibnamefont
  {Krasnov}}, \ and\ \bibinfo {author} {\bibfnamefont {J.}~\bibnamefont
  {Schumacher}},\ }\bibfield  {title} {\enquote {\bibinfo {title} {Koopman
  analysis of the long-term evolution in a turbulent convection cell},}\
  }\href@noop {} {\bibfield  {journal} {\bibinfo  {journal} {J. Fluid Mech.}\
  }\textbf {\bibinfo {volume} {847}},\ \bibinfo {pages} {735--767} (\bibinfo
  {year} {2018})}\BibitemShut {NoStop}%
\bibitem [{\citenamefont {Vasiliev}\ \emph {et~al.}(2016)\citenamefont
  {Vasiliev}, \citenamefont {Sukhanovskii}, \citenamefont {Frick},
  \citenamefont {Budnikov}, \citenamefont {Fomichev}, \citenamefont
  {Bolshukhin},\ and\ \citenamefont {Romanov}}]{VSFBFBR16}%
  \BibitemOpen
  \bibfield  {author} {\bibinfo {author} {\bibfnamefont {A.}~\bibnamefont
  {Vasiliev}}, \bibinfo {author} {\bibfnamefont {A.}~\bibnamefont
  {Sukhanovskii}}, \bibinfo {author} {\bibfnamefont {P.}~\bibnamefont {Frick}},
  \bibinfo {author} {\bibfnamefont {A.}~\bibnamefont {Budnikov}}, \bibinfo
  {author} {\bibfnamefont {V.}~\bibnamefont {Fomichev}}, \bibinfo {author}
  {\bibfnamefont {M.}~\bibnamefont {Bolshukhin}}, \ and\ \bibinfo {author}
  {\bibfnamefont {R.}~\bibnamefont {Romanov}},\ }\bibfield  {title} {\enquote
  {\bibinfo {title} {High rayleigh number convection in a cubic cell with
  adiabatic sidewalls},}\ }\href {\doibase
  https://doi.org/10.1016/j.ijheatmasstransfer.2016.06.015} {\bibfield
  {journal} {\bibinfo  {journal} {International Journal of Heat and Mass
  Transfer}\ }\textbf {\bibinfo {volume} {102}},\ \bibinfo {pages} {201 -- 212}
  (\bibinfo {year} {2016})}\BibitemShut {NoStop}%
\bibitem [{\citenamefont {Vasiliev}\ \emph {et~al.}(2019)\citenamefont
  {Vasiliev}, \citenamefont {Frick}, \citenamefont {Kumar}, \citenamefont
  {Stepanov}, \citenamefont {Sukhanovskii},\ and\ \citenamefont
  {Verma}}]{VFKSSV19}%
  \BibitemOpen
  \bibfield  {author} {\bibinfo {author} {\bibfnamefont {A.}~\bibnamefont
  {Vasiliev}}, \bibinfo {author} {\bibfnamefont {P.}~\bibnamefont {Frick}},
  \bibinfo {author} {\bibfnamefont {A.}~\bibnamefont {Kumar}}, \bibinfo
  {author} {\bibfnamefont {R.}~\bibnamefont {Stepanov}}, \bibinfo {author}
  {\bibfnamefont {A.}~\bibnamefont {Sukhanovskii}}, \ and\ \bibinfo {author}
  {\bibfnamefont {M.~K.}\ \bibnamefont {Verma}},\ }\bibfield  {title} {\enquote
  {\bibinfo {title} {Transient flows and reorientations of large-scale
  convection in a cubic cell},}\ }\href@noop {} {\bibfield  {journal} {\bibinfo
   {journal} {International Communications in Heat and Mass Transfer}\ }\textbf
  {\bibinfo {volume} {108}},\ \bibinfo {pages} {104319} (\bibinfo {year}
  {2019})}\BibitemShut {NoStop}%
\bibitem [{\citenamefont {Ji}\ \emph {et~al.}(2020)\citenamefont {Ji},
  \citenamefont {Bai},\ and\ \citenamefont {Brown}}]{JBB20}%
  \BibitemOpen
  \bibfield  {author} {\bibinfo {author} {\bibfnamefont {D.}~\bibnamefont
  {Ji}}, \bibinfo {author} {\bibfnamefont {K.}~\bibnamefont {Bai}}, \ and\
  \bibinfo {author} {\bibfnamefont {E.}~\bibnamefont {Brown}},\ }\bibfield
  {title} {\enquote {\bibinfo {title} {Effects of tilt on the orientation
  dynamics of the large-scale circulation in turbulent rayleigh-b{\'e}nard
  convection},}\ }\href@noop {} {\bibfield  {journal} {\bibinfo  {journal}
  {Phys. Fluids}\ }\textbf {\bibinfo {volume} {32}},\ \bibinfo {pages} {075118}
  (\bibinfo {year} {2020})}\BibitemShut {NoStop}%
\bibitem [{\citenamefont {Ji}\ and\ \citenamefont
  {Brown}(2020{\natexlab{a}})}]{JB20b}%
  \BibitemOpen
  \bibfield  {author} {\bibinfo {author} {\bibfnamefont {D.}~\bibnamefont
  {Ji}}\ and\ \bibinfo {author} {\bibfnamefont {E.}~\bibnamefont {Brown}},\
  }\bibfield  {title} {\enquote {\bibinfo {title} {Oscillation in the
  temperature profile of the large-scale circulation of turbulent convection
  induced by a cubic container},}\ }\href@noop {} {\bibfield  {journal}
  {\bibinfo  {journal} {Phys. Rev. Fluids}\ }\textbf {\bibinfo {volume} {5}},\
  \bibinfo {pages} {063501} (\bibinfo {year} {2020}{\natexlab{a}})}\BibitemShut
  {NoStop}%
\bibitem [{\citenamefont {Lorenz}(1963)}]{Lo63}%
  \BibitemOpen
  \bibfield  {author} {\bibinfo {author} {\bibfnamefont {E.~N.}\ \bibnamefont
  {Lorenz}},\ }\bibfield  {title} {\enquote {\bibinfo {title} {Deterministic
  nonperiodic flow},}\ }\href@noop {} {\bibfield  {journal} {\bibinfo
  {journal} {J. Atmos. Sci}\ }\textbf {\bibinfo {volume} {20}},\ \bibinfo
  {pages} {130--141} (\bibinfo {year} {1963})}\BibitemShut {NoStop}%
\bibitem [{\citenamefont {Brown}\ and\ \citenamefont
  {Ahlers}(2007{\natexlab{a}})}]{BA07a}%
  \BibitemOpen
  \bibfield  {author} {\bibinfo {author} {\bibfnamefont {E.}~\bibnamefont
  {Brown}}\ and\ \bibinfo {author} {\bibfnamefont {G.}~\bibnamefont {Ahlers}},\
  }\bibfield  {title} {\enquote {\bibinfo {title} {Large-scale circulation
  model of turbulent {{Rayleigh-B\'enard}} convection},}\ }\href@noop {}
  {\bibfield  {journal} {\bibinfo  {journal} {Phys. Rev. Lett.}\ }\textbf
  {\bibinfo {volume} {98}},\ \bibinfo {pages} {134501--1--4} (\bibinfo {year}
  {2007}{\natexlab{a}})}\BibitemShut {NoStop}%
\bibitem [{\citenamefont {de~la Torre}\ and\ \citenamefont
  {Burguete}(2007)}]{TB07}%
  \BibitemOpen
  \bibfield  {author} {\bibinfo {author} {\bibfnamefont {A.}~\bibnamefont
  {de~la Torre}}\ and\ \bibinfo {author} {\bibfnamefont {J.}~\bibnamefont
  {Burguete}},\ }\bibfield  {title} {\enquote {\bibinfo {title} {Slow dynamics
  in a turbulent von k\'arm\'an swirling flow},}\ }\href@noop {} {\bibfield
  {journal} {\bibinfo  {journal} {Phy. Rev. Lett.}\ }\textbf {\bibinfo {volume}
  {99}},\ \bibinfo {pages} {054101} (\bibinfo {year} {2007})}\BibitemShut
  {NoStop}%
\bibitem [{\citenamefont {Thual}\ \emph {et~al.}(2014)\citenamefont {Thual},
  \citenamefont {Majda},\ and\ \citenamefont {Stechmann}}]{TMS14}%
  \BibitemOpen
  \bibfield  {author} {\bibinfo {author} {\bibfnamefont {S.}~\bibnamefont
  {Thual}}, \bibinfo {author} {\bibfnamefont {A.J.}\ \bibnamefont {Majda}}, \
  and\ \bibinfo {author} {\bibfnamefont {S.N.}\ \bibnamefont {Stechmann}},\
  }\bibfield  {title} {\enquote {\bibinfo {title} {A stochastic skeleton model
  for the mjo},}\ }\href@noop {} {\bibfield  {journal} {\bibinfo  {journal} {J.
  of the Atmospheric Sciences}\ }\textbf {\bibinfo {volume} {71}},\ \bibinfo
  {pages} {697} (\bibinfo {year} {2014})}\BibitemShut {NoStop}%
\bibitem [{\citenamefont {Rigas}\ \emph {et~al.}(2015)\citenamefont {Rigas},
  \citenamefont {Morgans}, , \citenamefont {Brackston},\ and\ \citenamefont
  {Morrison}}]{RMBM15}%
  \BibitemOpen
  \bibfield  {author} {\bibinfo {author} {\bibfnamefont {Georgios}\
  \bibnamefont {Rigas}}, \bibinfo {author} {\bibfnamefont {Aimee~S.}\
  \bibnamefont {Morgans}}, , \bibinfo {author} {\bibfnamefont {R.D.}\
  \bibnamefont {Brackston}}, \ and\ \bibinfo {author} {\bibfnamefont
  {Jonathan~F.}\ \bibnamefont {Morrison}},\ }\bibfield  {title} {\enquote
  {\bibinfo {title} {Diffusive dynamics and stochastic models of turbulent
  axisymmetric wakes},}\ }\href@noop {} {\bibfield  {journal} {\bibinfo
  {journal} {J. Fluid Mech.}\ }\textbf {\bibinfo {volume} {778}},\ \bibinfo
  {pages} {R2--1--10} (\bibinfo {year} {2015})}\BibitemShut {NoStop}%
\bibitem [{\citenamefont {Sreenivasan}\ \emph {et~al.}(2002)\citenamefont
  {Sreenivasan}, \citenamefont {Bershadski},\ and\ \citenamefont
  {Niemela}}]{SBN02}%
  \BibitemOpen
  \bibfield  {author} {\bibinfo {author} {\bibfnamefont {K.~R.}\ \bibnamefont
  {Sreenivasan}}, \bibinfo {author} {\bibfnamefont {A.}~\bibnamefont
  {Bershadski}}, \ and\ \bibinfo {author} {\bibfnamefont {J.J.}\ \bibnamefont
  {Niemela}},\ }\bibfield  {title} {\enquote {\bibinfo {title} {Mean wind and
  its reversals in thermal convection},}\ }\href@noop {} {\bibfield  {journal}
  {\bibinfo  {journal} {Phys. Rev. E}\ }\textbf {\bibinfo {volume} {65}},\
  \bibinfo {pages} {056306} (\bibinfo {year} {2002})}\BibitemShut {NoStop}%
\bibitem [{\citenamefont {Benzi}(2005)}]{Be05}%
  \BibitemOpen
  \bibfield  {author} {\bibinfo {author} {\bibfnamefont {R.}~\bibnamefont
  {Benzi}},\ }\bibfield  {title} {\enquote {\bibinfo {title} {Flow reversal in
  a simple dynamical model of turbulence},}\ }\href@noop {} {\bibfield
  {journal} {\bibinfo  {journal} {Phys. Rev. Lett.}\ }\textbf {\bibinfo
  {volume} {95}},\ \bibinfo {pages} {024502--1--4} (\bibinfo {year}
  {2005})}\BibitemShut {NoStop}%
\bibitem [{\citenamefont {{{Fontenele Araujo}}}\ \emph
  {et~al.}(2005)\citenamefont {{{Fontenele Araujo}}}, \citenamefont
  {Grossmann},\ and\ \citenamefont {Lohse}}]{FGL05}%
  \BibitemOpen
  \bibfield  {author} {\bibinfo {author} {\bibfnamefont {F.}~\bibnamefont
  {{{Fontenele Araujo}}}}, \bibinfo {author} {\bibfnamefont {S.}~\bibnamefont
  {Grossmann}}, \ and\ \bibinfo {author} {\bibfnamefont {D.}~\bibnamefont
  {Lohse}},\ }\bibfield  {title} {\enquote {\bibinfo {title} {Wind reversals in
  turbulent {{Rayleigh-B\'enard}} convection},}\ }\href@noop {} {\bibfield
  {journal} {\bibinfo  {journal} {Phys. Rev. Lett.}\ }\textbf {\bibinfo
  {volume} {95}},\ \bibinfo {pages} {084502} (\bibinfo {year}
  {2005})}\BibitemShut {NoStop}%
\bibitem [{\citenamefont {Chandra}\ and\ \citenamefont {Verma}(2011)}]{CV11}%
  \BibitemOpen
  \bibfield  {author} {\bibinfo {author} {\bibfnamefont {M.}~\bibnamefont
  {Chandra}}\ and\ \bibinfo {author} {\bibfnamefont {M.~K.}\ \bibnamefont
  {Verma}},\ }\bibfield  {title} {\enquote {\bibinfo {title} {Dynamics and
  symmetries of flow reversals in turbulent convection},}\ }\href@noop {}
  {\bibfield  {journal} {\bibinfo  {journal} {Physical Review E}\ }\textbf
  {\bibinfo {volume} {83}},\ \bibinfo {pages} {067303} (\bibinfo {year}
  {2011})}\BibitemShut {NoStop}%
\bibitem [{\citenamefont {Podvin}\ and\ \citenamefont {Sergent}(2015)}]{PS15}%
  \BibitemOpen
  \bibfield  {author} {\bibinfo {author} {\bibfnamefont {B.}~\bibnamefont
  {Podvin}}\ and\ \bibinfo {author} {\bibfnamefont {A.}~\bibnamefont
  {Sergent}},\ }\bibfield  {title} {\enquote {\bibinfo {title} {A large-scale
  investigation of wind reversal in a square rayleigh- b{\'e}nard cell},}\
  }\href@noop {} {\bibfield  {journal} {\bibinfo  {journal} {J. Fluid Mech.}\
  }\textbf {\bibinfo {volume} {766}},\ \bibinfo {pages} {172--201} (\bibinfo
  {year} {2015})}\BibitemShut {NoStop}%
\bibitem [{\citenamefont {Brown}\ and\ \citenamefont
  {Ahlers}(2008{\natexlab{a}})}]{BA08a}%
  \BibitemOpen
  \bibfield  {author} {\bibinfo {author} {\bibfnamefont {E.}~\bibnamefont
  {Brown}}\ and\ \bibinfo {author} {\bibfnamefont {G.}~\bibnamefont {Ahlers}},\
  }\bibfield  {title} {\enquote {\bibinfo {title} {A model of diffusion in a
  potential well for the dynamics of the large-scale circulation in turbulent
  {{Rayleigh-B\'enard}} convection},}\ }\href@noop {} {\bibfield  {journal}
  {\bibinfo  {journal} {Phys. Fluids}\ }\textbf {\bibinfo {volume} {20}},\
  \bibinfo {pages} {075101--1--16} (\bibinfo {year}
  {2008}{\natexlab{a}})}\BibitemShut {NoStop}%
\bibitem [{\citenamefont {Brown}\ and\ \citenamefont
  {Ahlers}(2008{\natexlab{b}})}]{BA08b}%
  \BibitemOpen
  \bibfield  {author} {\bibinfo {author} {\bibfnamefont {E.}~\bibnamefont
  {Brown}}\ and\ \bibinfo {author} {\bibfnamefont {G.}~\bibnamefont {Ahlers}},\
  }\bibfield  {title} {\enquote {\bibinfo {title} {Azimuthal asymmetries of the
  large-scale circulation in turbulent {{Rayleigh-B\'enard}} convection},}\
  }\href@noop {} {\bibfield  {journal} {\bibinfo  {journal} {Phys. Fluids}\
  }\textbf {\bibinfo {volume} {20}},\ \bibinfo {pages} {105105--1--15}
  (\bibinfo {year} {2008}{\natexlab{b}})}\BibitemShut {NoStop}%
\bibitem [{\citenamefont {Zhong}\ \emph {et~al.}(2017)\citenamefont {Zhong},
  \citenamefont {Li},\ and\ \citenamefont {Wang}}]{ZLW17}%
  \BibitemOpen
  \bibfield  {author} {\bibinfo {author} {\bibfnamefont {J.-Q.}\ \bibnamefont
  {Zhong}}, \bibinfo {author} {\bibfnamefont {H.-M.}\ \bibnamefont {Li}}, \
  and\ \bibinfo {author} {\bibfnamefont {X.-Y.}\ \bibnamefont {Wang}},\
  }\bibfield  {title} {\enquote {\bibinfo {title} {Enhanced azimuthal rotation
  of the large-scale flow through stochastic cessations in turbulent rotating
  convection with large rossby numbers},}\ }\href@noop {} {\bibfield  {journal}
  {\bibinfo  {journal} {Phys. Rev. Fluids.}\ }\textbf {\bibinfo {volume} {2}},\
  \bibinfo {pages} {044602} (\bibinfo {year} {2017})}\BibitemShut {NoStop}%
\bibitem [{\citenamefont {Sterl}\ \emph {et~al.}(2016)\citenamefont {Sterl},
  \citenamefont {Li},\ and\ \citenamefont {Zhong}}]{SLZ16}%
  \BibitemOpen
  \bibfield  {author} {\bibinfo {author} {\bibfnamefont {S.}~\bibnamefont
  {Sterl}}, \bibinfo {author} {\bibfnamefont {H.-M.}\ \bibnamefont {Li}}, \
  and\ \bibinfo {author} {\bibfnamefont {J.-Q.}\ \bibnamefont {Zhong}},\
  }\bibfield  {title} {\enquote {\bibinfo {title} {Dynamical and statistical
  phenomena of circulation and heat transfer in periodically forced rotating
  turbulent rayleigh-b{'e}nard convection},}\ }\href@noop {} {\bibfield
  {journal} {\bibinfo  {journal} {Phys. Rev. Fluids.}\ }\textbf {\bibinfo
  {volume} {1}},\ \bibinfo {pages} {084401} (\bibinfo {year}
  {2016})}\BibitemShut {NoStop}%
\bibitem [{\citenamefont {Ji}\ and\ \citenamefont
  {Brown}(2020{\natexlab{b}})}]{JB20a}%
  \BibitemOpen
  \bibfield  {author} {\bibinfo {author} {\bibfnamefont {D.}~\bibnamefont
  {Ji}}\ and\ \bibinfo {author} {\bibfnamefont {E.}~\bibnamefont {Brown}},\
  }\bibfield  {title} {\enquote {\bibinfo {title} {Low-dimensional model of the
  large-scale circulation of turbulent {{Rayleigh-B\'enard}} convection in a
  cubic container},}\ }\href@noop {} {\bibfield  {journal} {\bibinfo  {journal}
  {Phys. Rev. Fluids}\ }\textbf {\bibinfo {volume} {5}},\ \bibinfo {pages}
  {064606} (\bibinfo {year} {2020}{\natexlab{b}})}\BibitemShut {NoStop}%
\bibitem [{\citenamefont {Assaf}\ \emph {et~al.}(2011)\citenamefont {Assaf},
  \citenamefont {Angheluta},\ and\ \citenamefont {N.Goldenfeld}}]{AAG11}%
  \BibitemOpen
  \bibfield  {author} {\bibinfo {author} {\bibfnamefont {M.}~\bibnamefont
  {Assaf}}, \bibinfo {author} {\bibfnamefont {L.}~\bibnamefont {Angheluta}}, \
  and\ \bibinfo {author} {\bibnamefont {N.Goldenfeld}},\ }\bibfield  {title}
  {\enquote {\bibinfo {title} {Rare fluctuations and large-scale circulation
  cessations in turbulent convection},}\ }\href@noop {} {\bibfield  {journal}
  {\bibinfo  {journal} {Phys. Rev. Lett.}\ }\textbf {\bibinfo {volume} {107}},\
  \bibinfo {pages} {044502} (\bibinfo {year} {2011})}\BibitemShut {NoStop}%
\bibitem [{\citenamefont {Bailon-Cuba}\ \emph {et~al.}(2010)\citenamefont
  {Bailon-Cuba}, \citenamefont {Emran},\ and\ \citenamefont
  {Schumacher}}]{BES10}%
  \BibitemOpen
  \bibfield  {author} {\bibinfo {author} {\bibfnamefont {J.}~\bibnamefont
  {Bailon-Cuba}}, \bibinfo {author} {\bibfnamefont {M.S.}\ \bibnamefont
  {Emran}}, \ and\ \bibinfo {author} {\bibfnamefont {J.}~\bibnamefont
  {Schumacher}},\ }\bibfield  {title} {\enquote {\bibinfo {title} {Aspect ratio
  dependence of heat transfer and large-scale flow in turbulent convection},}\
  }\href@noop {} {\bibfield  {journal} {\bibinfo  {journal} {J. Fluid Mech.}\
  }\textbf {\bibinfo {volume} {655}},\ \bibinfo {pages} {152--173} (\bibinfo
  {year} {2010})}\BibitemShut {NoStop}%
\bibitem [{\citenamefont {van~der Pel}\ \emph {et~al.}(2012)\citenamefont
  {van~der Pel}, \citenamefont {Stevens}, \citenamefont {Sugiyama},\ and\
  \citenamefont {Lohse}}]{PSSL12}%
  \BibitemOpen
  \bibfield  {author} {\bibinfo {author} {\bibfnamefont {E.~P.}\ \bibnamefont
  {van~der Pel}}, \bibinfo {author} {\bibfnamefont {R.~J. A.~M.}\ \bibnamefont
  {Stevens}}, \bibinfo {author} {\bibfnamefont {K.}~\bibnamefont {Sugiyama}}, \
  and\ \bibinfo {author} {\bibfnamefont {D.}~\bibnamefont {Lohse}},\ }\bibfield
   {title} {\enquote {\bibinfo {title} {Flow states in two-dimensional
  rayleigh- b{\'e}nard convection as a function of aspect-ratio and rayleigh
  number},}\ }\href@noop {} {\bibfield  {journal} {\bibinfo  {journal} {Physics
  of Fluids}\ }\textbf {\bibinfo {volume} {24}},\ \bibinfo {pages}
  {085104--1--12} (\bibinfo {year} {2012})}\BibitemShut {NoStop}%
\bibitem [{\citenamefont {P{\'e}rez-Garc{\'i}a}\ \emph
  {et~al.}(2004)\citenamefont {P{\'e}rez-Garc{\'i}a}, \citenamefont {Madruga},
  \citenamefont {Echebarria}, \citenamefont {Lebon},\ and\ \citenamefont
  {Burguete}}]{PMELB04}%
  \BibitemOpen
  \bibfield  {author} {\bibinfo {author} {\bibfnamefont {Carlos}\ \bibnamefont
  {P{\'e}rez-Garc{\'i}a}}, \bibinfo {author} {\bibfnamefont {Santiago}\
  \bibnamefont {Madruga}}, \bibinfo {author} {\bibfnamefont {Blas}\
  \bibnamefont {Echebarria}}, \bibinfo {author} {\bibfnamefont {Georgy}\
  \bibnamefont {Lebon}}, \ and\ \bibinfo {author} {\bibfnamefont {Javier}\
  \bibnamefont {Burguete}},\ }\bibfield  {title} {\enquote {\bibinfo {title}
  {Hydrothermal waves and corotating rolls in laterally heated convection in
  simple liquids},}\ }\href@noop {} {\bibfield  {journal} {\bibinfo  {journal}
  {J. Non-Equilibrium Thermodynamics}\ }\textbf {\bibinfo {volume} {29}},\
  \bibinfo {pages} {377--388} (\bibinfo {year} {2004})}\BibitemShut {NoStop}%
\bibitem [{\citenamefont {Mercader}\ \emph {et~al.}(2019)\citenamefont
  {Mercader}, \citenamefont {Batiste}, \citenamefont {Alonso},\ and\
  \citenamefont {Knobloch}}]{MBAK19}%
  \BibitemOpen
  \bibfield  {author} {\bibinfo {author} {\bibfnamefont {I.}~\bibnamefont
  {Mercader}}, \bibinfo {author} {\bibfnamefont {O.}~\bibnamefont {Batiste}},
  \bibinfo {author} {\bibfnamefont {A.}~\bibnamefont {Alonso}}, \ and\ \bibinfo
  {author} {\bibfnamefont {E.}~\bibnamefont {Knobloch}},\ }\bibfield  {title}
  {\enquote {\bibinfo {title} {Effect of small inclination on binary convection
  in elongated rectangular cells},}\ }\href@noop {} {\bibfield  {journal}
  {\bibinfo  {journal} {Phys. Rev. E}\ }\textbf {\bibinfo {volume} {99}},\
  \bibinfo {pages} {023113--1--17} (\bibinfo {year} {2019})}\BibitemShut
  {NoStop}%
\bibitem [{\citenamefont {Zwirner}\ \emph {et~al.}(2020)\citenamefont
  {Zwirner}, \citenamefont {Tilgner},\ and\ \citenamefont {Shishkina}}]{ZTS20}%
  \BibitemOpen
  \bibfield  {author} {\bibinfo {author} {\bibfnamefont {Lukas}\ \bibnamefont
  {Zwirner}}, \bibinfo {author} {\bibfnamefont {Andreas}\ \bibnamefont
  {Tilgner}}, \ and\ \bibinfo {author} {\bibfnamefont {Olga}\ \bibnamefont
  {Shishkina}},\ }\bibfield  {title} {\enquote {\bibinfo {title} {Elliptical
  instability and multiple-roll flow modes of the large-scale circulation in
  confined turbulent rayleigh-b{\'e}nard convection},}\ }\href@noop {}
  {\bibfield  {journal} {\bibinfo  {journal} {Phys. Rev. Lett.}\ }\textbf
  {\bibinfo {volume} {125}},\ \bibinfo {pages} {054502} (\bibinfo {year}
  {2020})}\BibitemShut {NoStop}%
\bibitem [{\citenamefont {Xie}\ and\ \citenamefont {Xia}(2013)}]{XX13}%
  \BibitemOpen
  \bibfield  {author} {\bibinfo {author} {\bibfnamefont {Yi-Chao}\ \bibnamefont
  {Xie}}\ and\ \bibinfo {author} {\bibfnamefont {Ke-Qing}\ \bibnamefont
  {Xia}},\ }\bibfield  {title} {\enquote {\bibinfo {title} {Dynamics and flow
  coupling in two-layer turbulent thermal convection},}\ }\href@noop {}
  {\bibfield  {journal} {\bibinfo  {journal} {J. Fluid Mech.}\ }\textbf
  {\bibinfo {volume} {728}} (\bibinfo {year} {2013})}\BibitemShut {NoStop}%
\bibitem [{\citenamefont {Gao}\ \emph {et~al.}(2013)\citenamefont {Gao},
  \citenamefont {Sargent}, \citenamefont {Podvin}, \citenamefont {Xin},
  \citenamefont {Qu{\'e}r{\'e}},\ and\ \citenamefont {Tuckerman}}]{GSPXQT13}%
  \BibitemOpen
  \bibfield  {author} {\bibinfo {author} {\bibfnamefont {Zhenlan}\ \bibnamefont
  {Gao}}, \bibinfo {author} {\bibfnamefont {Anne}\ \bibnamefont {Sargent}},
  \bibinfo {author} {\bibfnamefont {Berengere}\ \bibnamefont {Podvin}},
  \bibinfo {author} {\bibfnamefont {Shihe}\ \bibnamefont {Xin}}, \bibinfo
  {author} {\bibfnamefont {Patrick~Le}\ \bibnamefont {Qu{\'e}r{\'e}}}, \ and\
  \bibinfo {author} {\bibfnamefont {Laurette~S.}\ \bibnamefont {Tuckerman}},\
  }\bibfield  {title} {\enquote {\bibinfo {title} {Transition to chaos of
  natural convection between two infinite differentially heated vertical
  plates},}\ }\href@noop {} {\bibfield  {journal} {\bibinfo  {journal} {Phys.
  Rev. E}\ }\textbf {\bibinfo {volume} {88}},\ \bibinfo {pages} {023010}
  (\bibinfo {year} {2013})}\BibitemShut {NoStop}%
\bibitem [{\citenamefont {Petry}\ and\ \citenamefont {Busse}(2003)}]{PB03}%
  \BibitemOpen
  \bibfield  {author} {\bibinfo {author} {\bibfnamefont {M.}~\bibnamefont
  {Petry}}\ and\ \bibinfo {author} {\bibfnamefont {F.~H.}\ \bibnamefont
  {Busse}},\ }\bibfield  {title} {\enquote {\bibinfo {title} {Theoretical study
  of flow coupling mechanisms in two-layer rayleigh-b{\'e}nard convection},}\
  }\href@noop {} {\bibfield  {journal} {\bibinfo  {journal} {Phys. Rev. E}\
  }\textbf {\bibinfo {volume} {68}},\ \bibinfo {pages} {016305--1--9} (\bibinfo
  {year} {2003})}\BibitemShut {NoStop}%
\bibitem [{\citenamefont {Brown}\ and\ \citenamefont
  {Ahlers}(2006{\natexlab{b}})}]{BA06b}%
  \BibitemOpen
  \bibfield  {author} {\bibinfo {author} {\bibfnamefont {E.}~\bibnamefont
  {Brown}}\ and\ \bibinfo {author} {\bibfnamefont {G.}~\bibnamefont {Ahlers}},\
  }\bibfield  {title} {\enquote {\bibinfo {title} {Effect of the earth's
  coriolis force on turbulent {Rayleigh-B\'enard} convection in the
  laboratory},}\ }\href@noop {} {\bibfield  {journal} {\bibinfo  {journal}
  {Phys. Fluids}\ }\textbf {\bibinfo {volume} {18}},\ \bibinfo {pages}
  {125108--1--15} (\bibinfo {year} {2006}{\natexlab{b}})}\BibitemShut {NoStop}%
\bibitem [{\citenamefont {Akashi}\ \emph {et~al.}(2022)\citenamefont {Akashi},
  \citenamefont {Yanagisawa}, \citenamefont {Sakuraba}, \citenamefont
  {Schindler}, \citenamefont {Horn}, \citenamefont {Vogt},\ and\ \citenamefont
  {Eckert}}]{AYSSHVE22}%
  \BibitemOpen
  \bibfield  {author} {\bibinfo {author} {\bibfnamefont {Megumi}\ \bibnamefont
  {Akashi}}, \bibinfo {author} {\bibfnamefont {Takatoshi}\ \bibnamefont
  {Yanagisawa}}, \bibinfo {author} {\bibfnamefont {Ataru}\ \bibnamefont
  {Sakuraba}}, \bibinfo {author} {\bibfnamefont {Felix}\ \bibnamefont
  {Schindler}}, \bibinfo {author} {\bibfnamefont {Susanne}\ \bibnamefont
  {Horn}}, \bibinfo {author} {\bibfnamefont {Tobias}\ \bibnamefont {Vogt}}, \
  and\ \bibinfo {author} {\bibfnamefont {Sven}\ \bibnamefont {Eckert}},\
  }\bibfield  {title} {\enquote {\bibinfo {title} {Jump rope vortex flow in
  liquid metal rayleigh-b{\'e}nard convection in a cuboid container of aspect
  ratio $\gamma = 5$},}\ }\href@noop {} {\bibfield  {journal} {\bibinfo
  {journal} {J. Fluid Mech.}\ }\textbf {\bibinfo {volume} {932}},\ \bibinfo
  {pages} {A27--1--26} (\bibinfo {year} {2022})}\BibitemShut {NoStop}%
\bibitem [{\citenamefont {Urban}\ \emph {et~al.}(2023)\citenamefont {Urban},
  \citenamefont {Kr{\'a}l{\'i}k}, \citenamefont {Musilov{\'a}}, \citenamefont
  {Schmoranzer},\ and\ \citenamefont {Skrbek}}]{UKMSS22}%
  \BibitemOpen
  \bibfield  {author} {\bibinfo {author} {\bibfnamefont {P.}~\bibnamefont
  {Urban}}, \bibinfo {author} {\bibfnamefont {T.}~\bibnamefont
  {Kr{\'a}l{\'i}k}}, \bibinfo {author} {\bibfnamefont {V.}~\bibnamefont
  {Musilov{\'a}}}, \bibinfo {author} {\bibfnamefont {D.}~\bibnamefont
  {Schmoranzer}}, \ and\ \bibinfo {author} {\bibfnamefont {L.}~\bibnamefont
  {Skrbek}},\ }\bibfield  {title} {\enquote {\bibinfo {title} {Propagation and
  interference of thermal waves in turbulent thermal convection},}\ }\href@noop
  {} {\bibfield  {journal} {\bibinfo  {journal} {Phys. Rev. Fluids}\ }\textbf
  {\bibinfo {volume} {8}},\ \bibinfo {pages} {063501} (\bibinfo {year}
  {2023})}\BibitemShut {NoStop}%
\bibitem [{\citenamefont {Brown}\ \emph {et~al.}(2005)\citenamefont {Brown},
  \citenamefont {Funfschilling}, \citenamefont {Nikolaenko},\ and\
  \citenamefont {Ahlers}}]{BFNA05}%
  \BibitemOpen
  \bibfield  {author} {\bibinfo {author} {\bibfnamefont {E.}~\bibnamefont
  {Brown}}, \bibinfo {author} {\bibfnamefont {D.}~\bibnamefont
  {Funfschilling}}, \bibinfo {author} {\bibfnamefont {A.}~\bibnamefont
  {Nikolaenko}}, \ and\ \bibinfo {author} {\bibfnamefont {G.}~\bibnamefont
  {Ahlers}},\ }\bibfield  {title} {\enquote {\bibinfo {title} {Heat transport
  by turbulent {{Rayleigh-B\'enard}} convection: Effect of finite top- and
  bottom conductivity},}\ }\href@noop {} {\bibfield  {journal} {\bibinfo
  {journal} {Phys. Fluids}\ }\textbf {\bibinfo {volume} {17}},\ \bibinfo
  {pages} {075108} (\bibinfo {year} {2005})}\BibitemShut {NoStop}%
\bibitem [{\citenamefont {Funfschiling}\ \emph {et~al.}(2005)\citenamefont
  {Funfschiling}, \citenamefont {Brown}, \citenamefont {Nikolaenko},\ and\
  \citenamefont {Ahlers}}]{FBNA05}%
  \BibitemOpen
  \bibfield  {author} {\bibinfo {author} {\bibfnamefont {D.}~\bibnamefont
  {Funfschiling}}, \bibinfo {author} {\bibfnamefont {E.}~\bibnamefont {Brown}},
  \bibinfo {author} {\bibfnamefont {A.}~\bibnamefont {Nikolaenko}}, \ and\
  \bibinfo {author} {\bibfnamefont {G.}~\bibnamefont {Ahlers}},\ }\bibfield
  {title} {\enquote {\bibinfo {title} {Heat transport by turbulent
  {{Rayleigh-B\'enard}} convection in cylindrical cells with aspect ratio one
  and larger},}\ }\href@noop {} {\bibfield  {journal} {\bibinfo  {journal} {J.
  Fluid Mech.}\ }\textbf {\bibinfo {volume} {536}},\ \bibinfo {pages}
  {145--154} (\bibinfo {year} {2005})}\BibitemShut {NoStop}%
\bibitem [{Note1()}]{Note1}%
  \BibitemOpen
  \bibinfo {note} {For counter-offset states with $\Delta T_m /\delta _0
  \mathrel {\mathop {\sim }\limits ^{>}} 0.1$, $\theta _{p,1}$ and $\theta
  _{p,2}$ shift away from the corner, so that the forcing is not necessarily
  centered on the peak of $p(\theta _i)$. In these cases the apparent value of
  $b$ drops significantly even for small $\Delta T_m$. We attempted to add an
  extra constant offset term to be fit in the linear regression to fix this.
  While the values of $b$ do not systematically drop with small $\Delta T_m$
  with a constant offset in the linear regression, the extra regression
  parameter causes uncertainties to increase to be comparable to fit
  parameters. We do not report results for $b$ with $\Delta T_m \not =0$, and
  caution about this limitation of the algorithm for states where $\theta
  _{p,1}$ is offset from $\theta _{p,2}$.}\BibitemShut {Stop}%
\bibitem [{Note2()}]{Note2}%
  \BibitemOpen
  \bibinfo {note} {This value of $\omega _r^2$ is about half of the value
  reported in \cite {JB20a}, which was obtained less directly through
  measurements of the mean-square displacement of $\protect \mathaccentV
  {dot}05F\theta _0$ divided by the variance of the probability distribution of
  $\theta _0$}\BibitemShut {NoStop}%
\bibitem [{\citenamefont {Kaczorowski}\ and\ \citenamefont {Xia}(2013)}]{KX13}%
  \BibitemOpen
  \bibfield  {author} {\bibinfo {author} {\bibfnamefont {M.}~\bibnamefont
  {Kaczorowski}}\ and\ \bibinfo {author} {\bibfnamefont {K.-Q.}\ \bibnamefont
  {Xia}},\ }\bibfield  {title} {\enquote {\bibinfo {title} {Turbulent flow in
  the bulk of {{Rayleigh-B\'enard}} convection: small-scale properties in a
  cubic cell},}\ }\href@noop {} {\bibfield  {journal} {\bibinfo  {journal} {J.
  Fluid Mech.}\ }\textbf {\bibinfo {volume} {722}},\ \bibinfo {pages}
  {596--617} (\bibinfo {year} {2013})}\BibitemShut {NoStop}%
\bibitem [{\citenamefont {Ji}(2019)}]{Dandanthesis}%
  \BibitemOpen
  \bibfield  {author} {\bibinfo {author} {\bibfnamefont {Dandan}\ \bibnamefont
  {Ji}},\ }\emph {\bibinfo {title} {Modeling of the dynamics of large-scale
  coherent structures in the system of Rayliegh-B{\'e}nard convection}},\
  \href@noop {} {Ph.D. thesis},\ \bibinfo  {school} {Yale University} (\bibinfo
  {year} {2019})\BibitemShut {NoStop}%
\bibitem [{\citenamefont {Chill\`a}\ \emph {et~al.}(2004)\citenamefont
  {Chill\`a}, \citenamefont {Rastello}, \citenamefont {Chaumat},\ and\
  \citenamefont {Castaing}}]{CRCC04}%
  \BibitemOpen
  \bibfield  {author} {\bibinfo {author} {\bibfnamefont {F.}~\bibnamefont
  {Chill\`a}}, \bibinfo {author} {\bibfnamefont {M.}~\bibnamefont {Rastello}},
  \bibinfo {author} {\bibfnamefont {S.}~\bibnamefont {Chaumat}}, \ and\
  \bibinfo {author} {\bibfnamefont {B.}~\bibnamefont {Castaing}},\ }\bibfield
  {title} {\enquote {\bibinfo {title} {Long relaxation times and tilt
  sensitivity in {{Rayleigh-B\'enard}} turbulence},}\ }\href@noop {} {\bibfield
   {journal} {\bibinfo  {journal} {Euro. Phys. J. B}\ }\textbf {\bibinfo
  {volume} {40}},\ \bibinfo {pages} {223--227} (\bibinfo {year}
  {2004})}\BibitemShut {NoStop}%
\bibitem [{\citenamefont {Xi}\ and\ \citenamefont {Xia}(2008)}]{XX08}%
  \BibitemOpen
  \bibfield  {author} {\bibinfo {author} {\bibfnamefont {H.-D.}\ \bibnamefont
  {Xi}}\ and\ \bibinfo {author} {\bibfnamefont {K.-Q.}\ \bibnamefont {Xia}},\
  }\bibfield  {title} {\enquote {\bibinfo {title} {Azimuthal motion,
  reorientation, cessation, and reversal of the large-scale circulation in
  turbulent thermal convection: A comparative study in aspect ratio one and
  one-half geometries},}\ }\href@noop {} {\bibfield  {journal} {\bibinfo
  {journal} {Phys. Rev. E}\ }\textbf {\bibinfo {volume} {78}},\ \bibinfo
  {pages} {036326--1--11} (\bibinfo {year} {2008})}\BibitemShut {NoStop}%
\bibitem [{\citenamefont {Brown}\ and\ \citenamefont
  {Ahlers}(2007{\natexlab{b}})}]{BA07b}%
  \BibitemOpen
  \bibfield  {author} {\bibinfo {author} {\bibfnamefont {E.}~\bibnamefont
  {Brown}}\ and\ \bibinfo {author} {\bibfnamefont {G.}~\bibnamefont {Ahlers}},\
  }\bibfield  {title} {\enquote {\bibinfo {title} {Temperature gradients, and
  search for non-boussinesq effects, in the interior of turbulent
  {{Rayleigh-B\'enard}} convection},}\ }\href@noop {} {\bibfield  {journal}
  {\bibinfo  {journal} {Euro. Phys. Lett.}\ }\textbf {\bibinfo {volume} {80}},\
  \bibinfo {pages} {14001--1--6} (\bibinfo {year}
  {2007}{\natexlab{b}})}\BibitemShut {NoStop}%
\end{thebibliography}

\end{document}